\documentclass[structabstract]{aa} 
\usepackage{graphicx}
\usepackage{txfonts}
\usepackage{natbib}
\usepackage{longtable}
\usepackage{lscape}
\usepackage{multirow}
\usepackage{color}

\bibpunct{(}{)}{;}{a}{}{,}

\newcommand{\kms}{km\,s$^{-1}$}

\begin{document}
\title{A 1.3 cm line survey toward IRC +10216}

\author{Y. Gong\inst{1,2,3}
\and
C. Henkel\inst{2,4}
\and\
S. Spezzano\inst{5}
\and
S. Thorwirth\inst{5}
\and
K.~M. Menten\inst{2}
\and
F. Wyrowski\inst{2}
\and 
R.~Q. Mao\inst{1}
\and
B. Klein\inst{2,6}
}
\institute{Purple Mountain Observatory \& Key Laboratory for Radio Astronomy, Chinese Academy of Sciences, 2 West Beijing Road, 210008 Nanjing, PR China
\and Max-Planck Institute f\"ur Radioastronomy, Auf dem H\"ugel 69, 53121 Bonn, Germany
\and University of Chinese Academy of Sciences, No. 19A Yuquan Road, 100049 Beijing, PR China
\and Astronomy Department, King Abdulaziz University, P.O. Box 80203, Jeddah 21589, Saudi Arabia
\and I. Physikalisches Institut, Universit{\"a}t zu K{\"o}ln, 50937 K{\"o}ln, Germany
\and University of Applied Sciences Bonn-Rhein-Sieg, Grantham-Allee 20, 53757 Sankt Augustin, Germany
}
    
   \date{}

 \abstract{IRC +10216 is the prototypical carbon star exhibiting an extended molecular circumstellar envelope. Its spectral properties are therefore the template for an entire class of objects.}
{The main goal is to systematically study the $\lambda$ $\sim$1.3\,cm spectral line characteristics of IRC +10216.}
{We carried out a spectral line survey with the Effelsberg-100 m telescope toward IRC +10216. It covers the frequency range between 17.8~GHz and 26.3~GHz (K-band).}
{In the circumstellar shell of IRC +10216, we find 78 spectral lines, among which 12 remain unidentified. The identified lines are assigned to 18 different molecules and radicals. A total of 23 lines from species known to exist in this envelope are detected for the first time outside the Solar System and there are additional 20 lines first detected in IRC +10216. The potential orgin of ``U'' lines is also discussed. Assuming local thermodynamic equilibrium (LTE), we then determine rotational temperatures and column densities of 17 detected molecules. Molecular abundances relative to H$_{2}$ are also estimated. A non-LTE analysis of NH$_{3}$ shows that the bulk of its emission arises from the inner envelope with a kinetic temperature of 70$\pm$20~K. Evidence for NH$_{3}$ emitting gas with higher kinetic temperature is also obtained, and potential abundance differences between various $^{13}$C-bearing isotopologues of HC$_{5}$N are evaluated. Overall, the isotopic $^{12}$C/$^{13}$C ratio is estimated to be 49$\pm$9. Finally, a comparison of detected molecules in the $\lambda$ $\sim$1.3\,cm range with the dark cloud TMC-1 indicates that silicate-bearing molecules are more predominant in IRC +10216.}
{}
\keywords{astrochemistry -- circumstellar matter: molecules -- stars: individual (IRC +10216) -- radio lines: stars}

 \maketitle

\section{Introduction}
Molecular line surveys are a powerful tool for analyzing both physical and chemical parameters of astronomical objects. In fact, they are the only means of obtaining a complete, unbiased view of the molecular inventory of a given source. However, despite the long history of powerful large radio telescopes, few unbiased frequency surveys have been conducted in the cm-wave regime. A 4.4~GHz bandwidth survey from 17.6 to 22.0~GHz toward the massive star-forming region W51 has been reported by \citet{1993ApJS...86..211B} showing many previously unidentified lines. The dark cloud TMC-1 has been observed in selected bands from 4-6 and 8-10 GHz \citep{2004ApJ...610..329K}, as well as at 9-50~GHz \citep{2004PASJ...56...69K}. More recently, some systematic observations toward the galactic center have also been reported by \citet{2008ApJ...675L..85R} in the course of their GBT PRIMOS project\footnote[1]{http://www.cv.nrao.edu/$\sim$aremijan/PRIMOS/index.html}.

Spectroscopic features that are expected at long wavelengths (i.e. in the cm-wave regime) include recombination lines from H{\scriptsize I}, He{\scriptsize I}, and C{\scriptsize I}, which can be assigned in a straightforward manner. In addition, complex molecules have small rotational constants so that their lowest-energy transitions arise at cm-wavelenths. In this way, emission or absorption features from long carbon-chain molecules can be observed (e.g. HC$_{3}$N, HC$_{5}$N, HC$_{7}$N, HC$_{9}$N). Since systematic and highly sensitive surveys at these (low) frequencies that cover different sources have rarely been carried out, one can only speculate as to what else may be observed. The enormous potential of cm-wave astronomy for detecting new molecules in space has been highlighted in the recent past through GBT detections of complex organic molecules such as CH$_{2}$CHCHO (propenal), CH$_{3}$CH$_{2}$CHO (propanal), and CNCHO (cyanoformaldehyde)~\citep{2004ApJ...610L..21H,2008ApJ...675L..85R}. 

For this paper, we performed a 1.3 cm line survey toward IRC +10216, which is also known as CW Leonis. It is found to be the brightest object outside the Solar System at 5~$\mu$m \citep{1969ApJ...158L.133B}. It is a well-studied asymptotic giant branch (AGB) carbon star with a high mass loss rate (2.0~$\times~ 10^{-5}$M$_{\odot}$yr$^{-1}$) at a distance of $\sim$130\,pc~\citep{1997ApJ...483..913C,2012A&A...543A..73M}, which results in a dense circumstellar envelope (CSE). The CSE is the ideal target for studying gas-phase chemistry in a peculiar, namely carbon-rich environment. Existing surveys, summarized in Table~\ref{Tab:surveys}, cover parts of the frequency range from 4~GHz to 636.5~GHz for IRC +10216. These efforts have resulted in the discovery of over 70 species toward this object, including unusual carbon-chain molecules, metal cyanides such as MgNC, NaCN, and AlNC \citep{2000AAS..142..181C,2002ApJ...564L..45Z}, and even metal halides: NaCl, AlCl, KCl, and AlF \citep{1987A&A...183L..10C}. However, the frequency range between 18 and 26\,GHz is still largely unexplored. Here, we therefore systematically study the characteristics of IRC+10216 in the important $\lambda$ $\sim$1.3\,cm (K-band) spectral range, which is known to contain a number of lines from prominent molecules like H$_{2}$O, NH$_{3}$, and the cyanopolyynes.

\begin{table*}[!hbt]
\caption{Existing line surveys of IRC +10216.}\label{Tab:surveys}             
\normalsize
\centering                                      
\begin{tabular}{ccc}          
\hline\hline                        
Covered frequencies                 & Telescope       &  Reference                   \\
\hline
\multicolumn{3}{l}{IRC +10216}\\
\hline
4-6 GHz                            & Arecibo          &  \citet{2003ApJ...596..556A} \\
\textbf{17.8-26.3 GHz}         & \textbf{Effelsberg-100m} &  \textbf{This work}\\
28-50 GHz                           & Nobeyama-45m     &  \citet{1995PASJ...47..853K} \\
72-91 GHz                           & Onsala-20m      &  \citet{1984AA...130..227J,1985AAS...60..135J}  \\
129-172.5 GHz                       & IRAM-30m        &  \citet{2000AAS..142..181C} \\
131.2-160.3 GHz                     & ARO-12m         &  \citet{2008ApJS..177..275H} \\
215-285 GHz                         & SMT-10m         &  \citet{2010ApJS..190..348T} \\
219.5-245.5 GHz and 251.5-267.5 GHz   & SMT-10m         &  \citet{2008ApJS..177..275H} \\
293.9-354.8 GHz                     & SMA             &  \citet{2011ApJS..193...17P} \\
330-358 GHz                         & CSO-10m         &  \citet{1994ApJS...94..147G} \\                   
222.4-267.9 GHz and 339.6-364.6 GHz \tablefootmark{(1)}   & JCMT-15m        &  \citet{1992ApJS...83..363A} \\
554.5-636.5 GHz                     & Herschel/HIFI   &  \citet{2010AA...521L...8C}  \\
\hline
 \end{tabular}
 \tablefoot{\\
\tablefoottext{1}{Note that the frequency range from 222.4~GHz to 267.9 GHz was discontinuously covered.}\\
 }
 			        \normalsize
\end{table*}

\section{Observations and data reduction}\label{obs}
The observations were carried out in a position-switching mode with the primary focus $\lambda$ = 1.3 cm K-band receiver of the 100-m telescope at Effelsberg/Germany\footnote[2]{The 100-m telescope at Effelsberg is operated by the Max-Planck-Institut f{\"u}r Radioastronomie (MPIFR) on behalf of the Max-Planck-Gesellschaft (MPG).}, during 2012 in January, April, and 2013 in January, March, and May. On- and off-source integration times were two minutes per scan. The newly installed Fast Fourier Transform Spectrometer (FFTS) was used as backend. Each spectrometer, covering one of the two orthogonal linear polarizations with a bandwidth of 2 GHz provides 32768 channels, resulting in a channel spacing of 61 kHz, equivalent to 0.7 \kms at 26 GHz. The actual frequency and velocity resolutions are coarser by a factor of 1.16 \citep{2012A&A...542L...3K}. Several frequency setups were tuned to cover the entire frequency range from 17.8 GHz to 26.3 GHz with an overlap of at least 100 MHz between two adjacent frequency setups. 
Across the whole frequency range, the beamsize is 35\arcsec-50\arcsec ($\sim$40\arcsec at 23~GHz). The survey encompasses a total of $\sim$ 130 observing hours. The focus was checked every few hours, in particular after sunrise and sunset. Pointing was obtained every hour toward the quasar PG 0851+202 (OJ +287) and was found to be accurate to about 5\arcsec. Strong continuum sources (mostly 3C286) were also used to calibrate the spectral line flux according to their standard flux densities \citep{1994A&A...284..331O}. The typical rms noise is about $1-3$ mJy in $\sim$ 0.7 \kms\,wide channels. In view of the width of the IRC +10216 spectra, $\sim$ 30\,\kms, smoothed spectra typically reach rms noise levels of about 1~mJy. The conversion factor from Jy on a flux density scale ($S_{\nu}$) to K on a main beam brightness temperature scale ($T_{\rm mb}$) is $T_{\rm mb}/S_{\nu} \sim 1.5~{\rm K/Jy}$ at 22~GHz.

For the data reduction, the GILDAS\footnote[3]{http://www.iram.fr/IRAMFR/GILDAS} software package including CLASS and GREG was employed. During the data reduction, some inevitable defects occured at the edges of the spectra. Therefore 100 channels at each edge were excluded. Fourth- to tenth-order baselines were subtracted from each spectrum with 2000 to 3000 channels to avoid lines being truncated at the edge of subspectra, and then these subspectra were stitched together to resconstruct the complete spectrum. Since some of the subspectra suffer contamination from time variable radio frequency interference (RFI), the channels showing RFI signals have been ``flagged'', i.e. discarded from further analysis, resulting in some limited gaps in the averaged spectra. Furthermore, there were a few individual bad channels, which have also been eliminated. 

\section{Results}
\subsection{Line identifications}\label{line}
\begin{figure*}[!htbp]
\centering
\includegraphics[width = 0.8 \textwidth]{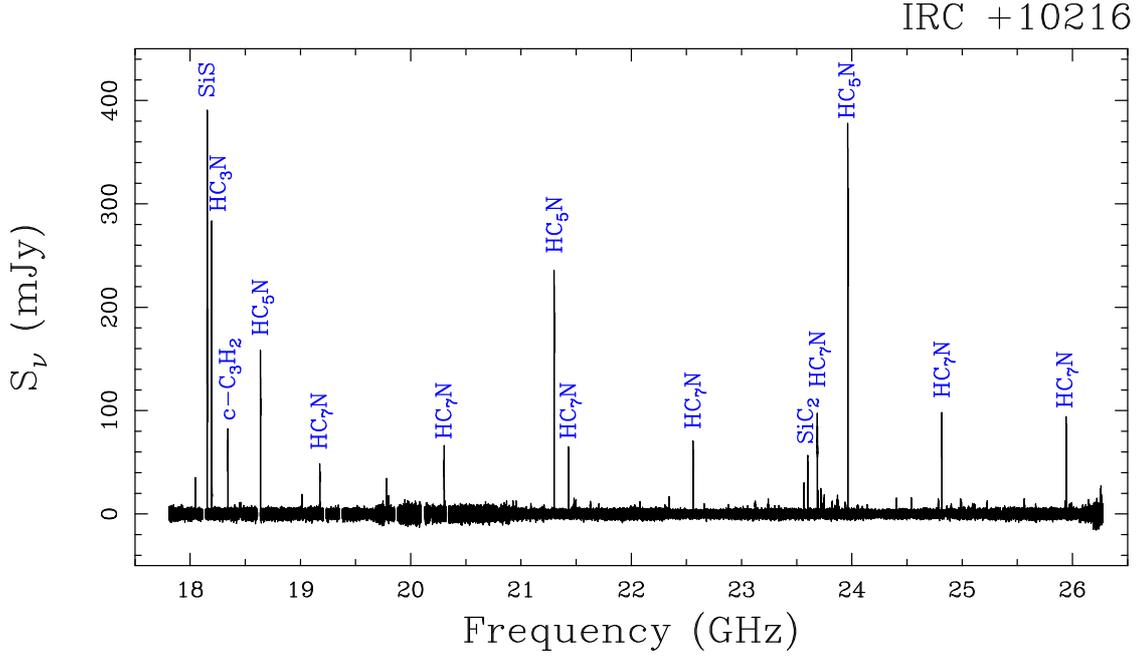}
\caption{{Overview of the 1.3 cm line survey toward IRC +10216 with strong lines marked.} \label{Fig:all}}
\end{figure*}

The line identification was performed on the basis of the JPL\footnote[4]{http://spec.jpl.nasa.gov}, CDMS\footnote[5]{http://www.astro.uni-koeln.de/cdms/catalog}, and splatalogue\footnote[6]{http://www.splatalogue.net} databases as well as on the online Lovas line list\footnote[7]{http://www.nist.gov/pml/data/micro/index.cfm} for astronomical spectroscopy \citep{2005JMoSt.742..215M,1998JQSRT..60..883P,2004JPCRD.33..177L}. A line is identified as real if it is characterized by a signal-to-noise ratio of at least five. Figure~\ref{Fig:all} presents an overview of our $\lambda \sim$1.3 cm line survey toward IRC +10216. We find 78 spectral lines, among which 12 remain unidentified. The identified lines are assigned to 18 different molecules. Except for SiS and NH$_{3}$, they are all C-bearing molecules. Twenty-three lines from already known species are detected for the first time outside the Solar System and there are additional 20 lines first detected in IRC +10216. Details are described in Sect.~\ref{new}. 

The spectrum is dominated by the strong profiles from six species: SiS, HC$_{3}$N, and HC$_{5}$N with $S_{\nu}>100$~mJy; HC$_{7}$N, c-C$_{3}$H$_{2}$, and SiC$_{2}$ with $S_{\nu}>20$~mJy. Figure~\ref{Fig:irc} shows the observed spectrum in more detail. Each panel covers $\sim$500\,MHz of the spectrum. Furthermore, there is a 10\,MHz overlap between adjacent panels so that lines truncated in one panel will not be truncated in the other panel. In addition, Fig.~\ref{Fig:zoomv1} shows zoom-in plots of all detected transitions.

We used the SHELL fitting routine in CLASS to derive the line parameters including peak intensity, integrated intensity, and expansion velocity, which is defined as the half-width at zero power. For the lines with hyperfine structure (hfs, like HC$_{3}$N, NH$_{3}$, etc.), only their main components are fitted with the routine to obtain their expansion velocities. For the lines that are blended or weak, the parameters are estimated directly by integrating the line profiles. The observed properties of the lines are displayed in Table~\ref{Tab:irclines}. The fitted lines reveal an expansion velocity of around 14.0 \kms, which is consistent with previous studies \citep[e.g.,][]{2000A&AS..142..181C,2008ApJS..177..275H}. There are some previously reported narrow lines with expansion velocities of $<$ 10\,\kms\, that are mainly from high-$J$ transitions in the ground vibrational state or from transitions in vibrationally excited states \citep{2009ApJ...692.1205P,2008ApJS..177..275H}, originating in the innermost, not 
yet fully accelerated shell. Lines with expansion velocities $<$ 10\,\kms~ are not seen in our study.

\subsection{Newly detected lines}\label{new}
In the following, we briefly describe most of the 23 newly detected lines (denoted by an ``N'' in the last column of Table~\ref{Tab:irclines}) and most of the other 20 lines newly detected in IRC +10216 (``NS''), together with related, already detected transitions from the same species. These transitions belong to HC$_{5}$N, HC$_{7}$N, HC$_{9}$N, SiC$_{4}$, NH$_{3}$, MgNC, ~l-C$_{5}$H, C$_{6}$H, C$_{6}$H$^{-}$, and C$_{8}$H. The $^{13}$C-bearing isotopologues of HC$_{5}$N are also briefly addressed (see also Sect.~\ref{ratio}). 

HC$_{5}$N. -- There are three transitions of cyanobutadiyne (HC$_{5}$N) in our observed frequency range. All of them have been detected, and the HC$_{5}$N line intensities rise with increasing frequency. HC$_{5}$N (9--8) has been reported by \citet{1978A&A....70L..37W}, but their peak intensity is almost twice the peak intensity measured by us even though their data were also obtained at Effelsberg. Significant variations in line intensities measured at submm and far-infrared wavelengths have recently been reported by \citet{2014arXiv1410.5852C}. However, unpublished HC$_{5}$N and HC$_{7}$N monitoring observations at Effelsberg about 25~yr ago (C. Henkel, priv. comm) did not reveal time variability in excess of 20\%. Furthermore, the $\lambda \sim 3$~mm data of \citet{2014arXiv1410.5852C} also show no strong variability. This, as well as other discrepancies outlined below, are therefore very likely caused by narrower bandwidths, lower quality baselines, and noise diodes with a more frequency-dependent output, thus resulting in a less accurate calibration than we have obtained here.

$^{13}$C-bearing HC$_{5}$N isotopologues. -- We have detected 11 transitions from the $^{13}$C-bearing isotopologues of HC$_{5}$N. Seven of them are first detected outside the Solar System, while four other transitions have already been studied with the NRAO-43m telescope \citep{1991ApJ...367L..33B}. 

\begin{figure*}[!htbp]
\centering
\includegraphics[width = 0.3 \textwidth]{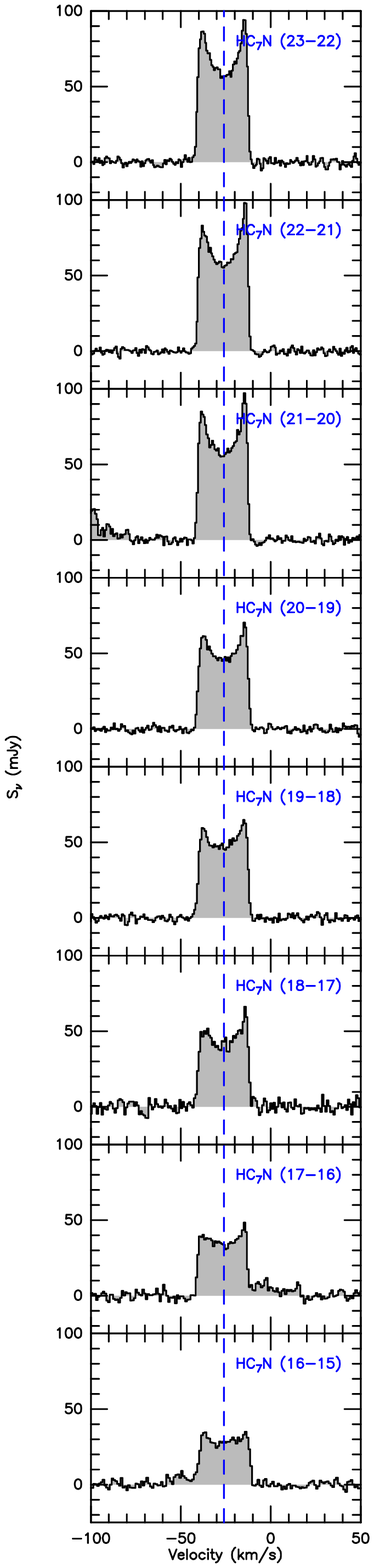}
\includegraphics[width = 0.3 \textwidth]{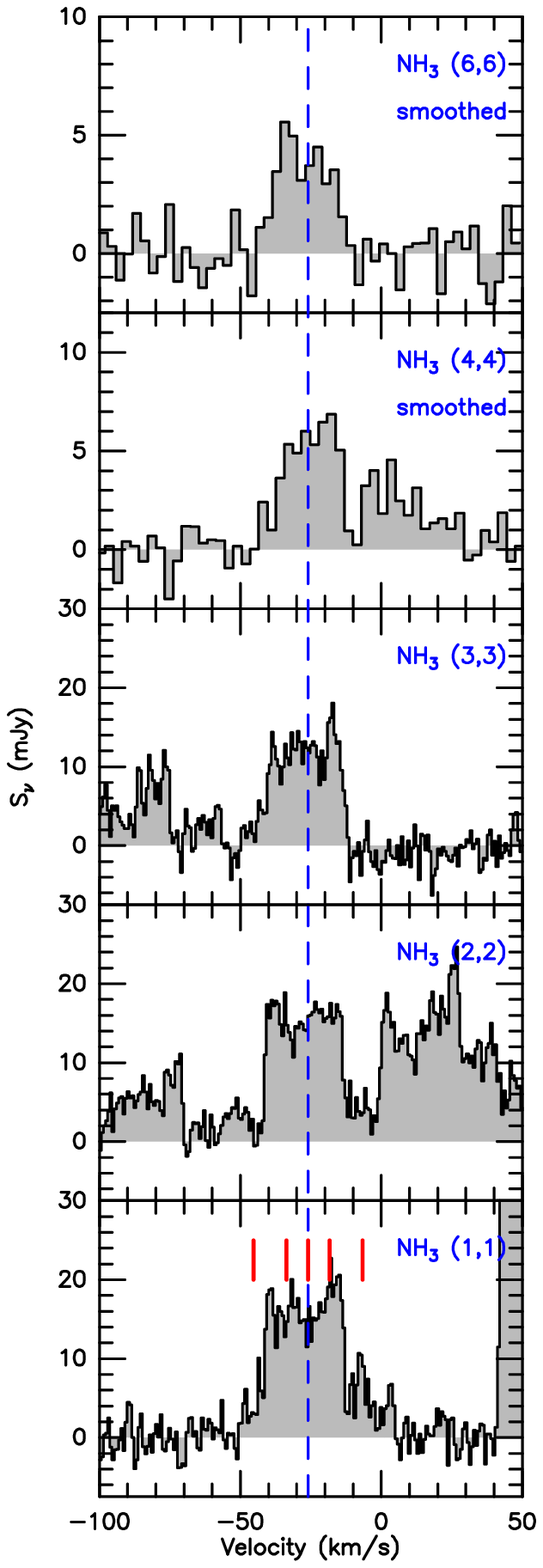}
\caption{{Detected spectra of HC$_{7}$N and NH$_{3}$ with transitions marked in the upper right of each panel. The NH$_{3}$ (4,4) and (6,6) lines have been smoothed to have a channel width of $\sim$3.3~\kms, while the channel width of other lines is $\sim$0.8~\kms. In the NH$_{3}$ (1,1) panel, the positions of the satellites are indicated by red solid lines. The blue dashed line traces the systemic LSR velocity ($-$26.0~\kms) of IRC +10216.}\label{Fig:linecomp}}
\end{figure*}

HC$_{7}$N. -- Eight transitions of HC$_{7}$N from $J=16-15$ to $J= 23-22$ fall in our band and are all detected. From Fig.~\ref{Fig:linecomp}, their line intensities rise with increasing frequency. Furthermore, we find that higher $J$ transitions have deeper dips than lower $J$ transitions in the center of line profiles. This is because the higher angular resolution of the higher $J$ transitions ends up lowering the contribution of emission from the CSE's outer regions to the line profiles. Apparently, the possible decrease in source size owing to higher excitation requirements of higher $J$ lines is less pronounced than effects caused by the decreasing beam size. HC$_{7}$N (21--20) has been reported before \citep{1978A&A....70L..37W}, but the peak intensity also obtained for it with the 100-m telescope is almost twice the peak intensity measured by us.
												
HC$_{9}$N. -- There are 14 transitions of HC$_{9}$N in our band. Five of them have been reported before \citep{1992ApJ...400..551B,1993A&A...277..133T}. The intensity of HC$_{9}$N (43--42) measured by us is roughly twice the intensity reported, also from Effelsberg, by \citet{1993A&A...277..133T}.

SiC$_{4}$. -- There are three lines of SiC$_{4}$ in this band and two of them are detected. They have lower upper level energies than previously reported transitions that span a frequency range from $\sim 30$ to $\sim 100$~GHz \citep{1995PASJ...47..853K,2004A&A...426..219M}. Although SiC$_{4}$ ($6$--$5$) falls into our frequency range as well, it is apparently too weak and remained undetected. This is consistent with the fact that the $J$=$7$--$6$ line is found to be significantly weaker than the $J$=$8$--$7$ line.


NH$_{3}$. -- We measured the five metastable inversion transitions ($J$, $K$) = (1,1), (2,2), (3,3), (4,4), (6,6) of ammonia (NH$_{3}$). The (5,5) line, also inside the measured band, is not seen (the non-detection of NH$_{3}$ (5,5) is discussed in Sect.~\ref{sec:lvg}). NH$_{3}$ (1,1) and NH$_{3}$ (2,2) have been reported by \citet{1984A&A...138L...5N} with the 100-m telescope, and their peak intensities are similar to ours. Inspecting Fig.~\ref{Fig:linecomp} and Table~\ref{Tab:irclines}, we find that the metastable inversion lines become slightly narrower with increasing $J$. Furthermore, we do not find clear evidence for hfs components in the NH$_{3}$ lines. For example, the (1,1) transition has one main and four satellite groups of hfs components, the centroid velocities of the groups of hfs components are symmetrically spaced by $\approx \pm 7.6$~\kms\, and $\pm 19.4$~\kms\, from the main group \citep{1967PhRv..156...83K}. Had the satellite groups appreciable optical depths this would lead to a significant (and symmetrical) broadening of this line. Assuming that all components of NH$_{3}$ (1,1) have the same excitation temperature, the limits on integrated intensity ratios ($>$8) between the main component and the hyperfine structure components indicate that the optical depth of NH$_{3}$ (1,1) is very low and might even depart from local thermodynamic equilibrium (LTE ratios on the order of 4). Meanwhile, there are no non-metastable NH$_{3}$ lines ($J$$>$$K$) detected. Non-metastable NH$_{3}$ lines require extreme densities (spontaneous radiative decay for rotational lines is $\sim 10^{5}$ times faster than for pure inversion transitions) or radiation fields to get excited and thus should only originate in the innermost shell. Any emission from this region would suffer from significant beam dilution effects.
																		
MgNC. -- The radical's higher $J$ transitions have already been studied before \citep{1995PASJ...47..853K,1993A&A...280L..19G}. This confirms our assignment about the $N$=2--1, $J$=5/2--3/2 transition of MgNC at 23875.0~MHz while the $N$=2--1, $J$=3/2--1/2 transition, also located within our frequency range, remains undetected. This is expected, since its integrated intensity may be $\sim$90~mJy~\kms\, (corresponding to a peak intensity of 3~mJy assuming an expansion velocity of 14~\kms\, and a flat-topped line profile) according to the expected intensity ratio between $N$=2--1, $J$=5/2--3/2 and $N$=2-1, $J$=3/2--1/2 lines under conditions of LTE and optically thin emission (see Sect.~\ref{dis}) and there are large uncertainties related to weak transitions.


l-C$_{5}$H. -- Five lines are detected in this survey, and they are from both the $^{2}\Pi_{1/2}$ and $^{2}\Pi_{3/2}$ ladders. Transitions of l-C$_{5}$H at higher frequencies have already been reported from IRC +10216 \citep{1995PASJ...47..853K,1986A&A...167L...5C}. 

C$_{6}$H. -- The first identification of the C$_{6}$H radical toward IRC +10216 has been reported by \citet{1987A&A...175L...5G}. Six lines of C$_{6}$H are observed here, stemming from both the $^{2}\Pi_{1/2}$ and the $^{2}\Pi_{3/2}$ ladders.

C$_{6}$H$^{-}$. -- C$_{6}$H$^{-}$ was the first detected interstellar anion, and its identification has been confirmed by observations toward IRC +10216 and TMC-1 \citep{2006ApJ...652L.141M}. Three transitions of C$_{6}$H$^{-}$ fall into our band, and two of them are detected. C$_{6}$H$^{-}$ (7--6) is also inside the frequency range covered by this survey, but is expected to be weaker than C$_{6}$H$^{-}$ (8--7).

C$_{8}$H. -- C$_{8}$H is the longest C$_{2n}$H molecule detected so far. It was first discovered toward IRC +10216 by \citet{1996A&A...309L..27C}. Four lines of C$_{8}$H are detected in this survey, all from the $^{2}\Pi_{3/2}$ ladder. Their hfs is not resolved. C$_{8}$H $J$=39/2--37/2 has been detected in IRC +10216 by \citet{2007ApJ...664L..47R}, and the intensity agrees well with our measurement. 

\subsection{Unidentified lines}
We do not see the 22.2 GHz line of water vapor \citep[see][for detections of higher frequency transitions]{2001Natur.412..160M,2011A&A...526L..11C,2011ApJ...727L..29N,2013ApJ...767L...3N}. This may be due to a coherent gain path of insufficient length or incorrect density-temperature combinations. For the 22.2 GHz line to mase, $T_{\rm kin} \sim$ 400~K and $10^{8}~\rm{cm^{-3}} < n < 10^{9}~\rm{cm^{-3}}$ may be required \citep[e.g.,][]{1991ApJ...373..525K}. This could be met in the inner envelope. If the density were too high in the $T \sim$ 400~K region, the maser action would be quenched. However, in that case, thermal emission would be expected. Assuming optically thick emission at $T_{\rm kin} =$ 600~K for the 22.2 GHz line and an emitting region of ten stellar radii or 0.8\arcsec\, \citep[roughly consistent with the $T_{\rm kin}$-radius profile in Fig.~7 of][]{1997ApJ...483..913C}, we get in our 43\arcsec\,beam (Sect.~\ref{obs}) $T_{\rm mb}$ = 600$\times$ 10$^3$ $\times$ (1.6/43)$^{2}$~mK, which gives an approximate $T_{\rm mb}\sim$ 800~mK. We clearly do not see this at a 5~$\sigma$ noise level of $T_{\rm mb}\sim$ 10~mK and conclude that the assumption of optically thick emission must be incorrect. This is consistent (E. Gonzalez-Alfonson, priv. comm) with the model that reproduces the H$_{2}$O data from Herschel, published by \citet{2011ApJ...727L..29N}. The C$_{5}$S lines are also not seen in our K-band spectral range \citep[see][for a report about the discovery of C$_{5}$S in IRC +10216]{2014arXiv1408.6306A} while \citet{1993ApJ...417L..37B} claimed detecting one C$_{5}$S line at 23990.2~MHz with a peak flux density of almost 20\,mJy. With expected rotational temperatures of $18-44$~K and expected column densities of $(2-14)\times 10^{12}$~cm$^{-2}$ \citep{2014arXiv1408.6306A}, we can estimate the peak intensity of the C$_{5}$S 23990.2~MHz line by assuming a low opacity, a flat-topped line profile, an expansion velocity of 14~\kms, a source size of 30\arcsec\, and a beamsize of 40\arcsec. The derived peak intensity is estimated to be $0.3-4$~mJy, corresponding to an integrated intensity of 8.4--112~mJy~\kms, where its electric dipole moment is taken to be 5.12 D \citep{B303753N}. Thus, our non-detection of the C$_{5}$S lines is expected and the detection reported by \citet{1993ApJ...417L..37B} remains unconfirmed.  

When referring to detected spectral features, we find 12 unidentified lines in this survey that cannot be assigned to any transitions in the spectroscopic catalogs available to us. Their high percentage ($\sim$15\% of the total number of lines) is probably due to the high sensitivity of the survey since their peak intensities are all less than 10 mJy. All of them show U-shaped or flat-topped profiles so they may represent isotopologues of well-known molecules or vibrationally excited lines or new species, and are marked  with ``U'' in Table~\ref{Tab:irclines}. The frequencies of unidentified lines are determined by eye assuming a source velocity of $V_{\rm lsr}=-$26.0 \kms. The accuracy of the frequencies of the U lines is estimated to lie within $\sim$180~kHz which is equivalent to three raw channels. The ``U'' line at 24901.4~MHz is close to C$_{4}$H~($^{2}\Pi_{1/2}$) ($\nu_{7}=1$, $J=5/2-3/2$, $l=e$), and their frequency difference is 454~kHz, corresponding to $\sim$5.5\,\kms.  Another ``U'' line at 25111.8~MHz is close to CH$_{2}$CHCN (10$_{1,9}$--10$_{1,10}$) with a rest frequency at 25111.4~MHz, but this line as observed would not be expected from this molecule because its fitted expansion velocity is about 19~\kms, if we use a single component to fit it. This is higher than the typical expansion velocity deduced from other detected transitions of CH$_{2}$CHCN in IRC +10216 \citep{2008A&A...479..493A}. Furthermore, the integrated intensity of CH$_{2}$CHCN (10$_{1,9}$--10$_{1,10}$) is expected to be around 0.3~mJy~\kms\, under the LTE assumption and assuming the column density and rotational temperature determined by \citet{2008A&A...479..493A} , which is much less than that of this ``U'' line.

Among the ``U'' lines, we find that there are three doublets, at 22304.854~MHz and 22323.3~MHz (pair-1, 18.5~MHz apart), at 25094.3~MHz and 25111.8~MHz (pair-2, 17.5~MHz apart), and at 25976.2~MHz and 25992.8~MHz (pair-3, 16.6~MHz apart). Their separations (16~MHz to 19~MHz) could be caused by hfs or different $^{13}$C substitutions. Estimates of their carrier's rotational constant ($B$) and centrifugal stretching constant ($D$) can help to make guesses about their identity. Assuming that two or three of them are from the same linear molecule, $B$ and $D$ can be derived via the formula \citep{2009tra..book.....W}
\begin{equation}\label{f1}
\nu (J) = 2B(J+1)-4D(J+1)^{3},
\end{equation}
where $\nu$ is the rest frequency of the $J$ -- ($J-1$) transition. We follow two steps to fit the parameters $B$ and $D$. Firstly, we obtain the number of $J$ with $D=0$. $J$ has to be an integer. Secondly, $B$ and $D$ can be derived with $J$ fixed. We find that the three doublets cannot be attributed to one species. If they belonged to one species, the three doublets would have to be assigned to the $J$, $J+3$, and $J+4$ levels but the $J+1$ and $J+2$ levels are not detected in our band. Furthermore, the fitted $D$ would be negative. Thus, we assume that two of them belong to the $J$ and $J+1$ levels of one species to derive the parameters $J$, $B$, and $D$. Since $D$ is usually a very small number, results are dismissed when the derived $D$ is larger than 0.5 MHz. The results are given in Table.~\ref{Tab:uline}. 

We find that the derived rotational constants are very small, which indicates a heavy molecule. When we assume that pair-1 and pair-2 come from the same species, the derived rotational constants are similar to that (1391.2~MHz) of C$_{6}$H. We note, however, that these rotational constants allow us to predict additional lower $J$ lines within our surveyed frequency range, which are not seen. It is probably because the lower $J$ level transitions are weaker than higher ones and cannot be detected by this survey.

\section{Discussion}\label{dis}
\subsection{Rotational temperatures and column densities}\label{phy}
 Line shapes can be used to evaluate the optical depth of individual transitions in IRC +10216. Parabolic lines arise from optically thick emission like that of CO (1--0), while flat-topped and U-shaped lines arise from optically thin spatially unresolved and resolved emission, respectively \citep{1982A&A...107..128O,1988A&A...190..167K}. We find no lines that show parabolic profiles among the identified spectral features in this survey. Also, lines with hfs should allow for estimates of their optical depth, if the hfs components are sufficiently displaced from each other (see the case of NH$_{3}$ discussed above). For such lines in our survey (NH$_{3}$, HC$_{3}$N, C$_{3}$N, etc.), we find low opacities. Thus, it is reasonable to assume optically thin emission for all the lines detected in this work. Assuming LTE, we use rotational diagrams to roughly estimate rotational temperatures and column densities. The standard formula used here is 
\begin{equation}\label{f2}
{\rm ln}(\frac{3kW}{8\pi^{3}\nu\mu^{2}S})~=~{\rm ln}(\frac{N_{\rm tot}}{Q(T_{\rm rot})})-\frac{E_{\rm u}}{kT_{\rm rot}},
\end{equation}
where $k$ is the Boltzmann constant, $W$ the integrated intensity, $S$ the transition's intrinsic strength, $\mu$ the permanent dipole moment, $N_{\rm tot}$ the total column density, $T_{\rm rot}$ the rotational temperature, $Q$ the partition function, and $E_{\rm u}$ the upper level energy of the transition. The values of $Q$ and $\mu$ are taken from the CDMS and JPL catalogs. 

At least two transitions of the same molecule with significant energy differences are required to determine rotational temperatures with this method. For those molecules with only one transition detected or without a wide dynamic range in upper level energies, we make use of the 28~GHz to 50~GHz data from \citet{1995PASJ...47..853K}, the C$_{8}$H data from \citet{2007ApJ...664L..47R}, and the C$_{3}$N data from \citet{2008ApJS..177..275H}, because those data have nearly the same resolution ($\sim 40$\arcsec) as our observations. The intensities of blended lines have large uncertainties, so are not included. SiS is also excluded, since SiS (1--0) is a maser in IRC +10216 \citep{1981A&A...101..238G,1983ApJ...267..184H} and the populations must deviate from LTE.

To derive the physical parameters, the intensities should be corrected for beam dilution by dividing by $\theta_{\rm s}^{2}/(\theta_{\rm s}^{2}+\theta_{\rm beam}^{2})$, where $\theta_{\rm s}$ is the source size, and $\theta_{\rm beam}$ is the beam size. Based on previous high resolution mapping of different molecules toward IRC +10216, we take the same source sizes as those listed in Table~\ref{Tab:irc_rd}. For species without high resolution mapping, their sizes are taken to be the same as chemically related species (details are described in the notes of Table~\ref{Tab:irc_rd}). We also used different source sizes $\theta_{\rm s}=15\arcsec, 18\arcsec, 30\arcsec, 50\arcsec$ to study the influence on the derived rotational temperatures and column densities of these molecules. It turns out that the uncertainties of the total column densities vary within a factor of four while the rotational temperatures nearly remain the same.

\begin{figure*}[!htbp]
\centering
\includegraphics[width = 0.4 \textwidth]{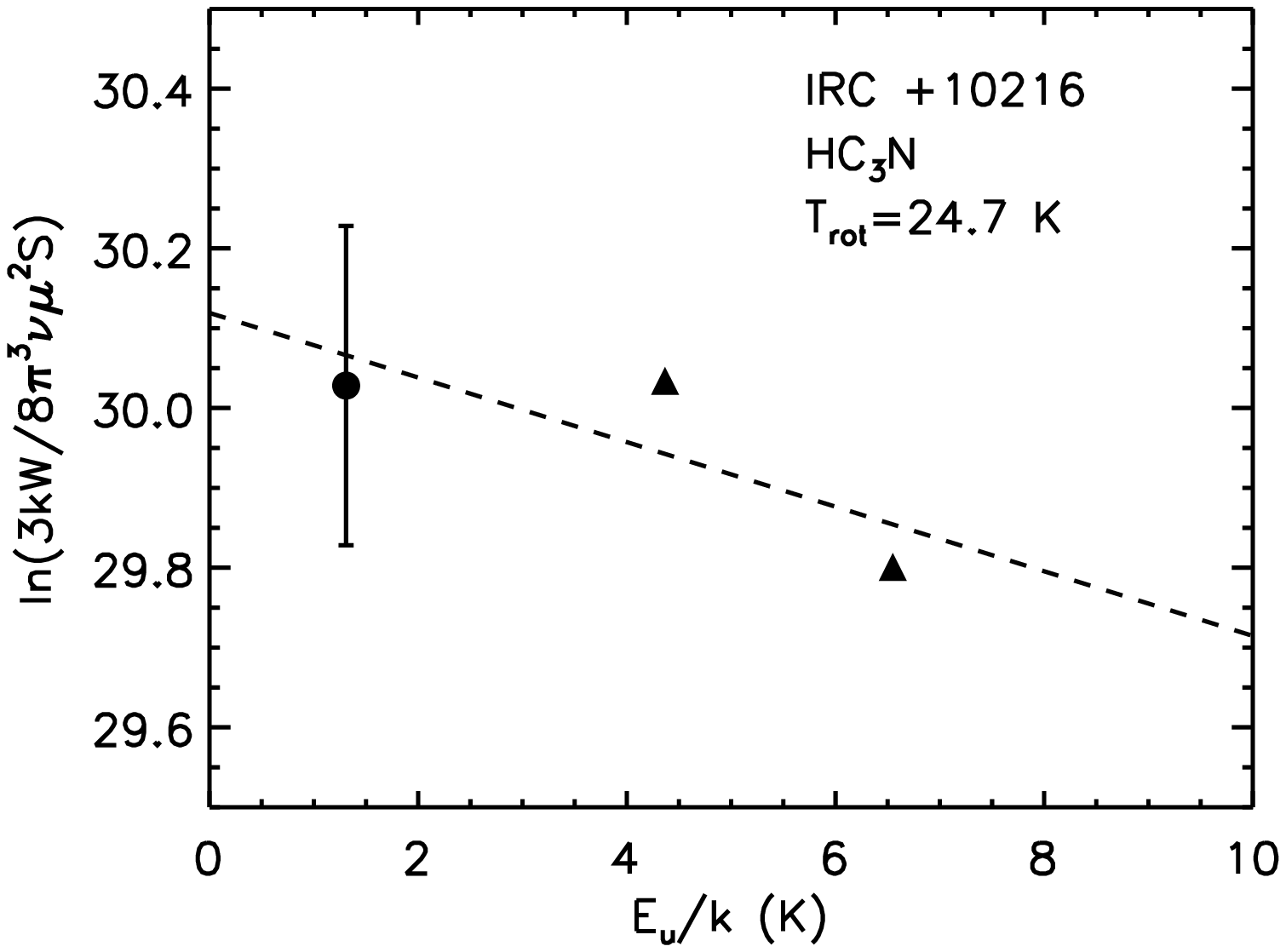}
\includegraphics[width = 0.4 \textwidth]{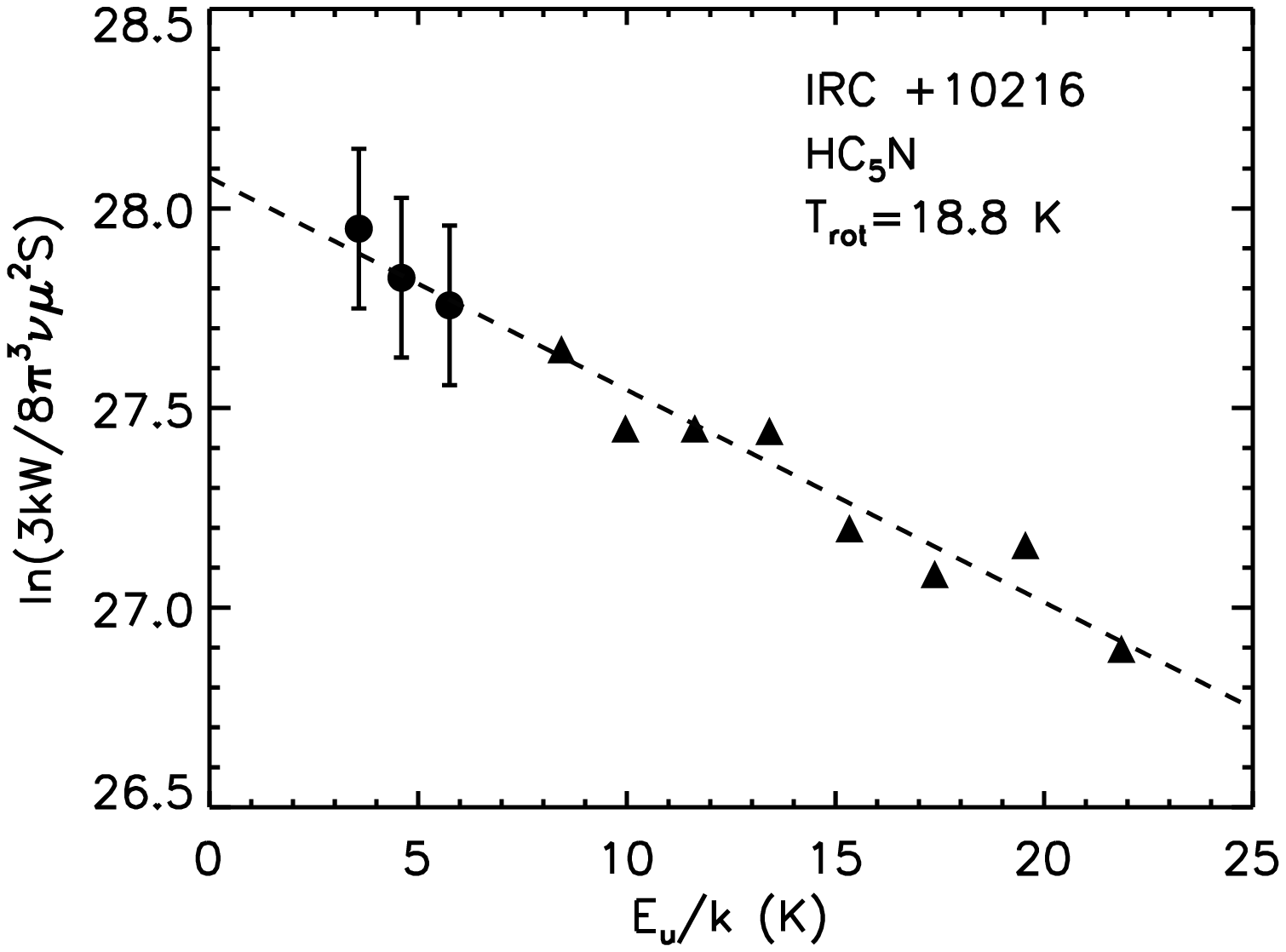}
\includegraphics[width = 0.4 \textwidth]{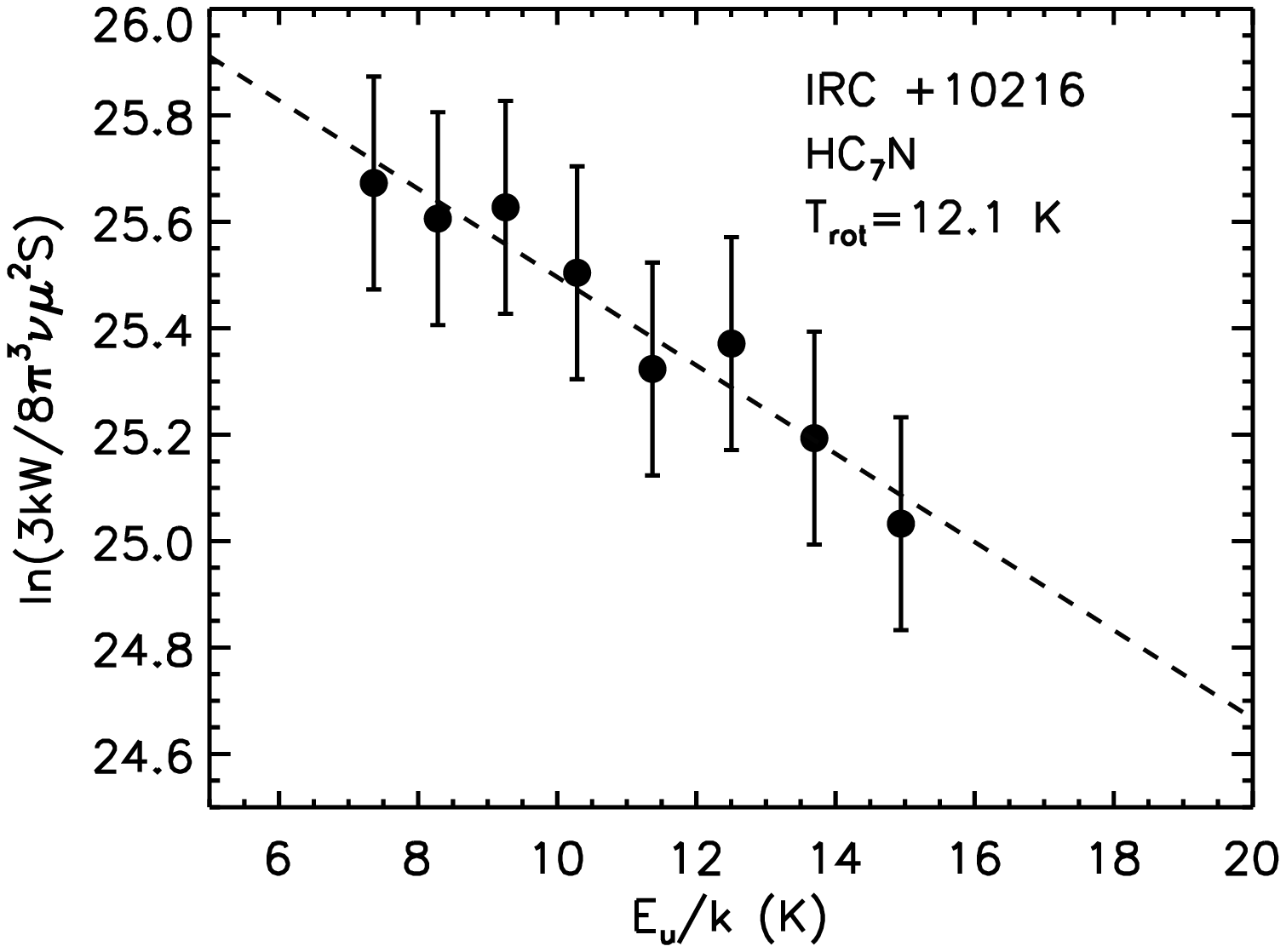}
\includegraphics[width = 0.4 \textwidth]{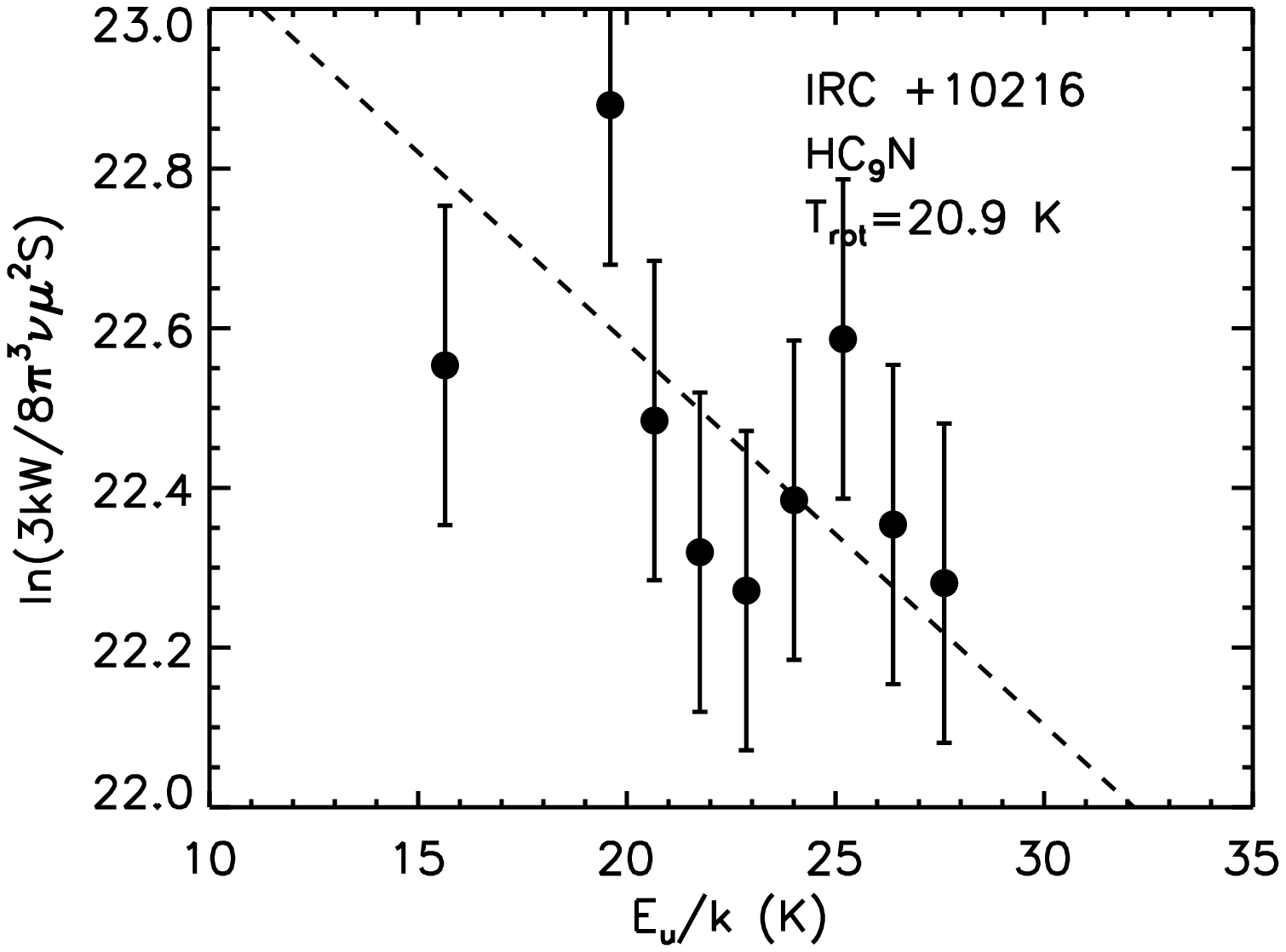}
\includegraphics[width = 0.4 \textwidth]{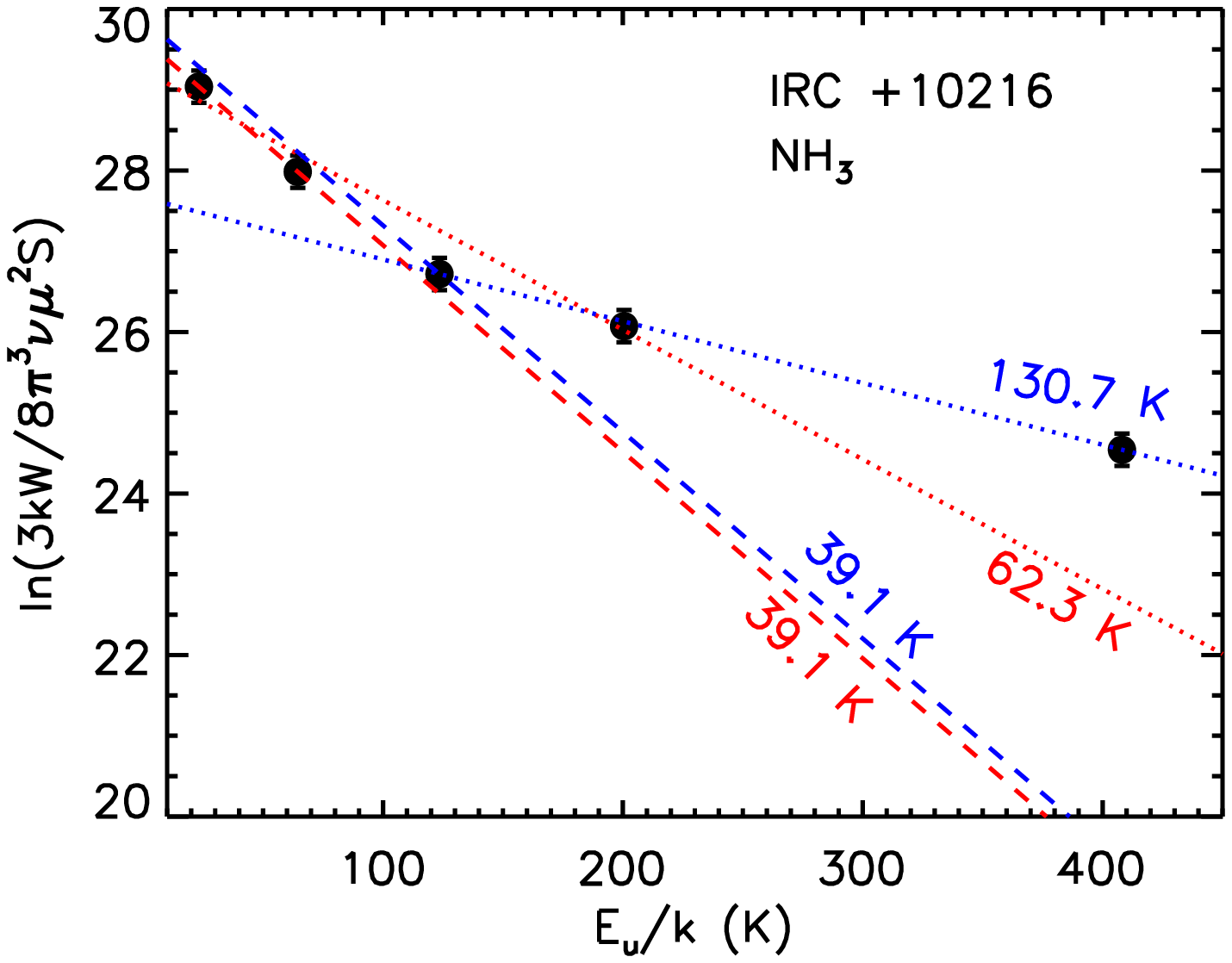}
\includegraphics[width = 0.4 \textwidth]{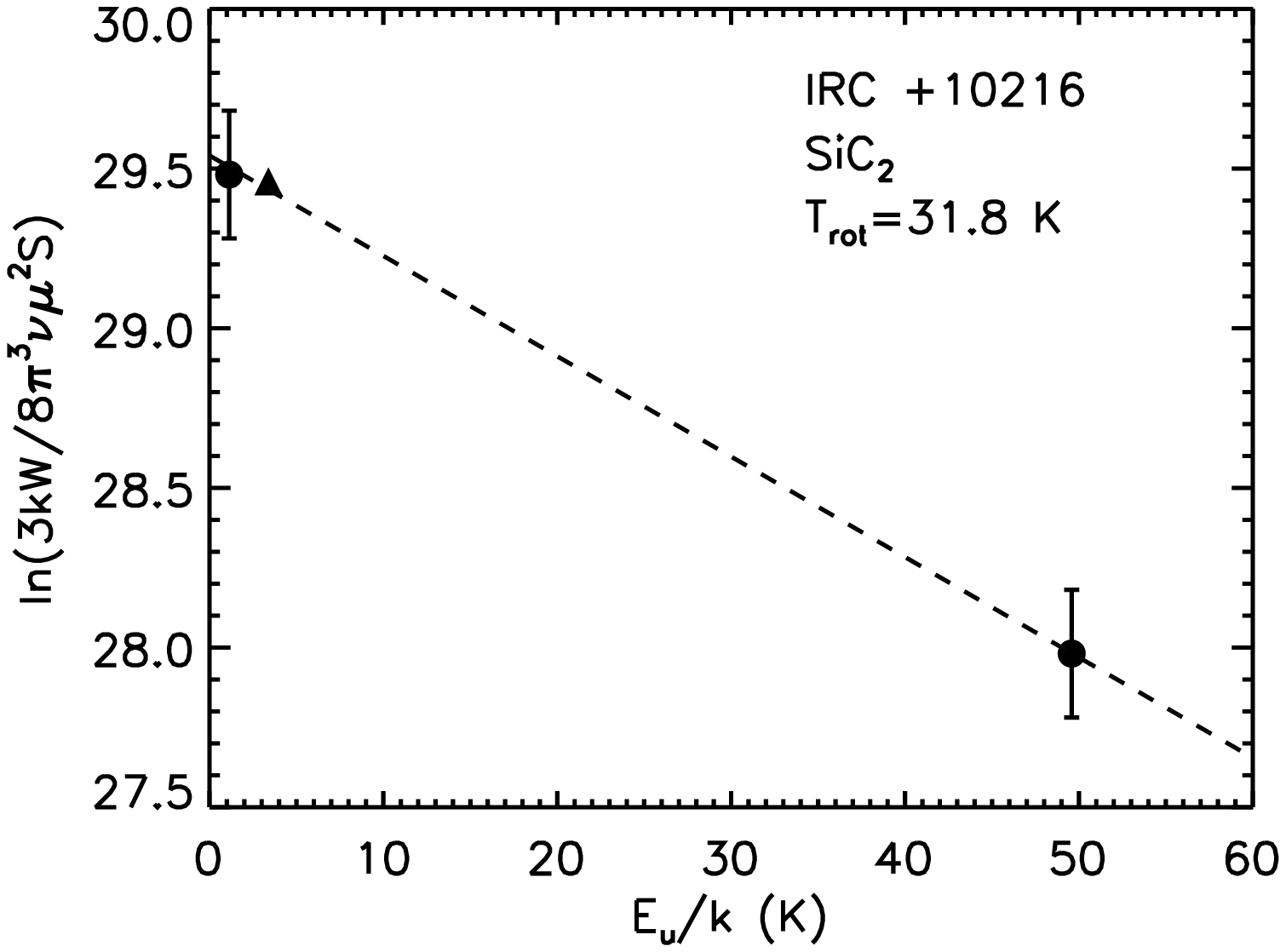}
\includegraphics[width = 0.4 \textwidth]{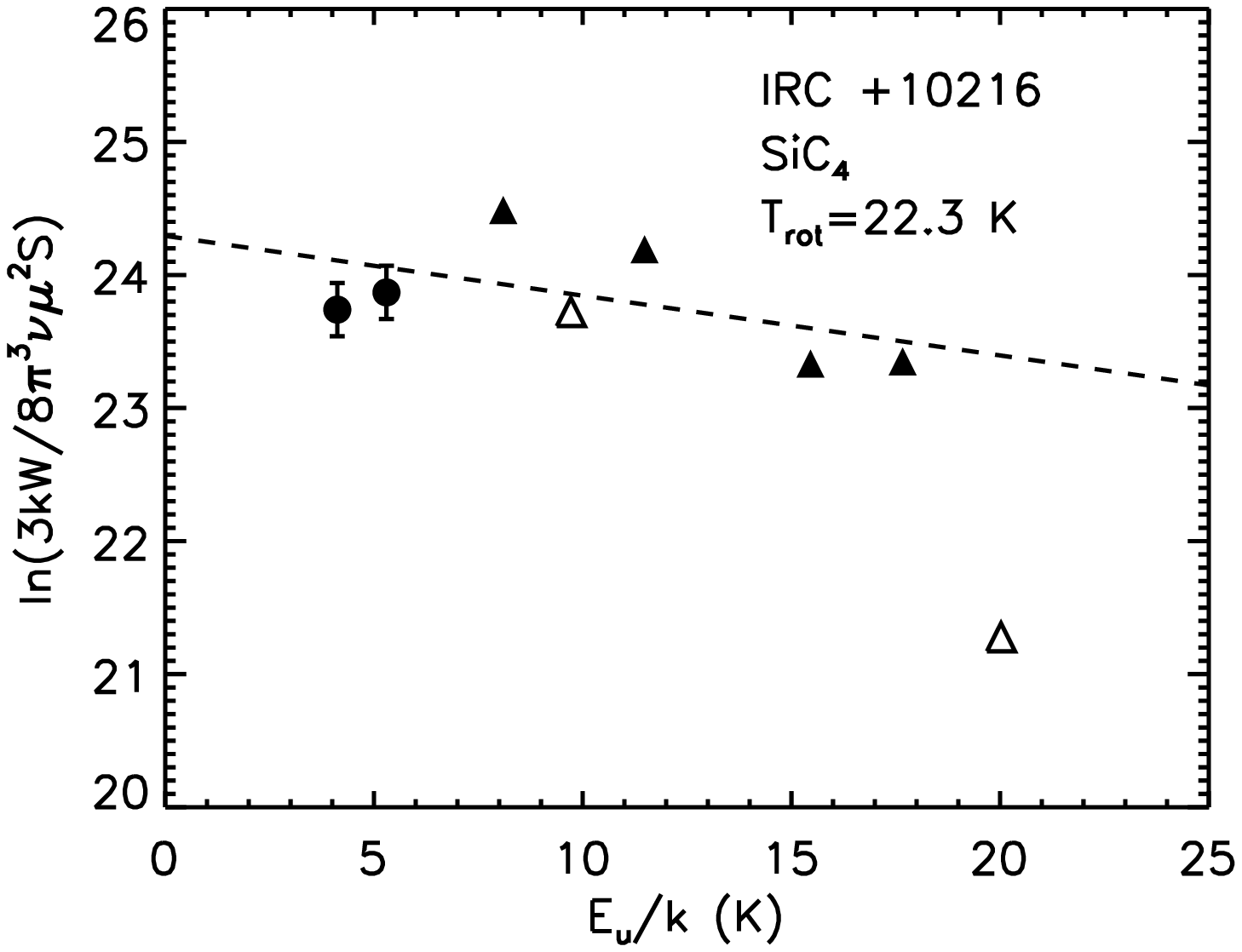}
\includegraphics[width = 0.4 \textwidth]{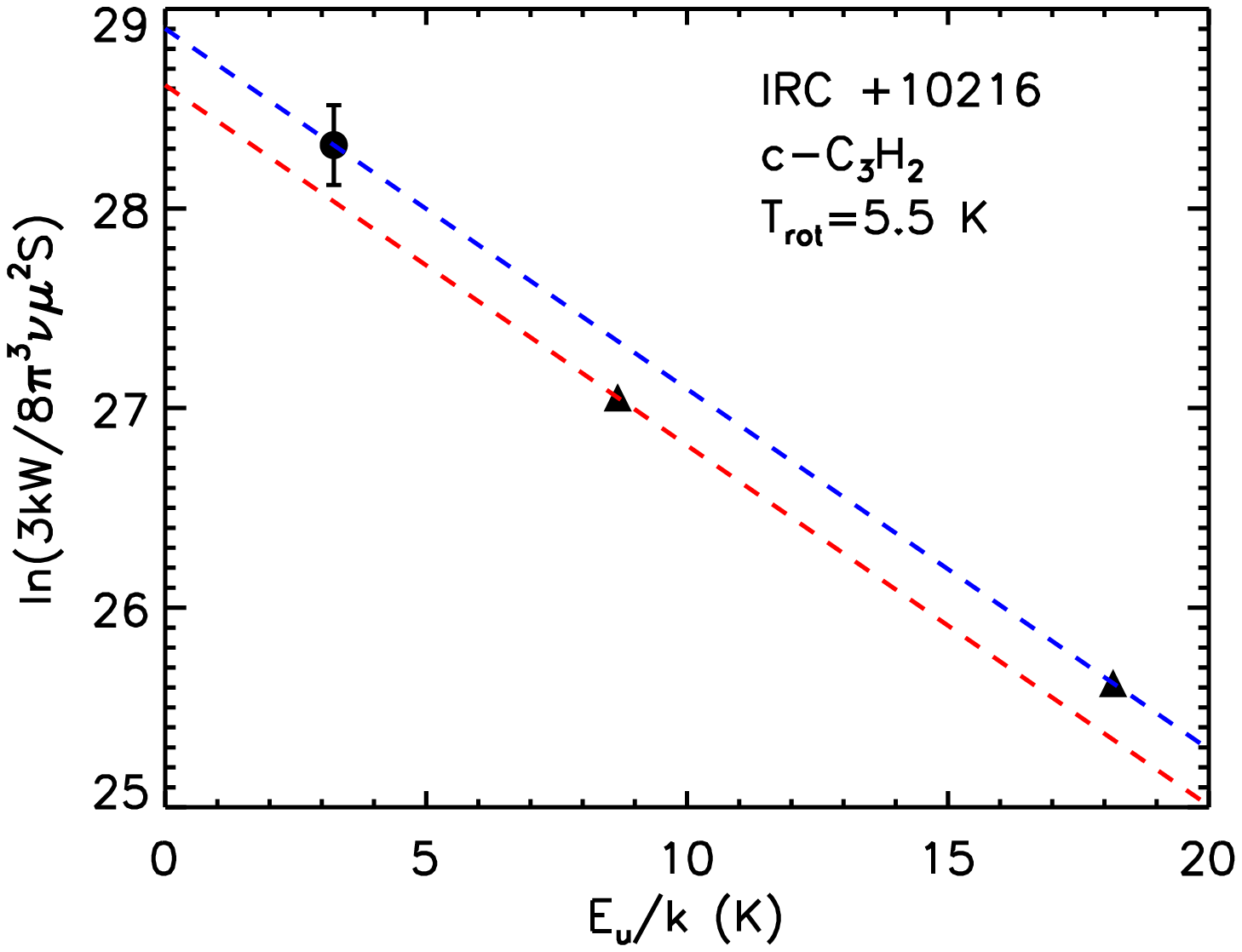}
\caption{{Rotational diagrams for the observed molecules in IRC +10216. For lines with hfs, all observed components have been taken into account. The circles with error bars are from our Effelsberg-100 m observations, where in addition to uncertainties in the fit to the integrated line intensities, a calibration error of 20\% has also been included. Filled and open symbols indicate signal-to-noise ratios that are larger than or less than five, respectively. The triangles without error bars are obtained from \citet{1995PASJ...47..853K}. The pentagrams with error bars for C$_8$H are obtained from \citet{2007ApJ...664L..47R}, and the squares for C$_3$N are from \citet{2008ApJS..177..275H}. Their values have been corrected for beam dilution. The molecules and their corresponding rotational temperatures, obtained from unweighted linear fits, are given in the upper right corner of each panel. Black dashed lines represent linear least-squares fit to the rotational diagram accounting for data presented by filled symbols, and the black dot-dashed lines represent the fit when the open circles are included. The red and blue dashed lines represent fits for the para and ortho states of NH$_{3}$ and c-C$_{3}$H$_{2}$ and the $^{2}\Pi_{1/2}$ and $^{2}\Pi_{1/2}$ state of l-C$_{5}$H and C$_{6}$H, respectively. Specifically for NH$_3$, the red dashed line shows the fit to the ($J,K$) = (1,1) and (2,2) para-lines, the blue dashed line uses the resulting excitation temperature (39.1\,K) and the ortho-(3,3) line, the dotted red line represents a fit to the para-(1,1), (2,2), and (4,4) transitions and the blue dotted line fits the ortho-(3,3) and (6,6) transitions.} \label{Fig:ircrd}}
\end{figure*}

We carried out linear least-square fits to the rotational diagrams of 17 species and four $^{13}$C isotopologues of HC$_{5}$N, which are shown in Fig.~\ref{Fig:ircrd}. For these fits, only data with signal-to-noise ratios higher than five are taken. The figures show that a few lines from the literature are outliers. In particular, SiC$_{4}$ (16--15) at $E_{\rm u}/k$= 20.0~K is much weaker than expected, and the data of C$_{6}$H ($^{2}\Pi_{3/2}$) from \citet{1995PASJ...47..853K} show significant scatter. This is due to weak lines with signal-to-noise ratios less than five, which, as mentioned above, have not been included in the fits. Therefore, only data points with higher signal-to-noise ratios than five are taken into account if there are enough points for the fits. Otherwise, data points with lower signal-to-noise ratios are also included and the fit to the data is represented not by a dashed but by a dash-dotted line. For example, those data points with signal-to-noise ratios less than five are used for the fits of $^{13}$C isotopologues of HC$_{5}$N and C$_{8}$H. The quality of the calibration of our observations is indicated by the fact that the data points for the HC$_{7}$N transitions show very little scatter about the fitted line in Fig.~\ref{Fig:ircrd}, while their frequencies cover nearly the entire frequency range surveyed by us. For l-C$_{5}$H ($^{2}\Pi_{3/2}$), we only detect one line that cannot be fitted with the data from \citet{1995PASJ...47..853K}, because this would result in a negative excitation temperature. We therefore use the rotational temperature derived from their data and our measured transition to obtain its column density (see the lower dashed line in the respective panel of Fig.~\ref{Fig:ircrd}; the upper dashed line refers solely to the data of \citealt{1995PASJ...47..853K}).

HC$_{9}$N shows large scatter, which may result from uncertainties related to its low line intensities. The HC$_{9}$N transitions from \citet{1995PASJ...47..853K} are not included since they show an even larger scatter. For NH$_{3}$ and c-C$_{3}$H$_{2}$, we fit their para and ortho states individually. If we only take NH$_{3}$ (1,1), (2,2), and (3,3) into account, the ortho/para ratio of NH$_{3}$ is estimated to be 1.3$\pm$0.6 in IRC +10216, which roughly agrees with the fact that the ortho/para abundance ratio approaches the statistical ratio of unity if NH$_{3}$ was formed in a medium with a kinetic temperature higher than $\sim$40~K \citep{2002PASJ...54..195T}. Including the NH$_{3}$ (4,4) and NH$_{3}$ (6,6) lines yields a higher rotational temperature and a lower ortho column density (see Table~\ref{Tab:irc_rd}). The ortho/para abundance ratio then becomes $0.6\pm0.2$. The derived ortho/para ratio of c-C$_{3}$H$_{2}$ is about 1.4$\pm$0.3, which is much less than the expected theoretical value of 3. We only have two points to fit the rotational temperature, and para and ortho $T_{\rm rot}$ values are assumed to be the same. We also fit the $^{2}\Pi_{1/2}$ and $^{2}\Pi_{3/2}$ states, separately, for l-C$_{5}$H and C$_{6}$H. However, the fits for l-C$_{5}$H ($^{2}\Pi_{3/2}$) shown in Fig.~\ref{Fig:ircrd} are tentative and should have large uncertainties because the data used here have an energy range ($\Delta E/k$) less than 10~K.

The derived rotational temperatures ($T_{\rm rot}$) and column densities ($N$), together with results from the literature, are listed in Table~\ref{Tab:irc_rd}. MgNC and C$_{2}$S show the largest uncertainties in the rotational temperature, which is due to the narrow energy ranges covered. From Table~\ref{Tab:irc_rd}, we find that the derived column densities agree roughly with previous studies. From Fig.~\ref{Fig:coltrot}, taking $T_{\rm rot}= 39.1\pm10.5$~K for the bulk of the para and ortho NH$_{3}$, we find rotational temperatures ranging from 5.5 to 47.2~K and molecular column densities ranging from 5.2$\times 10^{12}$ to 2.4$\times 10^{15}$~cm$^{-2}$. C$_{6}$H ($^{2}\Pi_{3/2}$) has the highest rotational temperature, while c-C$_{3}$H$_{2}$ has the lowest rotational temperatures indicating that they should be present in the cold outer molecular envelope. C$_{4}$H, SiC$_{2}$ and HC$_{3}$N are found to be the most abundant carbon-bearing molecules in our survey while C$_{6}$H$^{-}$, C$_{8}$H, and the $^{13}$C substitutions of HC$_{5}$N have the lowest column densities (see Table~\ref{Tab:irc_rd}). The column densities of all these species are spread over two orders of magnitude, and more complex molecules tend to be less abundant.

\subsection{Fractional abundances relative to H$_{2}$}\label{abu}
To obtain molecular abundances relative to H$_{2}$, one has to know the H$_{2}$ column density. \citet{2002A&A...387..624K} employed optical absorption spectroscopy toward background stars lying behind the envelope and found that the H+2H$_{2}$ column density at an offset of 37\arcsec ~from IRC +10216 is $\sim$2$\times 10^{21}$~cm$^{-2}$. Based on the optically thin lines of $^{13}$CO (1--0), (2--1), and (3--2), \citet{1994ApJS...94..147G} determined a beam-averaged ($\sim$20\arcsec) CO column density of $\sim$2$\times 10^{18}$~cm$^{-2}$ toward IRC +10216 by taking $^{12}$C/$^{13}$C to be 44. Previous studies have shown the fractional abundance CO/H$_{2}$ to be $6\times 10^{-4}$ in IRC +10216 \citep{1982ApJ...254..587K,1988ApJ...332.1009H}, resulting in an H$_{2}$ column density of 3.3$\times 10^{21}$~cm$^{-2}$. Furthermore, the average H$_{2}$ column density can also be calculated from the mass loss rate via the formula
\begin{equation}\label{f3}
N_{\rm H_{2}} = \frac{\dot{M}R/V_{\rm exp}}{\pi R^{2} m_{\rm H_{2}}} = \frac{\dot{M}}{\pi R V_{\rm exp}\, \mu m_{\rm H}} ~,
\end{equation}
where $\dot{M}$ is the mass loss rate which is believed to be 2.0$\times 10^{-5} M_{\odot}$yr$^{-1}$ \citep{1997ApJ...483..913C}; $R$ is the radius; $V_{\rm exp}$ is the expansion velocity of the CSE which is equal to 14~\kms; $\mu$ is the mean molecular weight, which we assume to be equal to 2.8; and $m_{\rm H}$ is the mass of a hydrogen atom. Of course, reference H$_{2}$ column densities to be used for abundance calculations depend on the location of their emission region in the CSE (see Formula~\ref{f3}). For example, if we take $R$ as the CO radius \citep[100\arcsec;][]{2006ApJ...652.1626F} and the HC$_{3}$N radius of 15\arcsec (Table~\ref{Tab:irc_rd}), the H$_{2}$ average column densities inside these radii are 3.2$\times 10^{20}$~cm$^{-2}$ and 2.1$\times 10^{21}$~cm$^{-2}$, respectively. While each line has a different distribution and therefore an individual $N_{\rm H_{2}}$, the choice of a specific $N_{\rm H_{2}}$ for all molecules is needed for a meaningful comparison with other studies. Thus, we use the 
H$_{2}$ average column density (2.1$\times 10^{21}$~cm$^{-2}$) within a typical radius of 15\arcsec\,to calculate molecular fractional abundances relative to H$_{2}$. The uncertainty of the average H$_{2}$ column density is within a factor of two since most detected molecules have a characteristic radius ranging from 7.5\arcsec~ to 30\arcsec \, in IRC +10216. The results are given in the fourth column of Table~\ref{Tab:irc_rd}. We find the fractional abundances relative to H$_{2}$ range for K-band detected species from 2.5$\times$10$^{-9}$ to $1.1 \times$10$^{-6}$ in IRC +10216.

\subsection{The RADEX non-LTE modeling for NH$_{3}$: the kinetic temperature}\label{sec:lvg}
Here we present a non-LTE analysis of NH$_{3}$ to determine the envelope's kinetic temperature using the RADEX expanding sphere code for IRC +10216 \citep{2007A&A...468..627V}. NH$_{3}$ is well suited to temperature determinations even with the limited frequency range covered by our survey, since its transitions arise from a wide range of energies above the ground state. Moreover, the metastable inversion lines are widely used as a galactic and extragalactic molecular temperature tracer \citep[e.g.,][]{1983A&A...122..164W,1983ARA&A..21..239H,2003A&A...403..561M}. 

Assuming that NH$_{3}$ has a typical expanding velocity of 14~\kms and a size of 18\arcsec\,(see Sect.~\ref{phy}), we obtain a velocity gradient of about 2468~\kms~pc$^{-1}$. Adopting a fixed para-NH$_{3}$ abundance of 1.1$\times 10^{-7}$ (see Table~\ref{Tab:irc_rd} and, for chemical considerations, \cite{2010A&A...521L...7M}) and a velocity gradient of 2500 \kms~pc$^{-1}$, we arrive at a para-NH$_{3}$ abundance per velocity gradient $[X]/(dv/dr)$ of $4.4\times10^{-11}$\,pc\,(\kms)$^{-1}$. The modeled kinetic temperatures range from 10 to 300 K with a step size of 5 K. The H$_{2}$ number density log($\frac{n(\rm{H}_{2})}{\rm cm^{-3}}$) varies from 2.0 to 7.0 with a step size of 0.1. For an adopted size of the NH$_{3}$ region of 18\arcsec, the density is around 10$^{5}$~cm$^{-3}$ according to the density profile toward IRC +10216 \citep{1997A&A...322L..21G}. We modeled the integrated intensity ratios $\frac{\rm NH_{3} (2,2)}{\rm NH_{3} (1,1)}$ in this study, which are shown in Fig.~\ref{Fig:lvg}a, because (a) the ratio between NH$_{3}$ (1,1) and NH$_{3}$ (2,2) is not affected by an ortho-to-para abundance ratio, because (b) NH$_3$ rotation temperatures provide lower limits to the kinetic temperature \citep[e.g.,][]{1983A&A...122..164W} with the ($J$,$K$) = (1,1), (2,2), and (3,3) transitions revealing the lowest $T_{\rm rot}$ values, and because (c) beam-filling factors (the (1,1) and (2,2) frequencies are less than 30~MHz apart) should not play a role. Assuming that both lines arise from the same source, we obtain a kinetic temperature of 70$\pm$20~K for NH$_{3}$ in the density range from $10^{4.5}$ to $10^{5.5}$ cm$^{-3}$ with an integrated intensity ratio $\frac{\rm NH_{3} (2,2)}{\rm NH_{3} (1,1)}$ of $0.73\pm0.1$ (see Fig.~\ref{Fig:lvg}a). 

The value of the para-NH$_{3}$ abundance per velocity gradient $[X]/(dv/dr)$, which was given at the start of the previous paragraph, may have a large uncertainty, potentially also leading to a significant error bar in our $T_{\rm kin}$ estimate. Thus, we carried out a non-LTE analysis with a fixed density of $10^{5}$ cm$^{-3}$ (see previous paragraph) and the para-NH$_{3}$ abundance per velocity gradient $[X]/(dv/dr)$ ranging from $1.0\times 10^{-13}$ to $1.0\times10^{-9}$\,pc\,(\kms)$^{-1}$. The result is shown in Fig.~\ref{Fig:lvg}b. We find that the kinetic temperature does not depend on the para-NH$_{3}$ abundance per velocity gradient $[X]/(dv/dr)$ in the range of 10$^{-13}$ to 10$^{-9}$\,pc\,(\kms)$^{-1}$, because the lines are optically thin. Therefore, the kinetic temperature of 70$\pm$20~K derived from the integrated intensity ratio $\frac{\rm NH_{3} (2,2)}{\rm NH_{3} (1,1)}$ is reasonable for the NH$_{3}$ environment in IRC +10216. 

At the same time, we also modeled the integrated intensity ratios $\frac{\rm NH_{3} (4,4)}{\rm NH_{3} (2,2)}$ and $\frac{\rm NH_{3} (6,6)}{\rm NH_{3} (3,3)}$ (see Figs.~\ref{Fig:lvg}c and \ref{Fig:lvg}d). While modeling the integrated intensity ratios $\frac{\rm NH_{3} (6,6)}{\rm NH_{3} (3,3)}$, an ortho-NH$_{3}$ abundance per velocity gradient $[X]/(dv/dr)$ of $5.6\times10^{-11}$\,pc\,(\kms)$^{-1}$ is used with the same velocity gradient as for para-NH$_{3}$. With the observed integrated intensity ratio $\frac{\rm NH_{3} (4,4)}{\rm NH_{3} (2,2)}= (0.36\pm0.04)$ and the assumption that NH$_{3}$ (2,2) and (4,4) arise from the same region, we obtain a kinetic temperature of 120$^{+10}_{-5}$~K in the density range from $10^{4.5}$ to $10^{5.5}$ cm$^{-3}$. We note that NH$_{3}$ (4,4) might be blended (see Fig.~\ref{Fig:zoomv1}) by a feature of unknown origin, in which case the derived kinetic temperature would be slightly overestimated. In any case, the derived kinetic temperature is higher than obtained from the observed integrated intensity ratio $\frac{\rm NH_{3} (2,2)}{\rm NH_{3} (1,1)}$, which indicates a kinetic temperature gradient with NH$_{3}$ (4,4) originating in a warmer region. 

For ortho-NH$_{3}$, the observed line ratio $\frac{\rm NH_{3} (6,6)}{\rm NH_{3} (3,3)}$ = (0.28$\pm$0.08) is beyond the modeled range, which indicates that NH$_{3}$ (6,6) may arise from a very hot region with a kinetic temperature larger than 300~K. Alternatively, the (6,6) emission may be affected by population inversion and an optical depth not far below unity. Such NH$_{3}$ (6,6) masers have already been found in NGC~6334I \citep{2007A&A...466..989B}, W51--IRS2 \citep{2013A&A...549A..90H}, and perhaps even in the nearby galaxy IC~342 \citep{2011A&A...534A..56L}, all of which are sources where a strong infrared radiation field may significantly contribute to the molecule's excitation. The inner parts of IRC+10216's envelope also possess this strong a radiation field. On the other hand, NH$_{3}$ (5,5) is not detected in this survey. Adopting $T_{\rm rot}$= 62.3 K from a fit to the (1,1), (2,2), and (4,4) para-NH$_{3}$ lines, the integrated intensity of NH$_{3}$ (5,5) is predicted to be 35~mJy~\kms\,, which is too weak to be detected. The non-detection of the (5,5) and (7,7) lines does not provide evidence for NH$_3$ formation pumping, which has recently been suggested for a variety of molecules in star-forming regions (\citet{2014ApJ...785..135L}, see also \cite{2006A&A...460..533W} for a particularly compelling observational candidate).

\subsection{Abundance differences among $^{13}$C isotopologues of HC$_{5}$N and the $^{12}$C/$^{13}$C ratio in IRC +10216}\label{ratio}
Chemical fractionation of $^{13}$C-bearing species has been observed toward low-temperature regions. \citet{1998A&A...329.1156T} reported chemical fractionation between the $^{13}$C isotopologues of HC$_{3}$N toward TMC-1. \citet{2007ApJ...663.1174S} report a C$^{13}$CS/$^{13}$CCS ratio of 4.2$\pm$2.3 (3$\sigma$) in TMC-1. \citet{2010A&A...512A..31S} also find the C$^{13}$CH /$^{13}$CCH abundance ratios to be 1.6$\pm$0.4 (3$\sigma$) and 1.6$\pm$0.1 (3$\sigma$) for TMC-1 and L1527. Finally, \citet{2013JPCA..117.9831S} find abundance differences among $^{13}$C isotopologues of C$_{3}$S and C$_{4}$H in TMC-1, which are believed to be caused by different formation processes or $^{13}$C isotope exchange reactions. However, in IRC +10216, $^{13}$C isotope fractionation is still not very well studied.

In our band, the $^{13}$C isotopologues of HC$_{3}$N, HC$_{5}$N, C$_{2}$S, C$_{3}$S, and C$_{4}$H can be used to address this issue. However, H$^{13}$CCCN (2--1) at 17633.7~MHz lies outside of our frequency range. The bands related to the (2--1) transition of HC$^{13}$CCN and HCC$^{13}$CN at 18120.8~MHz and 18119.0~MHz are flagged. Meanwhile, the $^{13}$C isotopologues of C$_{2}$S, C$_{3}$S, and C$_{4}$H are too weak to be detected in this survey. However, many of the $^{13}$C-bearing isotopologues of HC$_{5}$N have been measured (see Sect.~\ref{new}) and are thus used to investigate potential abundance differences among the $^{13}$C species toward IRC +10216. The $J$=9--8 transitions of the $^{13}$C-bearing isotopologues are the strongest in our band, so we compare their integrated intensities, except for HC$^{13}$CCCCN ($9-8$), which is blended with C$_{6}$H at 23719.4~MHz. With the four other $J$=9--8 transitions, the integrated line intensity ratios, applying a factor $\nu^{2}$ correction between H$^{13}$CCCCCN, HCC$^{13}$CCCN, HCCC$^{13}$CCN, and HCCCC$^{13}$CN, are estimated to be 1.00:(1.07$\pm$0.08):(1.28$\pm$0.09):(1.02$\pm$0.07). The errors represent 1$\sigma$. Therefore, the results suggest that the abundances of the $^{13}$C isotopologues of HC$_{5}$N may not be all the same in IRC +10216. \citet{2008ApJS..177..275H} may also have found an abundance differentiation among the $^{13}$C isotopologues of HC$_{3}$N in IRC +10216 by obtaining the integrated line intensity ratios H$^{13}$CCCN:HC$^{13}$CCN:HCC$^{13}$CN=1.00:(1.19$\pm$0.14):(1.31$\pm$0.15), applying mm-wave spectroscopy. Thus, we suggest that abundance differences possibly exist between the $^{13}$C isotopologues of HC$_{5}$N, but this has to be further checked by deeper integrations.

The $^{12}$C/$^{13}$C ratio is an important tool for tracing processes of stellar nucleosynthesis \citep{1994ARA&A..32..191W}. Here the carbon isotope ratio is based on the ratios between the same $J$ transitions from HC$_{5}$N and its $^{13}$C isotopologues. (The blended lines are not included.) The transitions of HC$_{5}$N and its $^{13}$C substitutions are optically thin as mentioned in Sect.~\ref{phy}. Thus, the isotopic ratio can be directly obtained from their integrated intensity ratio. Figure~\ref{Fig:cr} shows the $^{12}$C/$^{13}$C ratios determined with eight pairs of unblended transitions from HC$_{5}$N and its $^{13}$C isotopologues. The derived $^{12}$C/$^{13}$C ratios are $35\pm6$ from $J=7-6$, $56\pm9$, $56\pm9$, and $38\pm3$ from $J$=8--7, and $55\pm 3$, $51\pm3$, $43\pm 2$, and $54\pm 3$ from $J=9-8$. The unweighted average $^{12}$C/$^{13}$C ratio is $49\pm 9$, where the error is the standard deviation of the mean. This is greater than was derived from [SiCC]/[Si$^{13}$CC] ($34.7\pm4.3$) by \citet{2008ApJS..177..275H} and less than the 71 obtained from [$^{12}$CO]/[$^{13}$CO] by \citet{2014A&A...566A.145R}, but these results may suffer from opacity effects. Our value agrees well with the 47$^{+6}_{-5}$ of \citet{1988A&A...190..167K} derived from HC$_{3}$N, CS, HNC, CCH, SiCC, and their isotopologues. Therefore, we confirm that the $^{12}$C/$^{13}$C ratio in IRC +10216 is much smaller than the terrestrial ratio (89) and is also smaller than the value found for the local interstellar medium ($\sim$70) \citep{1994ARA&A..32..191W}.

\subsection{Comparison of the IRC +10216 and TMC-1 chemistry}\label{aburatio}
IRC +10216 and TMC-1, the dark cloud in the Taurus region, are the two prototypical sources for detailed studies of carbon chain molecules in outer space. Seen at distances of $D$ $\sim$ 130\,pc (Sect.\,1) and 140\,pc \citep[e.g.,][]{1997ApJ...486..862P}, respectively, they can be studied with similar linear resolution and sensitivity. \citet{2004PASJ...56...69K} carried out a line survey toward the cyanopolyyne peak of the dark cloud TMC-1 that includes the frequency range (17.8-26.3 GHz) discussed in this work. This allows us to compare chemical compositions between two physically different, prototypical astronomical environments characterized by large amounts of carbon chain molecules. We note that such a comparison may be biased by (1) the use of different telescopes, (2) by different source morphologies, and (3) by differences in line shapes, which all will affect relative sensitivities. While the TMC-1 study, carried out with the Nobeyama 45-m telescope, is characterized by a larger beam size, emission from TMC-1 is more extended than from IRC +10216, so that our Effelsberg 100-m data may actually suffer from smaller beam-filling factors. On the other hand, lines from IRC +10216, affected by an expansion velocity of $\sim$14\,\kms, are much broader than those from TMC-1, thus allowing us to smooth contiguous channels and to compensate for this disadvantage. This implies that sensitivities match each other by a factor of a few. Below, we constrain ourselves mainly to K-band data, but referring also to lines from other spectral ranges if needed.

Comparing the detected molecules in the same $\lambda$ $\sim$1.3\,cm spectral range (see Fig.~\ref{Fig:irc-tmc}), we find that five molecules are only detected in IRC +10216, nine molecules are only detected in TMC-1, and 13 molecules are detected in both sources. The five K-band molecules only detected in IRC +10216 are SiS, SiC$_{2}$, SiC$_{4}$, MgNC, and C$_{6}$H$^{-}$. While C$_{6}$H$^{-}$ is not detected in the Nobeyama survey \citep{2004PASJ...56...69K} toward TMC-1, it has been reported in later studies \citep[e.g.][]{2006ApJ...652L.141M,2007ApJ...664L..43B}. In contrast, MgCN could not be detected in a more recent sensitive study with the Green Bank 100-m telescope \citep{2005ApJ...621..817T}. This also holds for Si-bearing molecules \citep{1989ApJ...343..201Z}. The nine K-band molecules only detected in TMC-1 are HNCO, HNCCC, CH$_{3}$CN, CH$_{2}$CHCN, c-C$_{3}$HD, HCCNC, CH$_{2}$CN, H$_{2}$CCC, and DC$_{3}$N. While HNCCC, CH$_{3}$CN, CH$_{2}$CHCN, HCCNC, CH$_{2}$CN, and H$_{2}$CCC are not detected in our survey, they have been reported by other studies \citep[e.g.,][]{1997Ap&SS.251..199G,2008A&A...479..493A,2000AAS..142..181C}. The other K-band molecules, HC$_{\rm 2n+1}$N (n = 1, 2, 3, 4), C$_{\rm 2n}$H (n = 2, 3, 4), c-C$_{3}$H$_{2}$, l-C$_{5}$H, C$_{3}$N, C$_{2}$S, C$_{3}$S, and NH$_{3}$, are detected in both sources which indicates, as already mentioned, that the two targets are both effective in forming long carbon-chain molecules. The ratio $\frac{\rm NH_{3} (2,2)}{\rm NH_{3} (1,1)}$ is less than 0.1 in TMC-1, indicating that the NH$_{3}$ emitting region in TMC-1 is much colder ($\sim$ 10~K) than in IRC +10216, consistent with its more pronounced deuteration of molecules. The cyanopolyyne abundance ratios are estimated to be HC$_{5}$N:HC$_{7}$N:HC$_{9}$N=(18.4$\pm$10.3):(14.8$\pm$8.4):1 in IRC +10216 while they are HC$_{5}$N:HC$_{7}$N:HC$_{9}$N=13.1:4.8:1 in TMC-1 \citep{2008ApJ...672..371S}. The difference is probably related to different physical conditions or time scales for the formation and destruction of the cyanopolyynes in IRC +10216 and TMC-1. Meanwhile, there is no trend of the rotational temperatures of the cyanopolyynes increasing with the C-chain length that is found in TMC-1 \citep{1998ApJ...508..286B}. This indicates that the cyanopolyynes are probably emitting line photons from nearly the same region in IRC +10216. In IRC +10216, our C$_{6}$H$^{-}$/C$_{6}$H ratio is estimated to be (3.6$\pm$0.8)\%. (The C$_{6}$H column density is the sum of the column densities of the $^{2}\Pi_{1/2}$ and the $^{2}\Pi_{3/2}$ states.) The value is lower than the 8.6\%\, derived in a previous study of IRC +10216 \citep{2007ApJ...661L..61K}, because they obtained a higher rotational temperature (33~K) for C$_{6}$H, resulting in a lower column density. To summarize, both C$_{6}$H$^{-}$/C$_{6}$H ratios in IRC +10216 are at least as high as the value (2.5\%) reported for TMC-1 \citep{2006ApJ...652L.141M}.

\section{Summary and conclusions}
We performed a 1.3 cm line survey toward IRC +10216 using the 100-m telescope at Effelsberg. In total, 78 spectral lines were detected, among which 12 remain unidentified. The identified lines are assigned to 18 different molecules and radicals. No new species was found, but 23 lines from already known species were detected for the first time outside the Solar System and there are an additional 20 lines first detected in IRC +10216. We also discuss the origin of the ``U'' lines.

Assuming LTE, we derived rotational temperatures and column densities of 17 detected molecules. Their rotational temperatures range from 5.5 to 39.1~K, and molecular column densities range from 5.2$\times 10^{12}$ to 2.4$\times 10^{15}$~cm$^{-2}$. Molecular abundances relative to H$_{2}$ range between 2.5$\times 10^{-9}$ and 1.1$\times 10^{-6}$. A non-LTE analysis of NH$_{3}$ shows that its (1,1) and (2,2) emission arises from the inner envelope with a kinetic temperature of 70$\pm$20~K, but there may be warmer gas as is indicated by the detected (4,4) line. The (6,6) inversion line shows surprisingly strong emission and might indicate inverted populations and an optical depth not far below unity. Comparing transitions from the $^{13}$C isotopologues of HC$_{5}$N, we find there might be abundance differences between them. The isotopic $^{12}$C/$^{13}$C ratio is estimated to be 49$\pm$9 from the integrated intensity ratios of transitions from HC$_{5}$N and its $^{13}$C isotopologues. Finally, we compare the
chemical compositions of IRC +10216 and TMC-1 in the $\lambda$ $\sim$1.3\,cm spectral range. Apparently, Si-bearing molecules are favored in IRC +10216.

\section*{ACKNOWLEDGMENTS}\label{sec.ack}
We would like to thank the anonymous referee for a helpful report that led to improvements in the paper. We greatly thank Denise Keller for providing source size information from her new JVLA observations. We also wish to thank J. Kauffmann, T. Pillai, and E. Gonzalez-Alfonso for discussions or an exchange of emails and appreciate the assistance of the Effelsberg 100-m operators during the observations. Y. Gong acknowledges support by the MPG-CAS Joint Doctoral Promotion Program (DPP), and NSFC Grants 11127903, 11233007, and 10973040. S. Thorwirth gratefully acknowledges funding by the Deutsche Forschungsgemeinschaft (DFG) through grant TH 1301/3-2.

\bibliographystyle{aa}
\bibliography{references}

\begin{table*}[!hbt]
\caption{Fitted results for the ``U'' line doublets.}\label{Tab:uline}           
\normalsize
\centering                                      
\begin{tabular}{ccccc}          
\hline\hline                        
 $\nu (J-1)$ &    $\nu(J)$     &  $B$      &   $D$              & $J$        \\
   (MHz)     &    (MHz)        & (MHz)     &   ($10^{-3}$MHz)    &            \\
\hline
\multicolumn{5}{l}{pair-1/pair-2\tablefootmark{(1)}}                           \\
\hline
22304.9      & 25094.3         & 1395.605  &  3.101             & 8          \\
\hline
\multicolumn{5}{l}{pair-1/pair-3}                                            \\
\hline
22304.9      & 25976.2         & 1867.854  & 126.620            &  6         \\
22323.3      & 25992.9         & 1870.374  & 140.234            &  6         \\
\hline
\multicolumn{5}{l}{pair-2/pair-3}                                            \\
\hline
25094.3      & 25976.2         & 451.506   & 2.164              &  28        \\
\hline
 \end{tabular}
\tablefoot{--(1) ``pair-1/pair-2'' means that we assume them to be from the same species.}
                                \normalsize
\end{table*}

\begin{table*}[!hbt]
\caption{Column densities and rotational temperatures of the molecules in IRC +10216.}\label{Tab:irc_rd}             
\normalsize
\centering                                      
\begin{tabular}{cccccccccc}          
\hline\hline                        
          & \multicolumn{5}{c}{This work}                                     & & \multicolumn{2}{c}{Other study} &              \\
\cline{2-5} \cline{7-9}
          & $T_{\rm rot}$        &   $N$                & X($N/N_{\rm H_{2}}$)   &  $\theta_{\rm s}$ &  & &$T_{\rm rot}$ &   $N$        &            \\
Species   & (K)              & (cm$^{-2}$ )            &   &  (\arcsec)        & Note  & &(K)         & (cm$^{-2}$ )  & Ref.      \\
\hline                                                                                
HC$_{3}$N  &  24.7$\pm$18.5   & $(1.4\pm 0.2)\times 10^{15}$& $6.7\times 10^{-7}$ & 30   &  1     & &    28      &  $8.0\times 10^{14}$  & 10        \\
          &                 &                              &  &          &       & &    26      &  $1.7\times 10^{15}$  &   11       \\
          &                 &                              &  &          &       & &    12.7    &  $7.9\times 10^{14}$  &   14       \\ 
HC$_{5}$N  &  18.8$\pm$1.3  & $(4.6\pm 0.2)\times 10^{14}$ & $2.2\times 10^{-7}$ &  30  &  1     & &   29       &  $2.3\times 10^{14}$  & 11       \\ 
HC$_{7}$N  &  12.1$\pm$1.3   & $(3.7\pm 0.4)\times 10^{14}$& $1.8\times 10^{-7}$ &  30  &  2     & &   26       &  $1.29\times 10^{14}$ & 11       \\
HC$_{9}$N  &  20.9$\pm$10.7  & $(2.5\pm 1.4)\times 10^{13}$& $1.2\times 10^{-8}$ &  30  &  2     & &   23       &  $2.7\times 10^{13}$  & 11       \\
          &                 &                            &    &          &       & &    12      &  $4.0\times 10^{13}$  &   17       \\
NH$_{3}$ (para)\tablefootmark{(a)}   & 39.1$\pm$10.5 & $(2.3\pm 0.8)\times 10^{14}$   & $1.1\times 10^{-7}$ &  18  &  3     & & 133        & $4.2\times 10^{13}$  & 20\\
NH$_{3}$ (para)\tablefootmark{(b)}   & 62.3$\pm$5.9  & $(2.7\pm 0.5)\times 10^{14}$   & $1.3\times 10^{-7}$ &  18  &  3     & &            &                     &   \\
NH$_{3}$ (ortho)\tablefootmark{(c)}  & 39.1$\pm$10.5 & $(2.9\pm 0.9)\times 10^{14}$   & $1.4\times 10^{-7}$ &  18  &  3     & &            &                     &   \\
NH$_{3}$ (ortho)\tablefootmark{(d)}  & 130.7$\pm$17.0& $(1.5\pm 0.4)\times 10^{14}$   & $7.1\times 10^{-8}$ &  18  &  3     & &            &                     &   \\
SiC$_{2}$  &  31.8$\pm$0.9   & $(1.2\pm 0.0)\times 10^{15}$ & $5.7\times 10^{-7}$ &  27  &  4     & &  16        &  $2.2\times 10^{14}$ & 11        \\
          &                 &                              &  &          &       & &    96      &  $2.35\times 10^{15}$  &   10       \\ 
          &                 &                              &  &          &       & &    14      &  $9.5\times 10^{14}$  &   16       \\
          &                 &                              &  &          &       & &    60      &  $1.9\times 10^{15}$  &   16       \\
SiC$_{4}$  &  22.3$\pm$17.5  & $(1.1\pm 0.4)\times 10^{13}$& $5.2\times 10^{-9}$  &  27     &  5  & &  15        &  $7\times 10^{12}$    &   12    \\
C$_{2}$S   &  39.2$\pm$135.3 & $(2.3\pm 0.8)\times 10^{14}$ & $1.1\times 10^{-7}$  &  30     &  2  & &  14        &  $1.5\times 10^{14}$  &   13    \\
C$_{3}$S   &  20.4$\pm$13.7   & $(2.2\pm 0.4)\times 10^{13}$ & $1.0\times 10^{-8}$  &  30     &  2  & &  33       &  $4.9\times 10^{13}$  &   11    \\
          &                 &                              &  &          &       & &  17        &  $3.6\times 10^{13}$  &   14      \\
c-C$_{3}$H$_{2}$ (para)  & 5.5$\pm$0.6  & $(8.9\pm 0.5)\times 10^{13}$  & $4.2\times 10^{-8}$   & 30   & 6   & &    8     &  $1.22\times 10^{14}$ &10\\
c-C$_{3}$H$_{2}$ (ortho) & 5.5$\pm$0.6  & $(1.2\pm 0.3)\times 10^{14}$  & $5.7\times 10^{-8}$   & 30   & 6   & &    5.7   &  $4.82\times 10^{14}$ &10\\
C$_{6}$H$^{-}$ &  26.9$\pm$4.0 & $(5.8\pm 0.5)\times 10^{12}$ & $2.8\times 10^{-9}$ & 30    &   7   & &            & $4\times 10^{12}$    &     15   \\
          &                 &                             &   &          &       & &  30        &  $(6.1-8.0) \times 10^{12}$   & 21          \\
MgNC      &  25.8$\pm$54.9  & $(3.9\pm 1.5)\times 10^{13}$& $1.9\times 10^{-8}$   & 30       & 8  & &  15        &$2.5\times 10^{13}$    &   11     \\
          &                 &                             &   &          &       & &  8.6       &$7.8\times 10^{13}$    &    10       \\
C$_{3}$N   &  20.2$\pm$1.1  & $(3.1\pm 0.3)\times 10^{14}$& $1.5\times 10^{-7}$   &  36     &  9     & &  35      &  $4.54\times 10^{14}$ &  10    \\
          &                 &                             &   &          &       & &  15        & $4.1\times 10^{14}$   &    11       \\
          &                 &                             &   &          &       & &  20        & $2.5\times 10^{14}$   &    18       \\
C$_{4}$H   &  18.5$\pm$7.6   & $(2.4\pm 0.2)\times 10^{15}$& $1.1\times 10^{-6}$   &  30     &  6     & &  53        & $8.1\times 10^{15}$   & 10  \\
          &                 &                              &  &          &       & &  35        & $3.0\times 10^{15}$   &    18       \\
          &                 &                              &  &          &       & &  48        & $5.6\times 10^{15}$   &    16       \\
          &                 &                              &  &          &       & &  15        & $2.4\times 10^{15}$   &    11       \\
l-C$_{5}$H ($^{2}\Pi_{1/2}$) & 8.3$\pm$2.0  & $(2.9\pm 0.6)\times 10^{13}$ & $1.4\times 10^{-8}$ &  30     &  6     & &  27 & $2.9\times 10^{14}$ & 11\\
l-C$_{5}$H ($^{2}\Pi_{3/2}$) & 20.9$\pm$19.9 & $(1.2\pm 2.3)\times 10^{14}$ & $5.7\times 10^{-8}$ & 30      &  6     & &   39 & $2.0\times 10^{14}$ & 11\\
          &                 &                                &         &  &       & &  25        & $4.4\times 10^{13}$   &    18       \\
C$_{6}$H ($^{2}\Pi_{1/2}$) & 20.7$\pm$2.4  & $(1.0\pm 0.2)\times 10^{14}$ & $4.8\times 10^{-8}$ &  30    & 6    & &  46  & $1.13\times 10^{14}$ & 11  \\
C$_{6}$H ($^{2}\Pi_{3/2}$) & 47.2$\pm$10.3 & $(1.0\pm 0.1)\times 10^{14}$& $4.8\times 10^{-8}$ & 30     &  6   & &  35  & $1.65\times 10^{14}$ & 11  \\
          &                 &                                &        &  &        & &  35        & $5.5\times 10^{13}$   &    18       \\
C$_{8}$H ($^{2}\Pi_{3/2}$) & 13.9$\pm$1.3  & $(8.4\pm 1.4)\times 10^{12}$ & $4.0\times 10^{-9}$ &  30     &  6     & &  13  & $8\times 10^{12}$ & 19  \\
          &                 &                                &          & &      & &  52        & $1.0\times 10^{13}$   &    18       \\
H$^{13}$CCCCCN & 27.6$\pm$3.5 & $(1.2\pm 0.1)\times 10^{13}$   & $5.7\times 10^{-8}$ & 30   &  2	  & &		 &		&	   \\
HCC$^{13}$CCCN & 9.4$\pm$1.6 & $(6.4\pm 1.1)\times 10^{12}$   & $3.0\times 10^{-9}$ & 30	 &  2	  & &		 &		&	   \\
HCCC$^{13}$CCN & 13.7$\pm$3.2 & $(7.2\pm 1.3)\times 10^{12}$   & $3.4\times 10^{-9}$ & 30	 &  2	  & &		 &		&	   \\
HCCCC$^{13}$CN & 19.6$\pm$16.1 & $(1.1\pm 0.4)\times 10^{13}$  & $5.2\times 10^{-8}$ & 30	 &  2	  & &		 &		&	   \\
\hline
 \end{tabular}
 \tablefoot{\\ To Col.\,1 -- (a) NH$_{3}$ (1,1) and (2,2) are included in the fit. (b) NH$_{3}$ (1,1), (2,2) and (4,4) are 
    included in the fit. (c) NH$_{3}$ (3,3) with $T_{\rm rot}$ derived from the (1,1) and (2,2) lines. (d) NH$_{3}$ (3,3) and NH$_{3}$ (6,6) are included in the fit.\\
    Notes on the source sizes ($\theta_{\rm s}$) of individual molecules (Col.\,4).-- (1) Source sizes are determined by new JVLA observations (Keller, D. in preparation); (2) Their sizes are taken to be the same as that of HC$_{5}$N; (3) NH$_{3}$ may originate from the same region as the high $J$ transitions of centrally peaked molecules like SiS (6--5) which has a typical size of 18\arcsec\,\citep{1993AJ....105..576B,1995Ap&SS.224..293L}; (4) \citet{1995Ap&SS.224..293L}; (5) the source size is assumed to be the same as that of SiC$_{2}$; (6) the size is taken to be the same as that of C$_{4}$H \citep{1993A&A...280L..19G}; (7) the size is based on the chemical model of \citet{2009ApJ...697...68C}; (8) \citet{1993A&A...280L..19G}; (9) \citet{1993AJ....105..576B}.\\
    References for rotational temperatures and column densities from the literature.-- (10) \citet{2008ApJS..177..275H}; (11) \citet{1995PASJ...47..853K}; (12) \citet{1989ApJ...345L..83O}; (13) \citet{1987A&A...181L...9C}; (14) \citet{1993ApJ...417L..37B}; (15) \citet{2006ApJ...652L.141M} for the column density only; (16) \citet{1992ApJS...83..363A}; (17) \citet{1992ApJ...400..551B}; (18) \citet{2000AAS..142..181C}; (19) \citet{2007ApJ...664L..47R}; (20) \citet{1984A&A...138L...5N}; (21) \citet{2007ApJ...661L..61K}.\\
 }
 			        \normalsize
\end{table*}

\clearpage

\clearpage

\begin{figure*}[!htbp]
\centering
\includegraphics[width = 0.4 \textwidth]{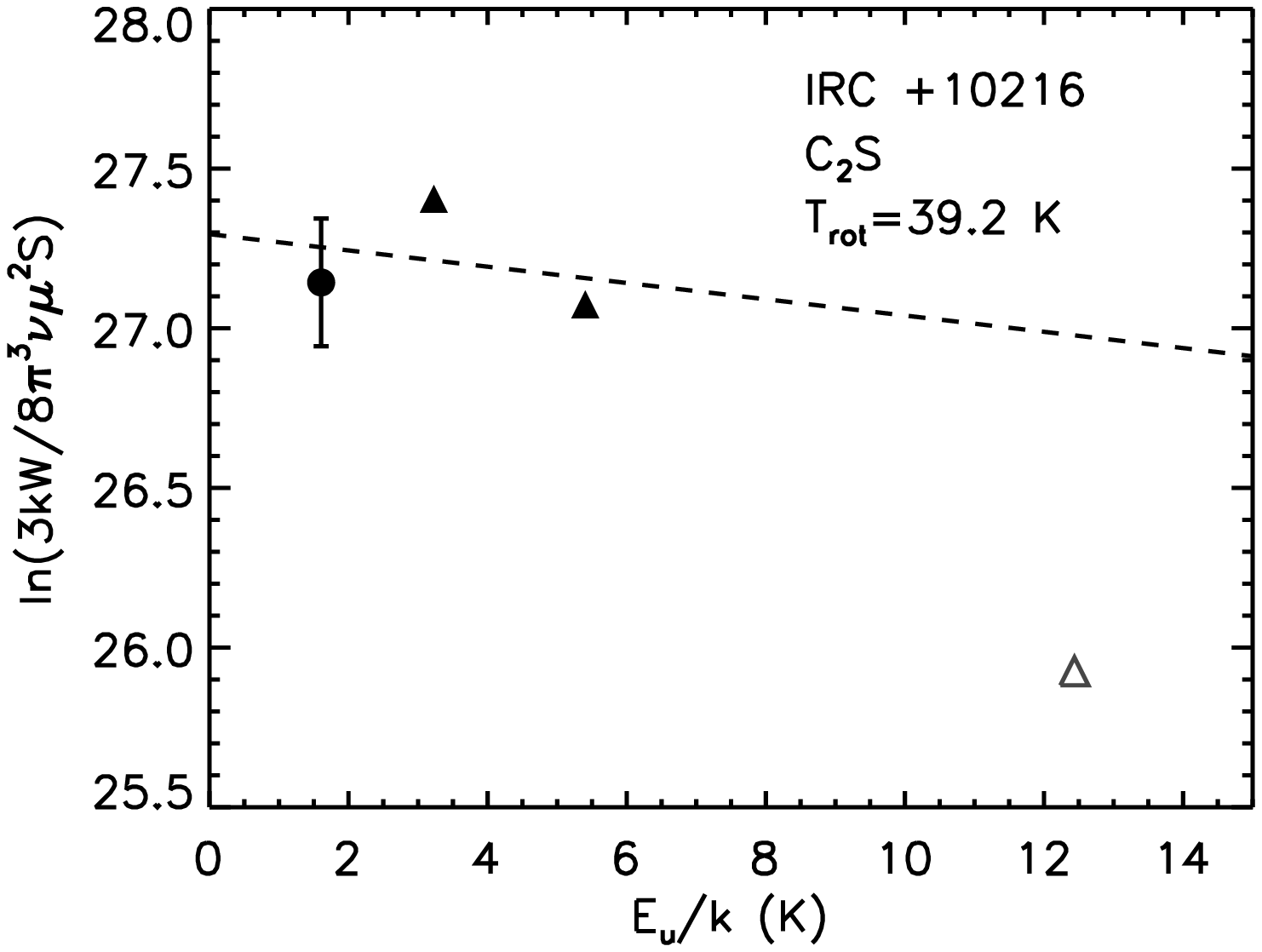}
\includegraphics[width = 0.4 \textwidth]{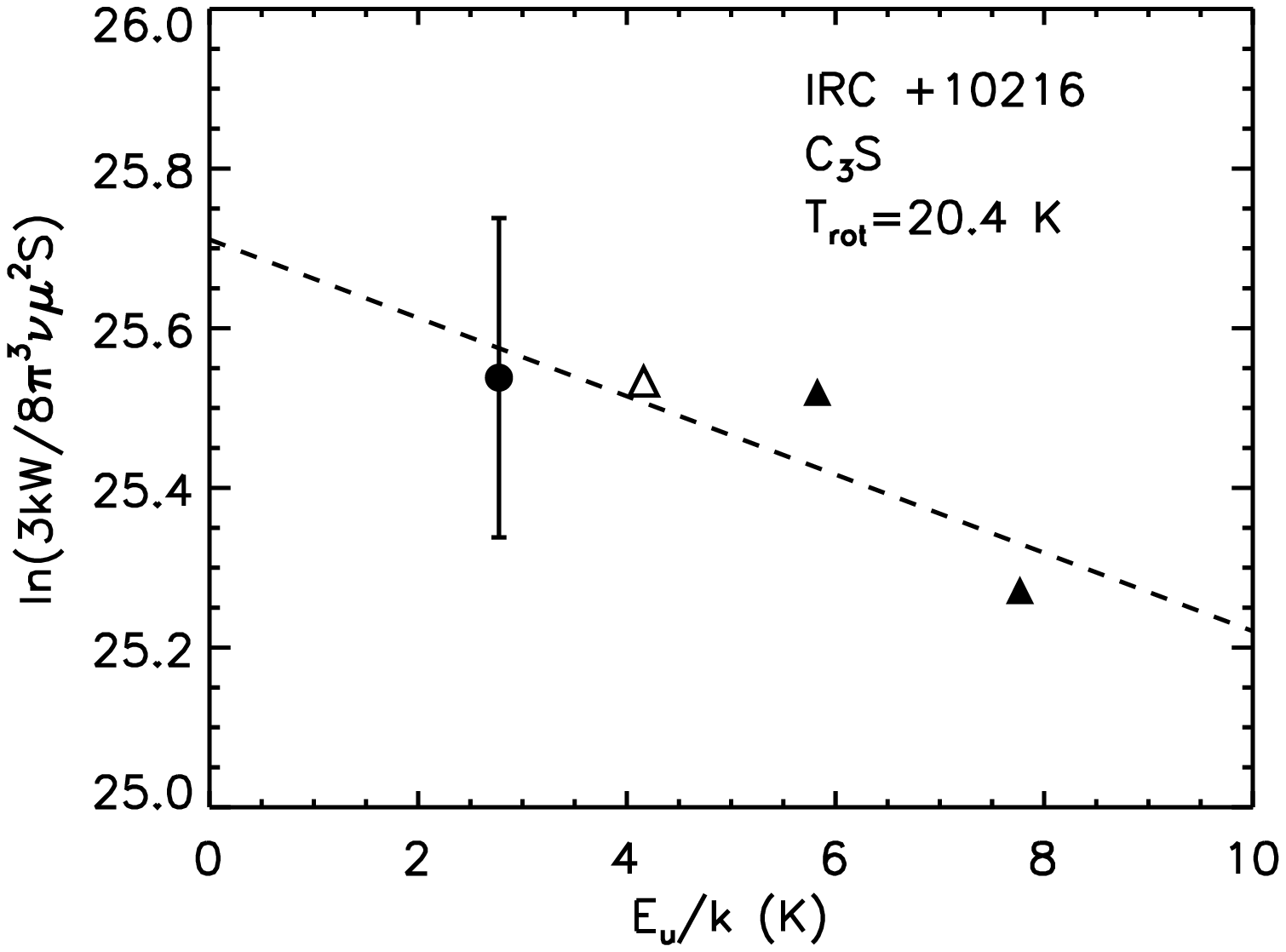}
\includegraphics[width = 0.4 \textwidth]{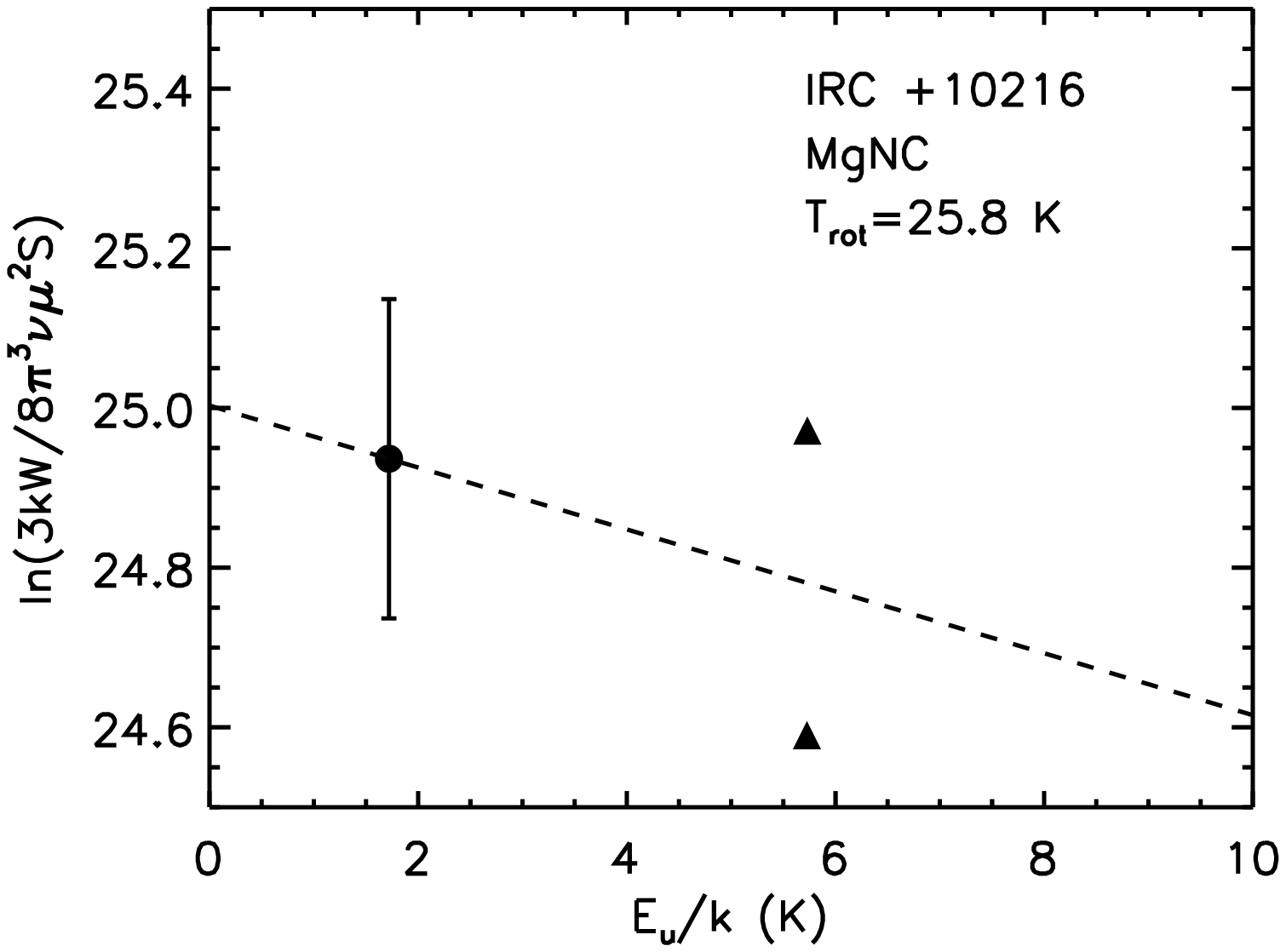}
\includegraphics[width = 0.4 \textwidth]{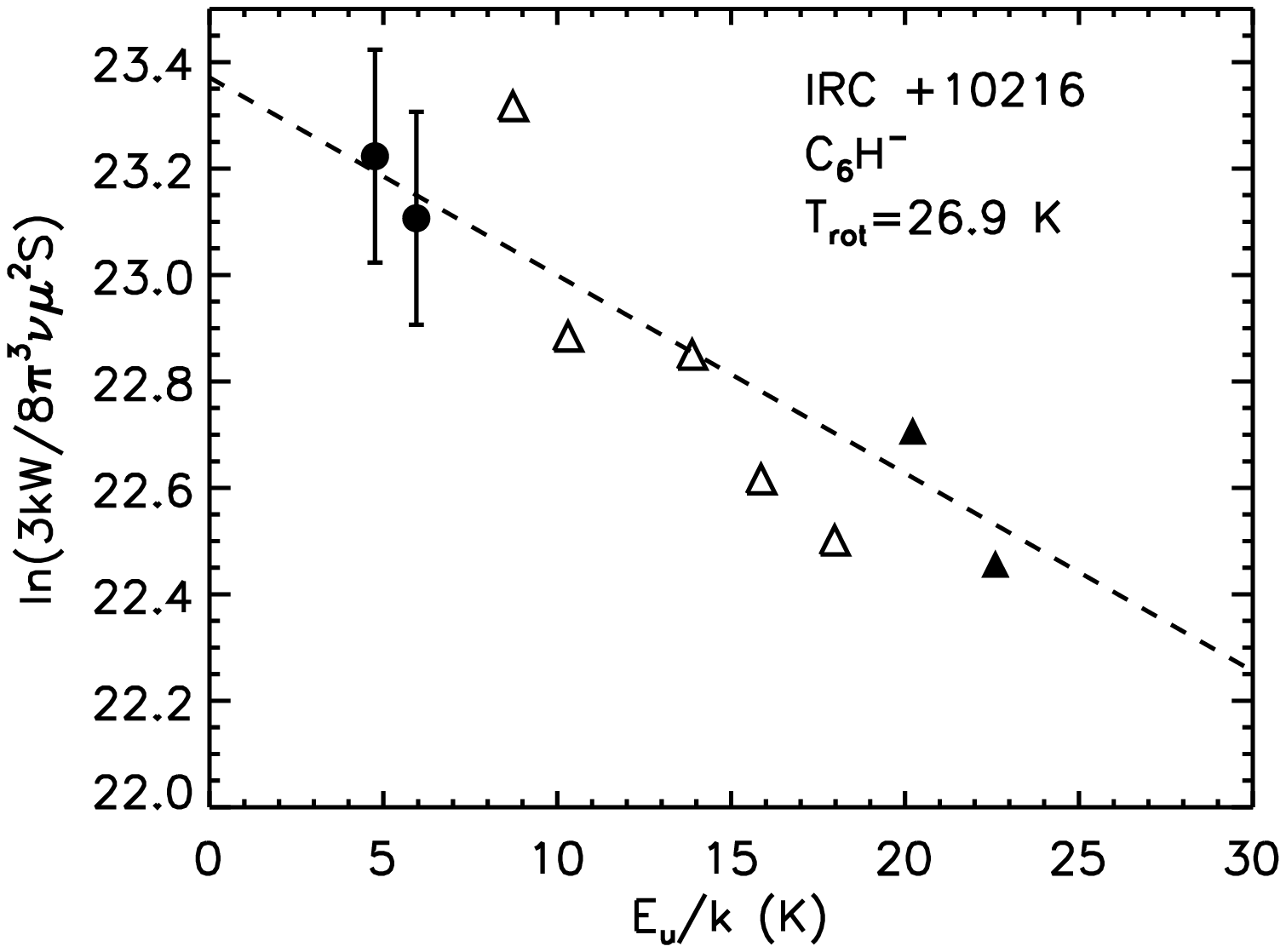}
\includegraphics[width = 0.4 \textwidth]{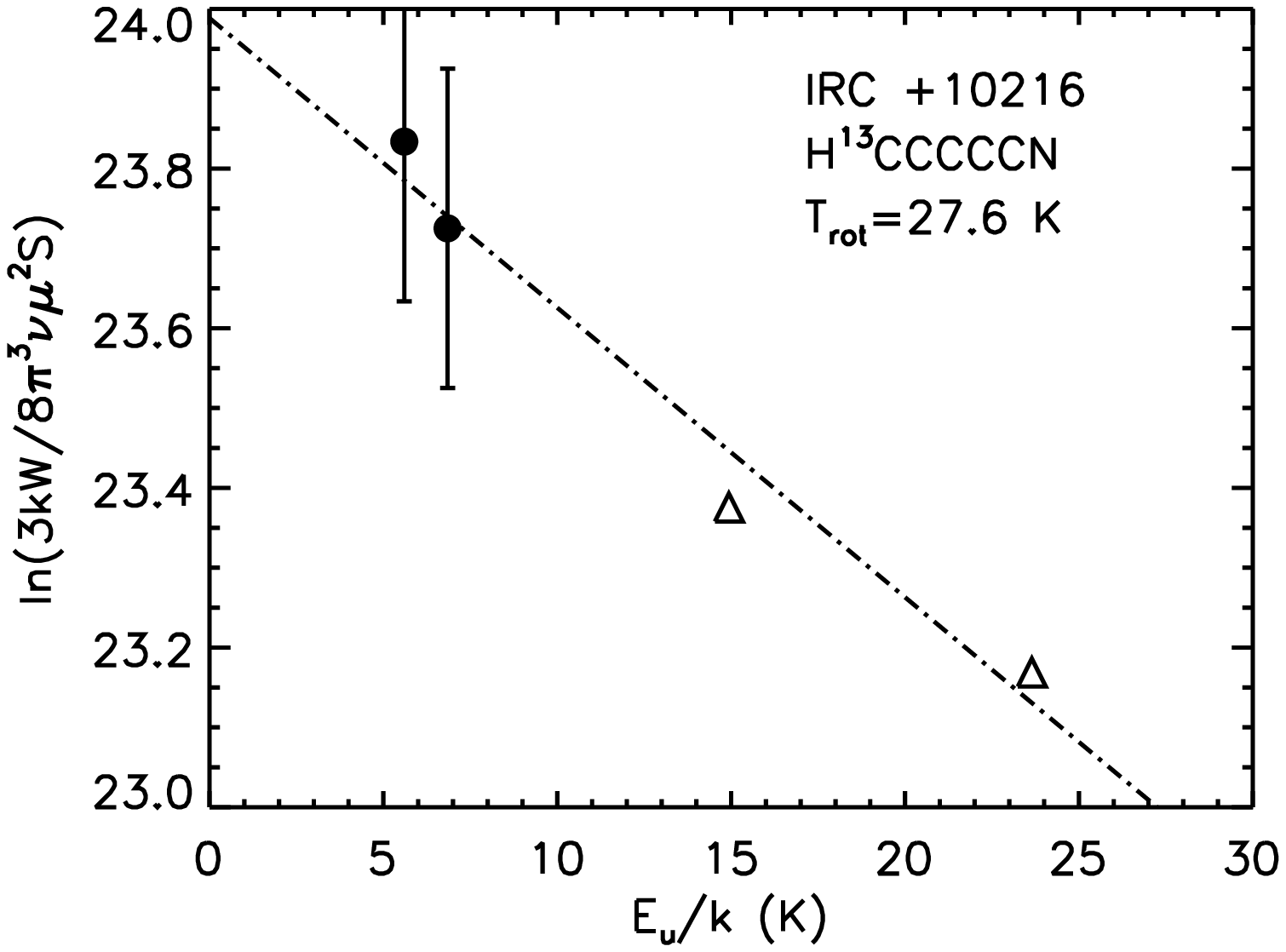}
\includegraphics[width = 0.4 \textwidth]{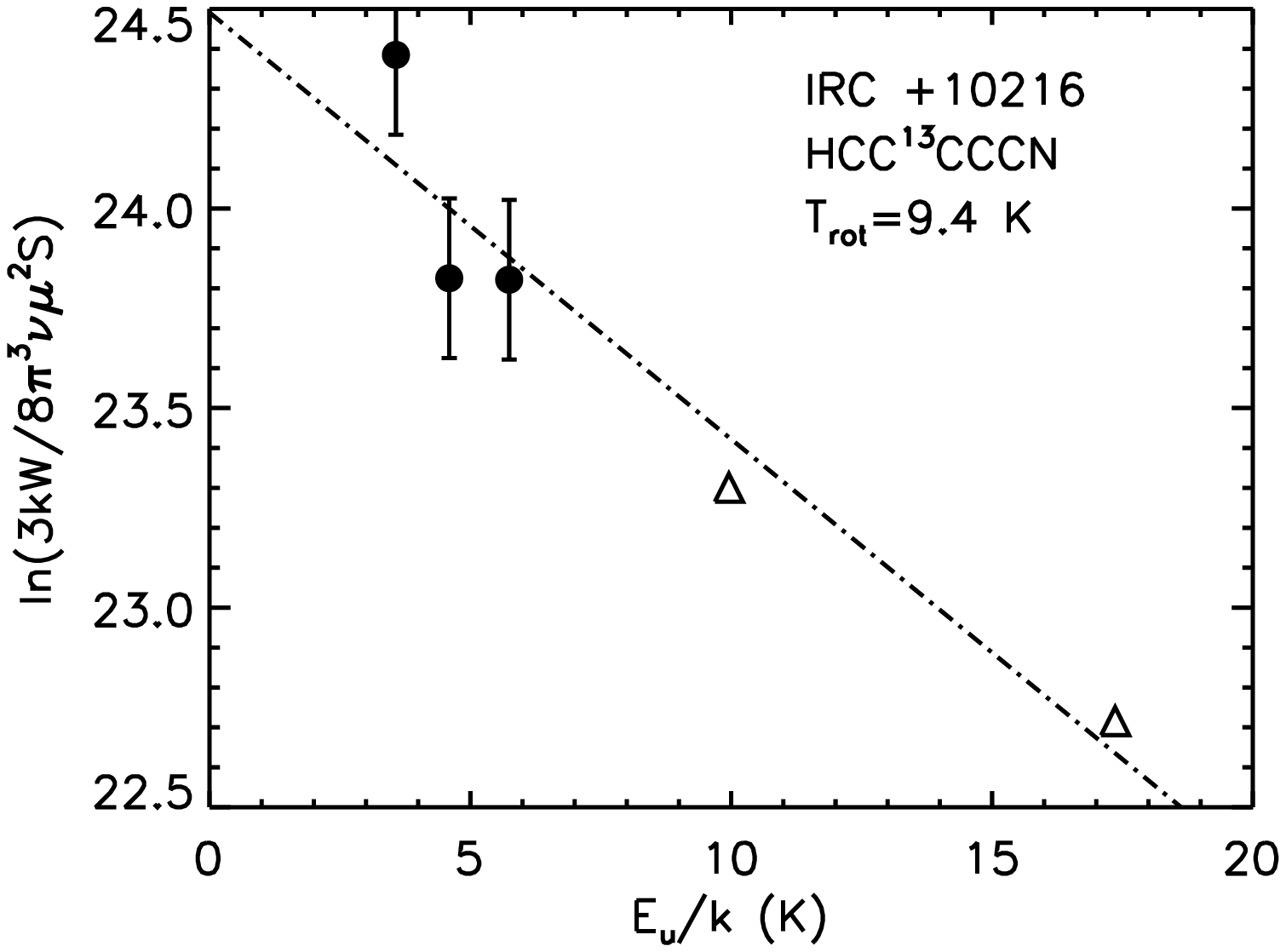}
\includegraphics[width = 0.4 \textwidth]{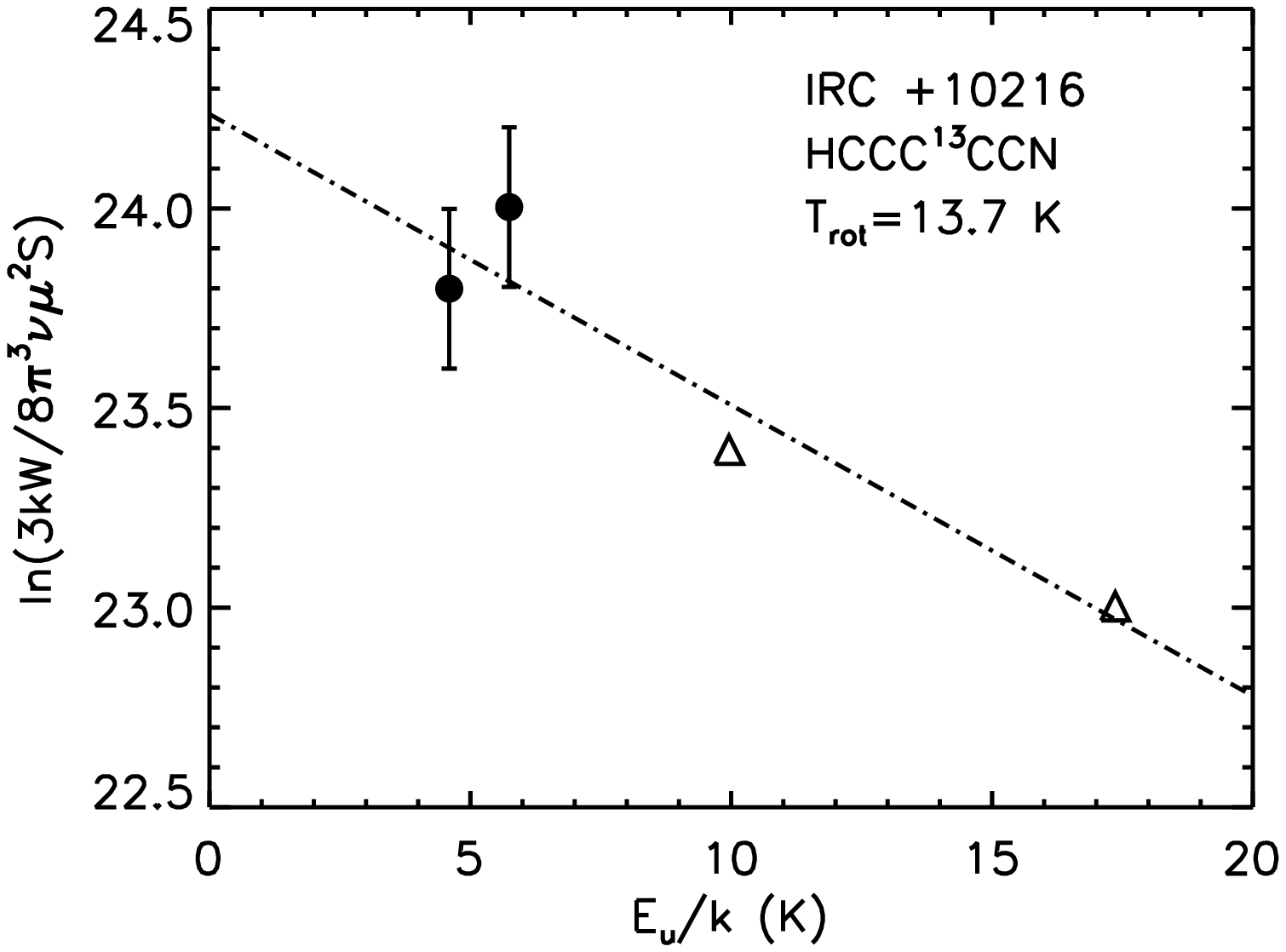}
\includegraphics[width = 0.4 \textwidth]{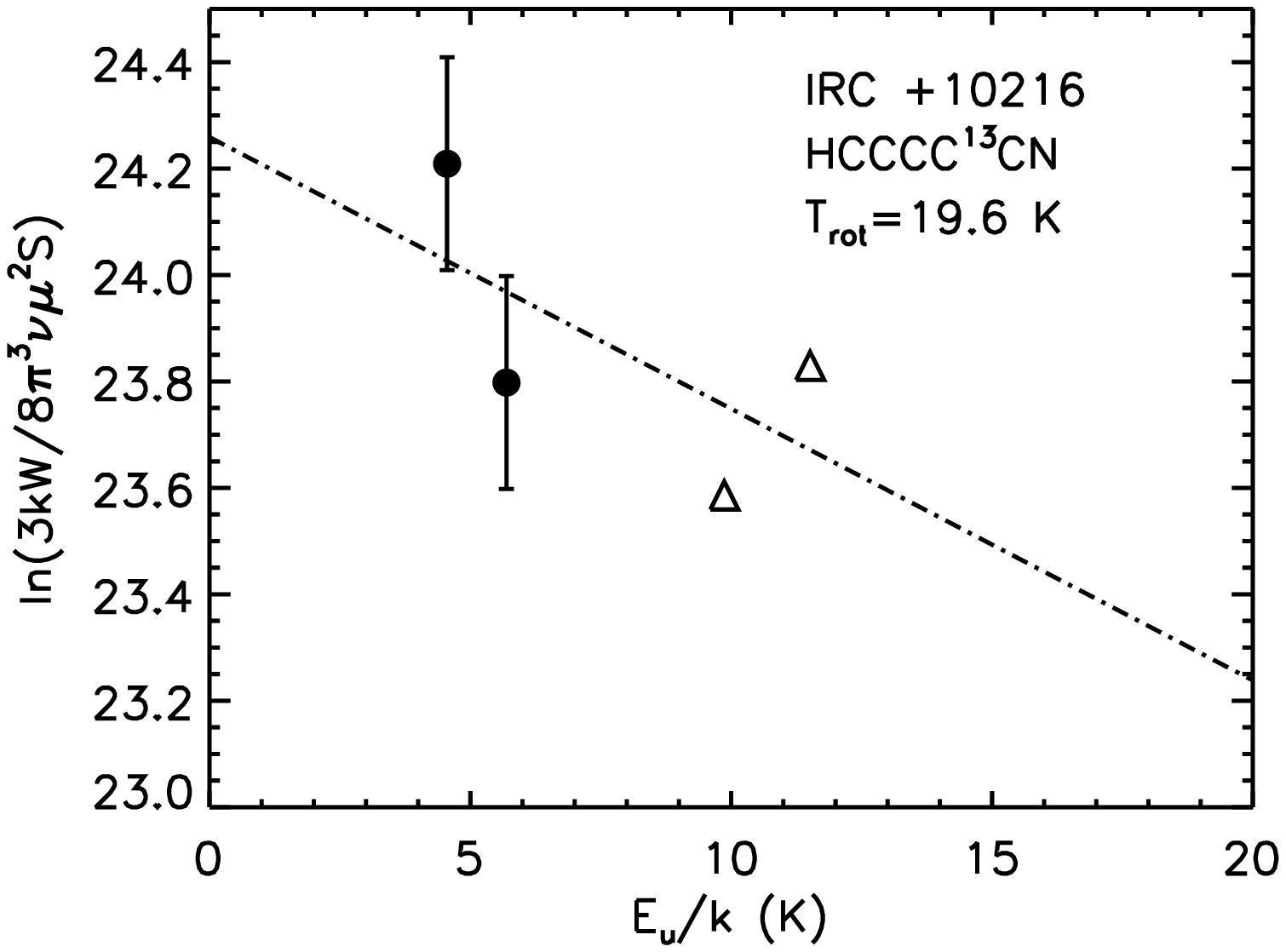}
\centerline{Fig. \ref{Fig:ircrd}. --- Continued.}
\end{figure*}

\begin{figure*}[!htbp]
\centering
\includegraphics[width = 0.4 \textwidth]{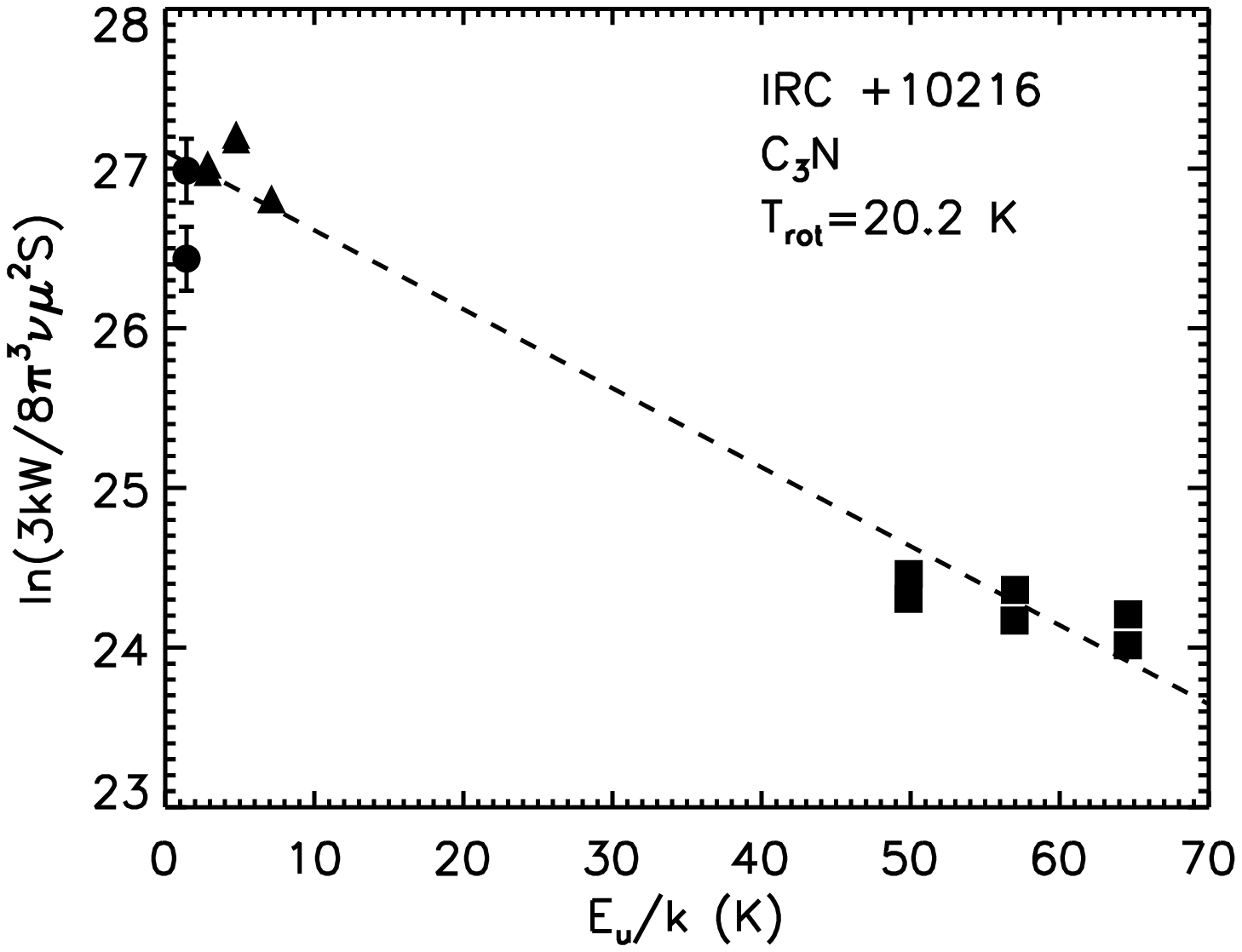}
\includegraphics[width = 0.4 \textwidth]{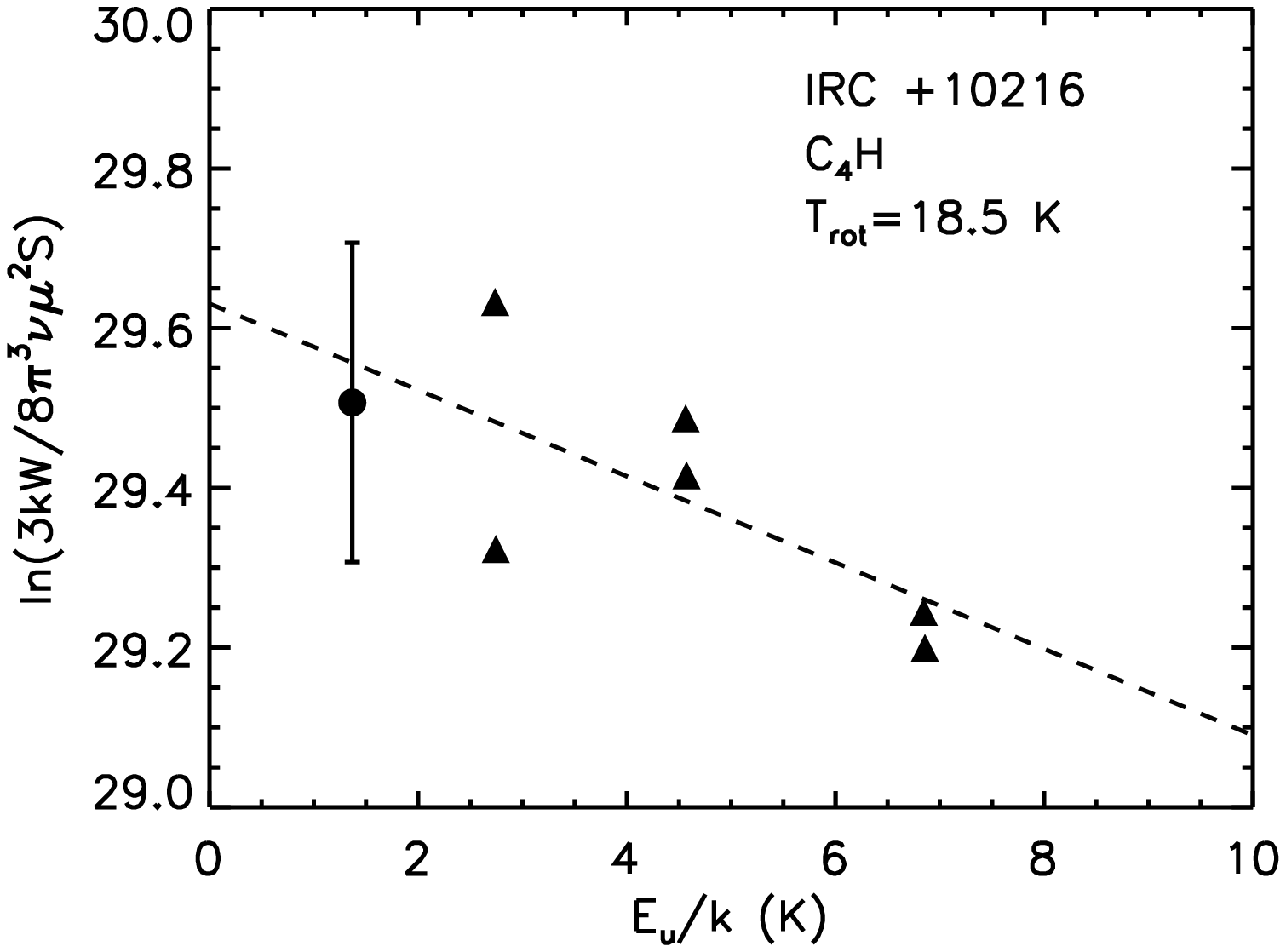}
\includegraphics[width = 0.4 \textwidth]{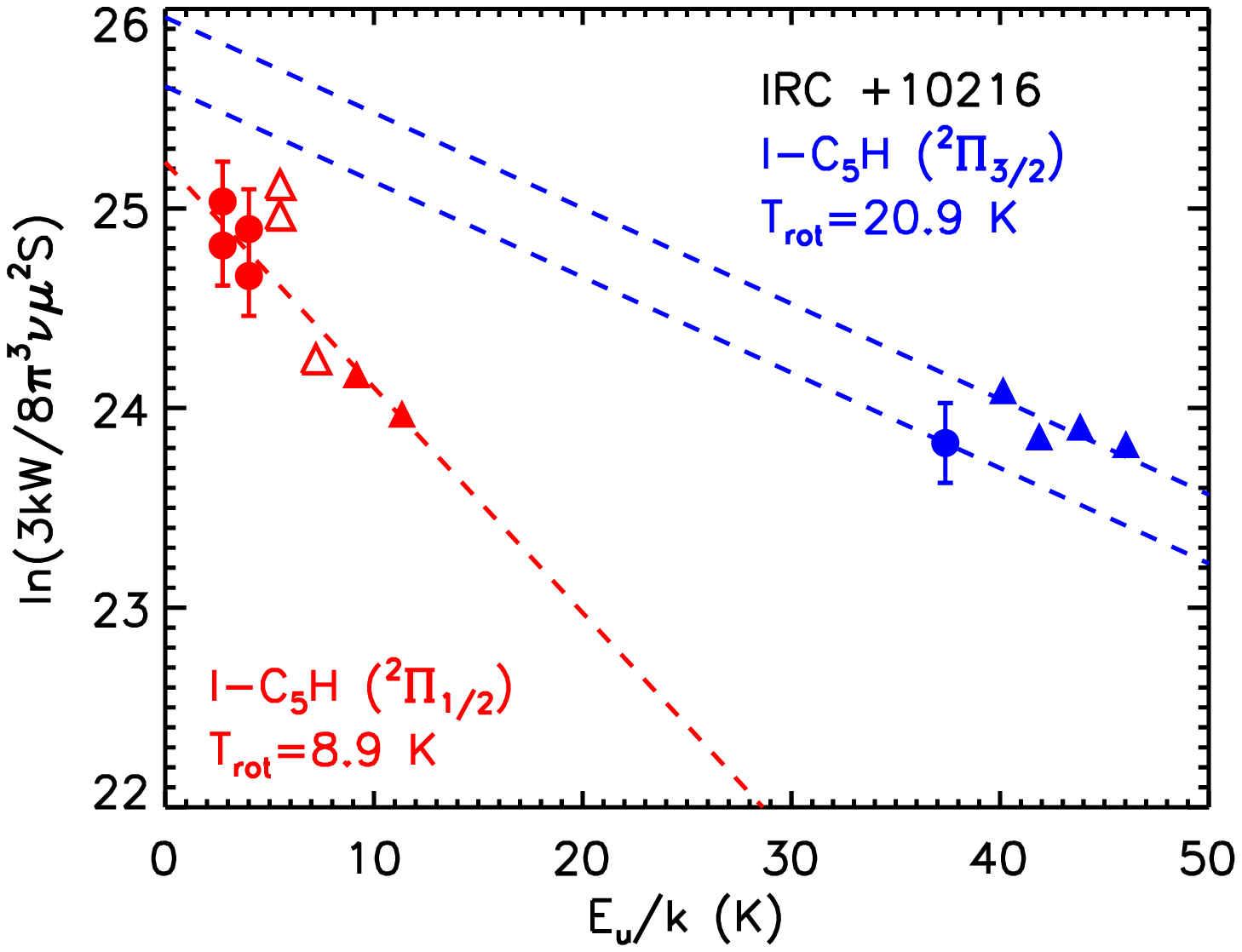}
\includegraphics[width = 0.4 \textwidth]{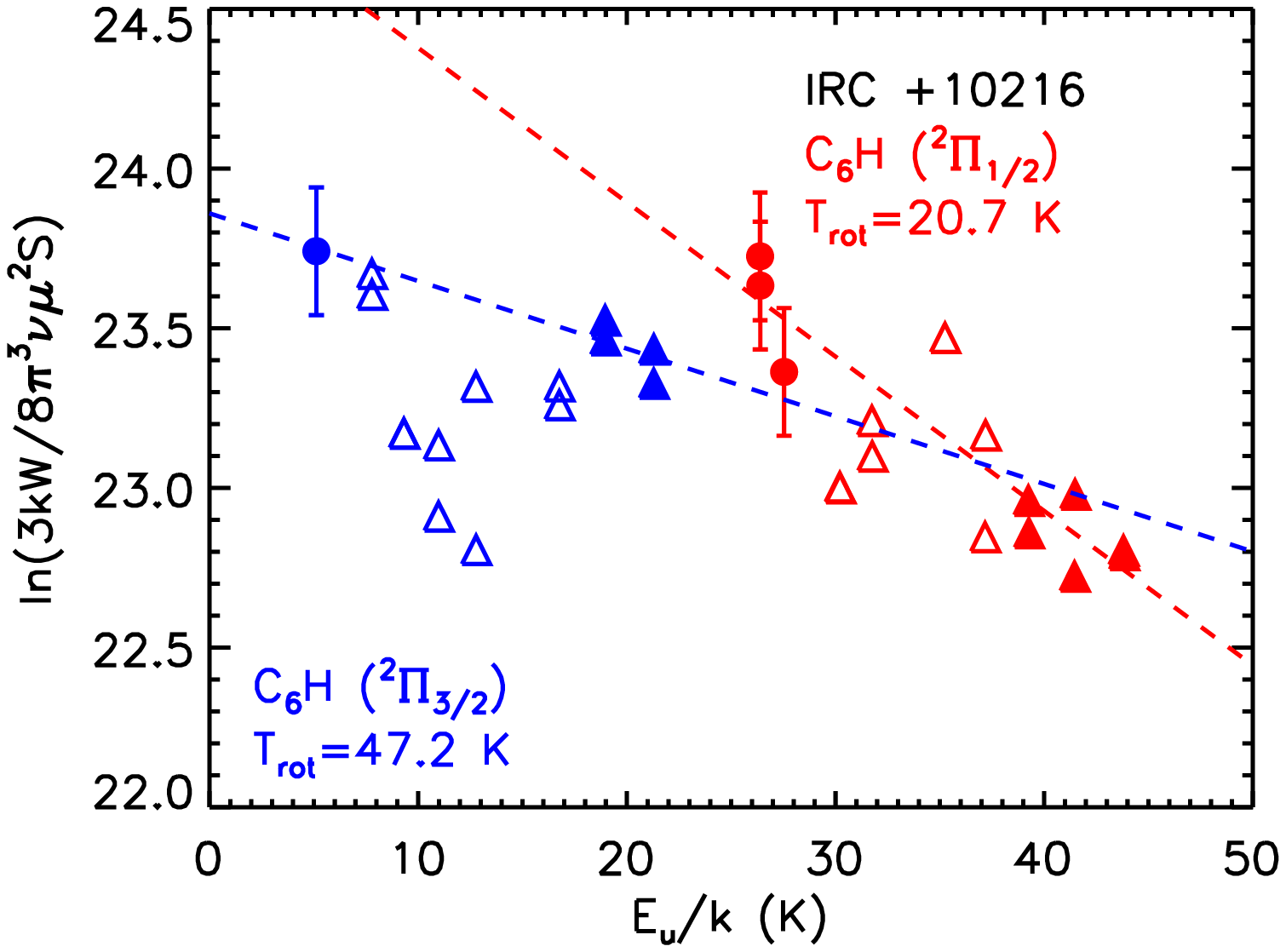}
\includegraphics[width = 0.4 \textwidth]{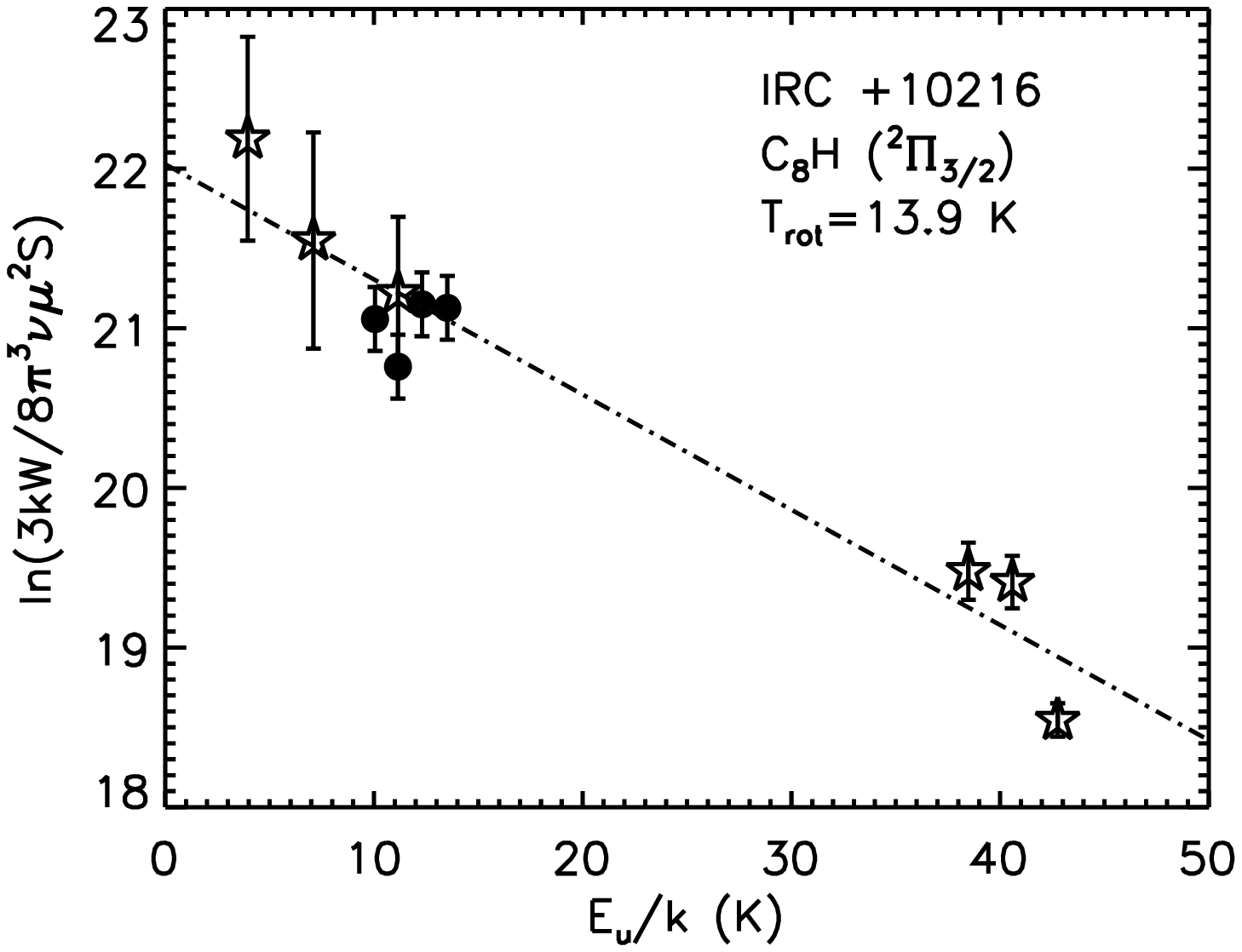}
\includegraphics[width = 0.4 \textwidth]{rd/irc_c8hw.eps}
\centerline{Fig. \ref{Fig:ircrd}. --- Continued.}
\end{figure*}

\begin{figure*}[!htbp]
\centering
\includegraphics[width = 1.0 \textwidth]{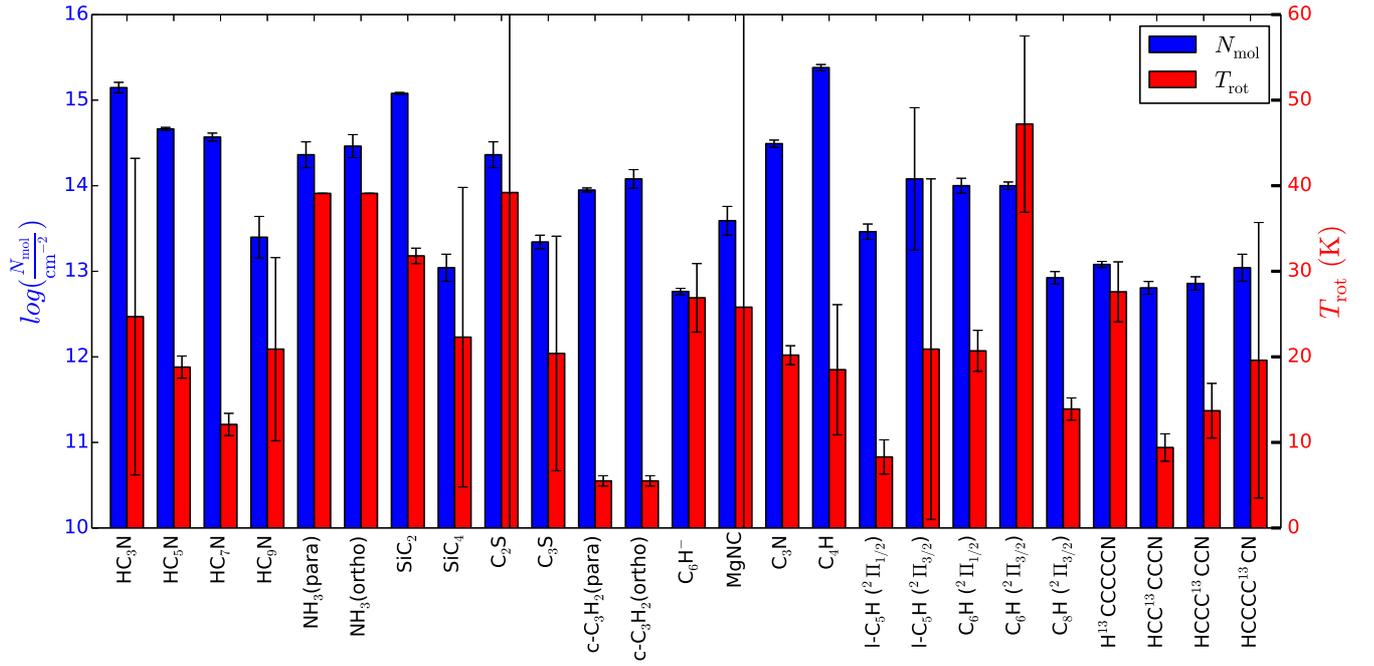}
\caption{{Comparison of the column densities and rotational temperatures of different molecules. The labels for column densities are blue on the left side, while those for rotational temperatures are red on the right side. Here,  a $T_{\rm rot}$ of 39.1~K has been taken for para and ortho NH$_{3}$ (see Sect.~\ref{phy} and Table~\ref{Tab:irc_rd})} \label{Fig:coltrot}}
\end{figure*}

\begin{figure*}[!htbp]
\centering
\includegraphics[width = 0.42 \textwidth]{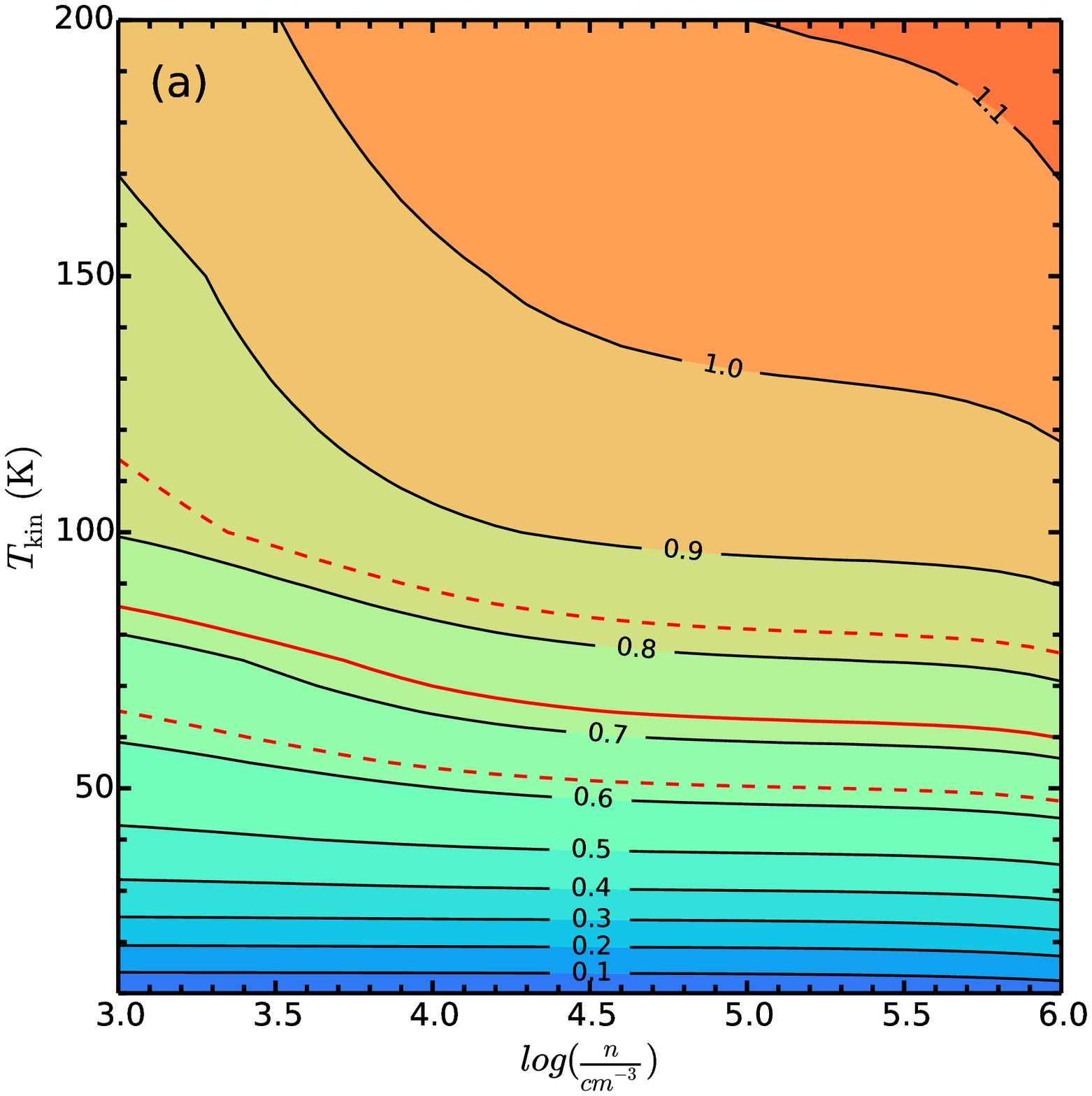}
\includegraphics[width = 0.42 \textwidth]{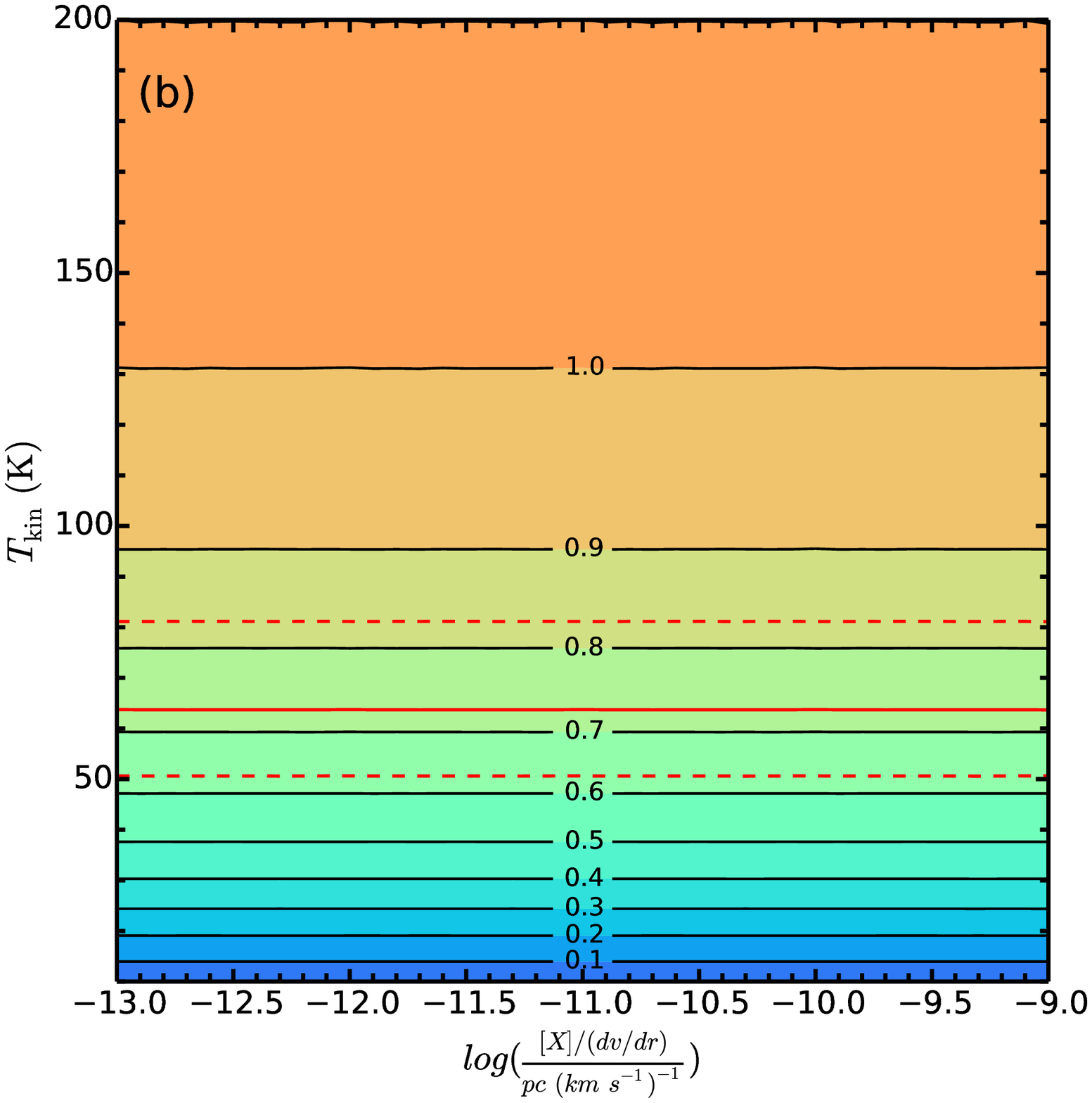}
\includegraphics[width = 0.42 \textwidth]{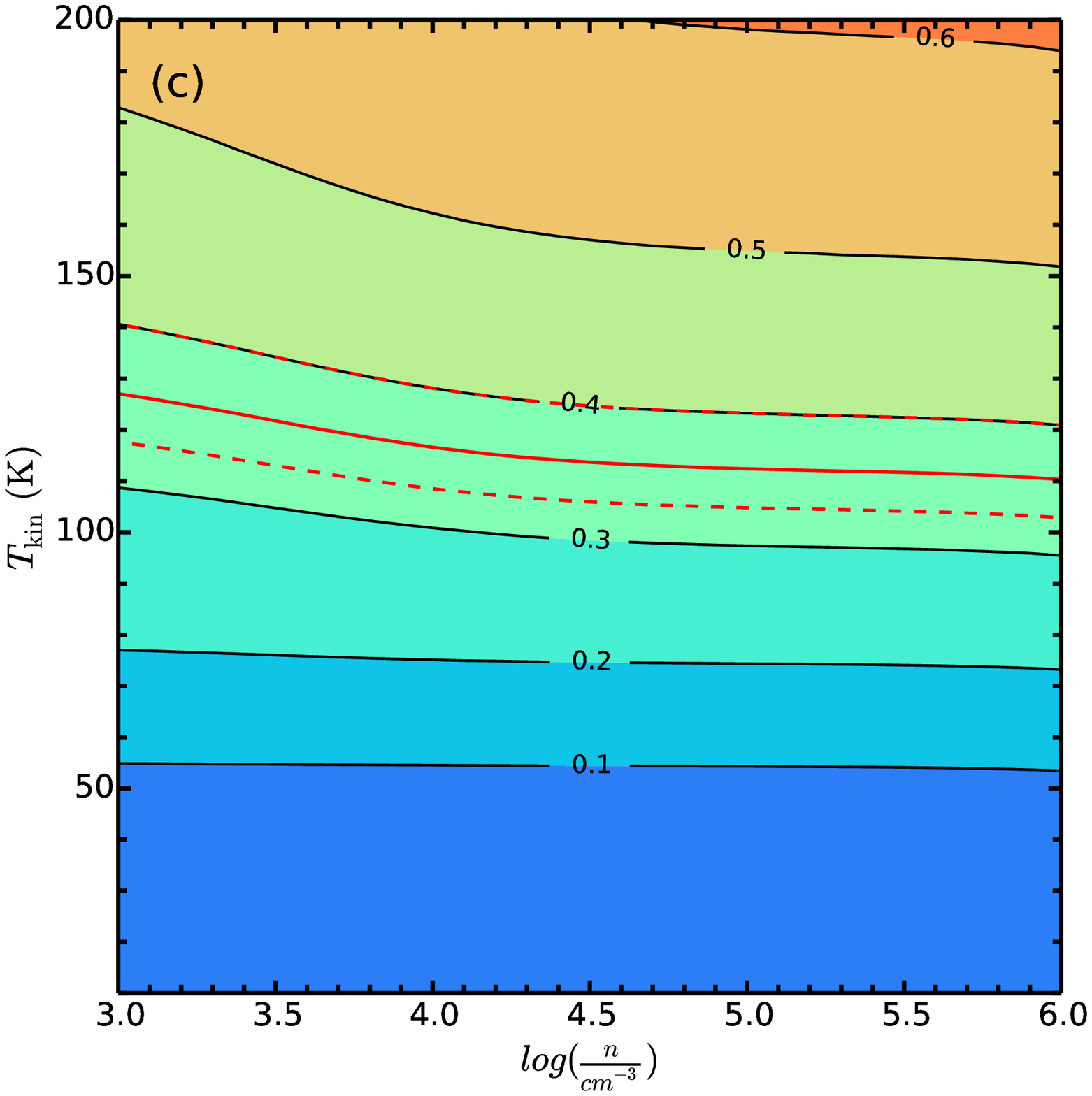}
\includegraphics[width = 0.42 \textwidth]{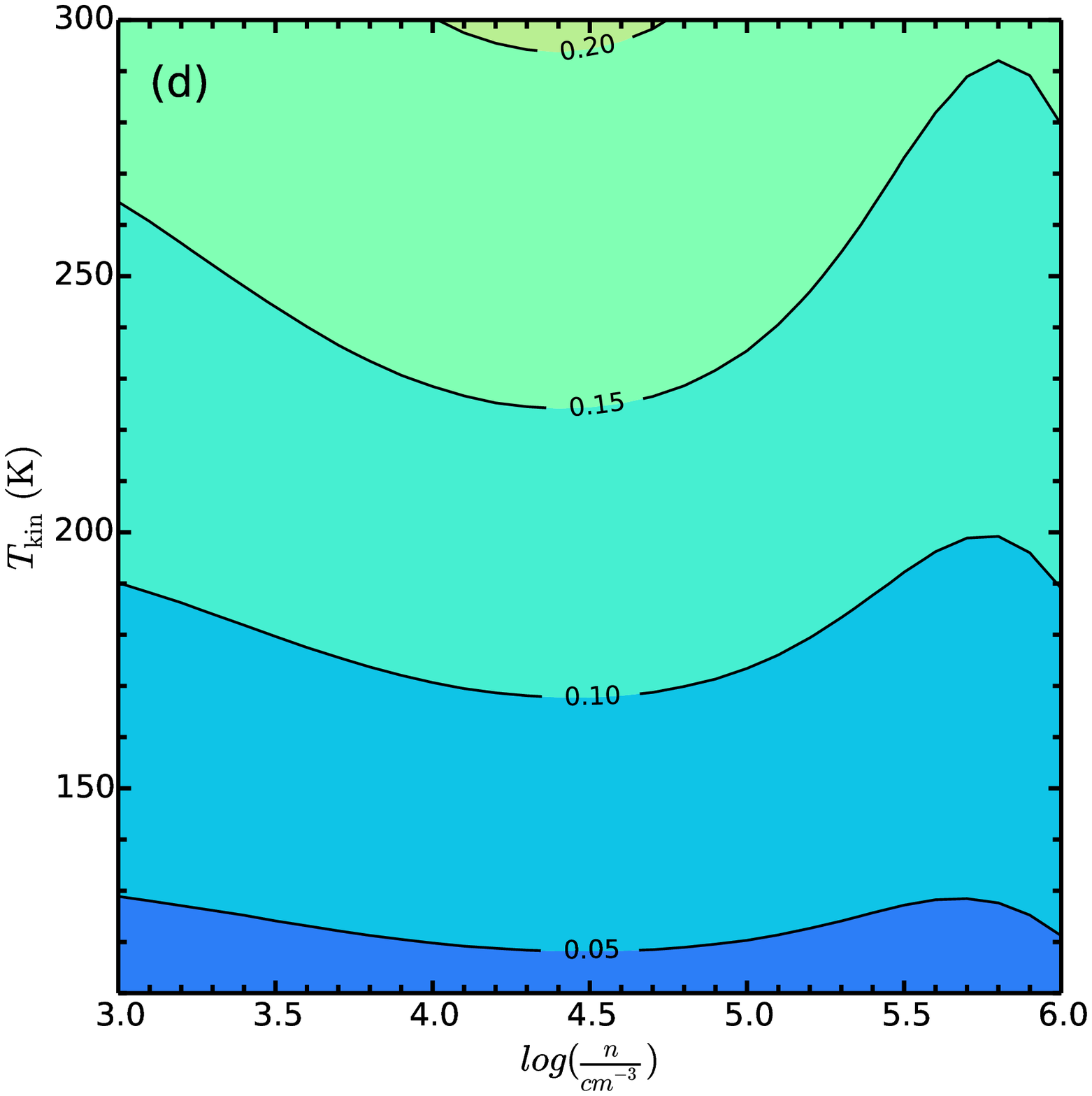}
\caption{{RADEX non-LTE modeling for NH$_{3}$. (a) the integrated intensity ratio $\frac{\rm NH_{3} (2,2)}{\rm NH_{3} (1,1)}$ as a function of $n_{\rm H_{2}}$ and $T_{\rm kin}$ for a given para-NH$_{3}$ abundance per velocity gradient [X]/(d$v$/d$r$) of $4.4\times 10^{-11}$~pc~(\kms)$^{-1}$. (b) the integrated intensity ratio $\frac{\rm NH_{3} (2,2)}{\rm NH_{3} (1,1)}$ as a function of [X]/(d$v$/d$r$) and $T_{\rm kin}$ for a given density of 10$^{5}$\,cm$^{-3}$. In both panels, the levels go from 0.1 to 1 by 0.1. (c) the integrated intensity ratio $\frac{\rm NH_{3} (4,4)}{\rm NH_{3} (2,2)}$ as a function of $n_{\rm H_{2}}$ and $T_{\rm kin}$ for a given para-NH$_{3}$ abundance per velocity gradient [X]/(d$v$/d$r$) of $4.4\times 10^{-11}$~pc~(\kms)$^{-1}$. The levels start from 0.1 to 0.6 by 0.1. The red solid line and the dashed lines represent the observed line ratio as well as its lower and upper limit in Fig.~\ref{Fig:lvg}a, b and c. (d) the integrated intensity ratio $\frac{\rm NH_{3} (6,6)}{\rm NH_{3} (3,3)}$ and as a function of $n_{\rm H_{2}}$ and $T_{\rm kin}$ for a given ortho-NH$_{3}$ abundance per velocity gradient [X]/(d$v$/d$r$) of $5.6\times 10^{-11}$~pc~(\kms)$^{-1}$. The levels start from 0.05 to 0.2 by 0.05. The observed line ratio $\frac{\rm NH_{3} (6,6)}{\rm NH_{3} (3,3)}$ = (0.28$\pm$0.08) is beyond the modeled range, indicating that the kinetic temperature is either higher than 300~K or that NH$_{3}$ (6,6) is affected by population inversion.} \label{Fig:lvg}}
\end{figure*}

\begin{figure*}[!htbp]
\centering
\includegraphics[width = 0.8 \textwidth]{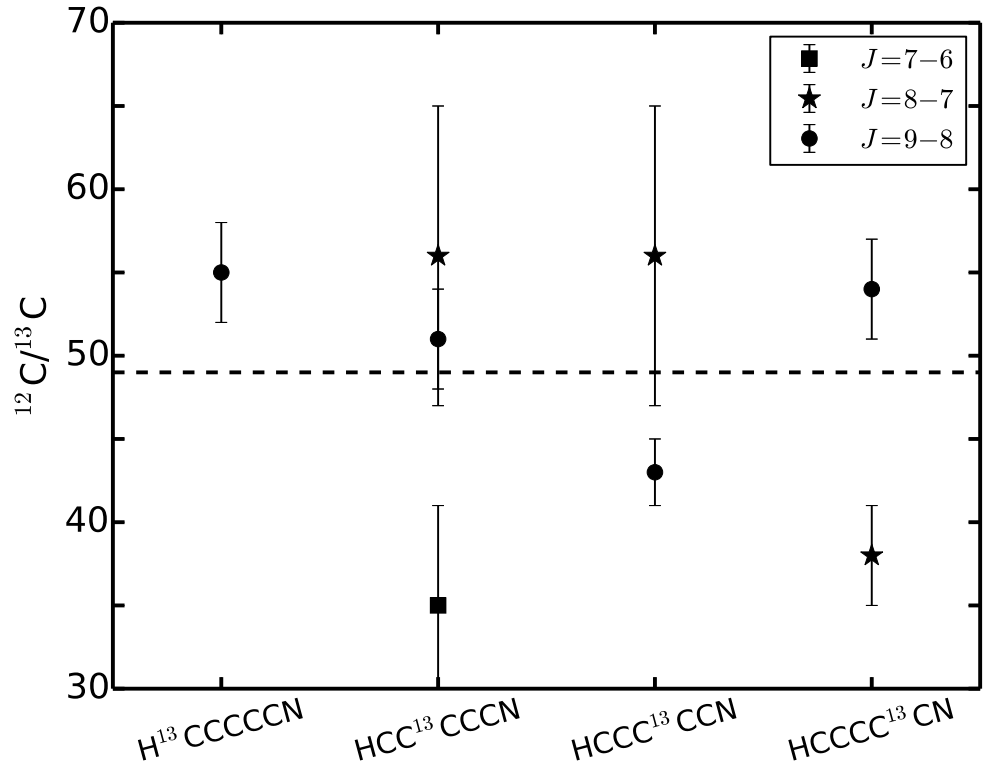}
\caption{{$^{12}$C/$^{13}$C isotopic abundance ratios from HC$_{5}$N. The ratios are based on the integrated intensity of the transitions of HC$_{5}$N and its $^{13}$C isotopologues. The ratios derived from the $J$=7--6, $J$=8--7, and $J$=9--8 transitions are denoted with filled squares, pentagrams, and circles, respectively. The dashed line denotes the unweighted average $^{12}$C/$^{13}$C value.} \label{Fig:cr}}
\end{figure*}

\begin{figure*}[!htbp]
\centering
\includegraphics[width = 1.0 \textwidth]{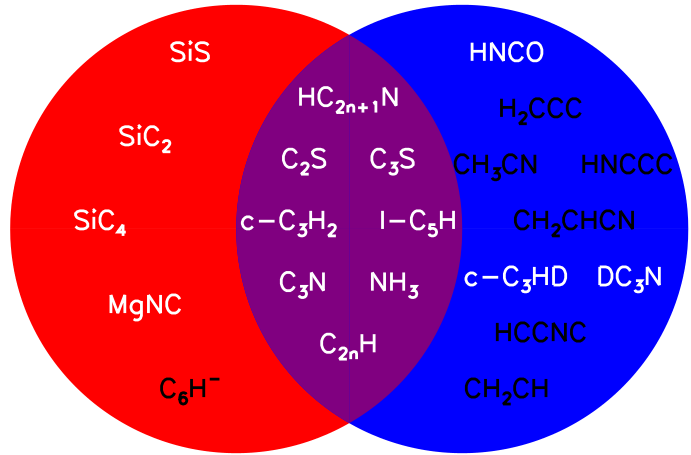}
\caption{{Schematic diagram of detected molecules in the same $\lambda\sim$ 1.3 cm spectral range toward IRC +10216 and TMC-1. Molecules in the red, blue, and purple regions indicate that they are detected in IRC +10216, TMC-1, and both sources, respectively. The molecules in black indicate that they have been seen in both sources but are not detected toward the other source in the $\lambda\sim$ 1.3 cm spectral range (see Sect.\,\ref{aburatio}).} \label{Fig:irc-tmc}}
\end{figure*}

\clearpage

\begin{appendix}\label{a}
\section{Observed lines in this survey}
\clearpage

\renewcommand{\tabcolsep}{0.15 cm}
\normalsize \longtab{1}{
\begin{longtable}{cccccccc}
	\caption{Lines detected in the survey of IRC+10216.}\label{Tab:irclines}\\
\hline
Frequency       & Species      & Transition   & $V_{\rm exp}$      &$S_{\nu}$ & $\int S_{\nu}dv$ & $\sigma$\tablefootmark{(a)} &   \\
(MHz)           &              &              & (\kms)        &(mJy)       & (mJy~km~s$^{-1}$)   & (mJy~km~s$^{-1}$)     & Notes \\
\hline
\endfirsthead
\caption{continued.} \\
\hline
Frequency       &  Species     & Transition   & $V_{\rm exp}$      &$S_{\nu}$ & $\int S_{\nu}dv$ & $\sigma$\tablefootmark{(a)} &      \\
(MHz)           &              &              &  (\kms)      &(mJy)       & (mJy~km~s$^{-1}$)   & (mJy~km~s$^{-1}$)    & Notes\\
\hline
\endhead
\hline  \\
\endfoot
\hline
\endlastfoot
 18048.0    & HC$_{7}$N 	        &  $J$=16--15			        & 14.2$\pm$0.1   & 33.6     & 862.5    &  11.6 & NS \\	   
 18154.9    & SiS			&  $J$=1--0				& 14.5$\pm$0.0	 & 301.2    & 3928.1   &  11.6 & 1   \\ 
 18196.2    & HC$_{3}$N			&  $J$=2--1				& 14.1$\pm$0.1	 & 275.0    & 5759.1   &  13.8 &    \\ 
 18343.1    & c-C$_{3}$H$_{2}$	        &  $J_{K_{\rm a},K_{\rm c}}$=1$_{1,0}$--1$_{0,1}$ & 14.0$\pm$0.0  &  82.1    & 1652.4   &  13.8  &     \\ 
 18447.6    & HCC$^{13}$CCCN             &  $J$=7--6                             &                &  6.5     & 92.3     &  14.8 & N	\\
 18454.5    & HCCCC$^{13}$CN		 &  $J$=7--6				 &		  &  6.5     & $<$631.6 & 14.8 & 2, N	  \\
 18458.0    & U				&					&		 &  9.9	    & $<$631.6 &  13.8 &     \\ 
 18638.6    & HC$_{5}$N			&  $J$=7--6				& 14.0$\pm$0.1	 & 154.1    & 3271.4   &  16.9 & NS	\\ 
 19015.1    & C$_{4}$H			&  $N$=2--1,$J$=5/2--3/2		& 13.9$\pm$0.3	 &  17.2    & 456.1    &  14.8 & NS	\\  
 19176.0    & HC$_{7}$N 	        &  $J$=17--16			        & 13.8$\pm$0.1   &  45.7    & 1030.2   &  14.8 & NS	\\ 
 19780.8    & C$_{3}$N			&  $N$=2--1,$J$=5/2--3/2		& 14.0$\pm$0.2	 &  34.1    & 773.6    &  17.7 &    \\ 
 19800.1    & C$_{3}$N			&  $N$=2--1,$J$=3/2--1/2		& 12.4$\pm$1.0	 &  16.6    & 272.7    &  17.7 &    \\
 20303.9    & HC$_{7}$N			&  $J$=18--17				& 14.2$\pm$0.1	 &  60.9    & 1325.5   &  20.1 & NS	\\ 
 20927.3    & C$_{6}$H($^{2}\Pi_{1/2}$)	&  $J$=15/2--13/2, $l=f$		& 14.7$\pm$0.2	 &  11.4    & 237.8    &  15.4 & N   \\ 
 20956.2    & C$_{6}$H($^{2}\Pi_{1/2}$)	&  $J$=15/2--13/2, $l=e$		& 13.4$\pm$0.2	 &  11.3    & 217.9    &  15.4 & N   \\ 
 21090.8    & HCCCC$^{13}$CN		&  $J$=8--7				& 17.8$\pm$1.0	 &   5.7    & 129.3    &  11.8 & N   \\
 21279.2    & HCC$^{13}$CCCN		 &  $J$=8--7				&		 &   5.0    & 90.5     &  14.2 & N    \\
 21281.8    & HCCC$^{13}$CCN		 &  $J$=8--7				&		 &   5.0    & 88.2     &  14.2 & N    \\
 21301.3    & HC$_{5}$N			&  $J$=8--7				& 13.8$\pm$0.0	 & 228.4    & 4963.3   &  11.8 & NS	\\ 
 21431.9    & HC$_{7}$N			&  $J$=19--18				& 13.8$\pm$0.0	 &  64.2    & 1459.3   &  13.0 & NS	\\ 
 21458.8    & U				&					&		 &   4.6    & 107.2    &  13.0 &     \\
 21472.7    & SiC$_{4}$			&  $J$=7--6				& 13.7$\pm$0.2	 &   6.8    & 142.0    &  13.0 & N   \\ 
 21480.8    & l-C$_{5}$H($^{2}\Pi_{1/2}$)  &  $J$=9/2--7/2, $F$= 5--4, $l=e$	& 13.3$\pm$2.2	 &  12.1    & 242.5    &  14.2 & 3, NS	  \\
 21481.3    & l-C$_{5}$H($^{2}\Pi_{1/2}$)  &  $J$=9/2--7/2, $F$= 4--3, $l=e$	&		 &	    &	       &       &	    \\
 21484.7    & l-C$_{5}$H($^{2}\Pi_{1/2}$)  &  $J$=9/2--7/2, $F$= 5--4, $l=f$	& 15.4$\pm$0.4	 &  10.0    & 194.6    &  14.2 & 3, NS \\ 
 21485.1    & l-C$_{5}$H($^{2}\Pi_{1/2}$)  &  $J$=9/2--7/2, $F$= 4--3, $l=f$	&		 &	    &	       &      &	    \\ 
 21498.2    & HC$_{9}$N			&  $J$=37--36				& 14.5$\pm$0.2	 &  11.9    & 254.0    &  13.0 & NS    \\ 
 21628.5    & l-C$_{5}$H($^{2}\Pi_{3/2}$)  &  $J$=9/2--7/2, $F$= 4--3, $l=f$	& 16.2$\pm$0.3	 &   8.8    & 241.7    &  13.0 & 3,N \\ 
 21628.6    & l-C$_{5}$H($^{2}\Pi_{3/2}$)  &  $J$=9/2--7/2, $F$= 4--3, $l=e$	&		 &	    &	       &      &     \\ 
 21629.6    & l-C$_{5}$H($^{2}\Pi_{3/2}$)  &  $J$=9/2--7/2, $F$= 5--4, $l=f$	&		 &	    &	       &      &     \\ 
 21629.7    & l-C$_{5}$H($^{2}\Pi_{3/2}$)  &  $J$=9/2--7/2, $F$= 5--4, $l=e$	&		 &	    &	       &      &     \\ 
 21706.6    & C$_{8}$H($^{2}\Pi_{3/2}$)    &  $J$=37/2--35/2, $F$=19--18, $l=e$    & 14.2$\pm$2.1   &   6.1    & 125.1    & 13.0 & 3,N \\ 
 21706.6    & C$_{8}$H($^{2}\Pi_{3/2}$)	&  $J$=37/2--35/2, $F$=18--17, $l=e$	&		 &	    &	       &      &     \\
 21706.8    & C$_{8}$H($^{2}\Pi_{3/2}$)	&  $J$=37/2--35/2, $F$=19--18, $l=f$	&		 &	    &	       &      &     \\ 
 21706.8    & C$_{8}$H($^{2}\Pi_{3/2}$)	&  $J$=37/2--35/2, $F$=18--17, $l=f$	&		 &	    &	       &      &     \\ 
 22029.7    & C$_{6}$H$^{-}$	        &  $J$=8--7			        & 14.2$\pm$0.3   &   6.4    & 175.6    &  14.2 & NS	\\ 
 22079.2    & HC$_{9}$N			&  $J$=38--37				& 13.9$\pm$0.2	 &  10.3    & 187.0    &  14.2 & NS	\\
 22304.9    & U				&					&		 &   7.4    & 182.1    &  11.8 &     \\
 22323.3    & U				&					&		 &   4.4    & 85.3     &  11.8 &     \\
 22344.0    & C$_{2}$S			&  $J_{N}$= 2$_{1}$-- 1$_{0}$		& 13.7$\pm$0.1	 &  15.4    & 318.1    &  11.8 & NS	\\ 
 22559.9    & HC$_{7}$N			&  $J$=20--19				& 13.9$\pm$0.0	 &  75.1    & 1501.0   &  11.8 & NS	\\ 
 22660.2    & HC$_{9}$N			&  $J$=39--38				& 13.9$\pm$0.2	 &   9.1    & 172.8    &  13.0 &     \\
 22879.9    & C$_{8}$H($\Pi_{3/2}$)	&  $J$=39/2--37/2, $F$=20--19, $l=e$	& 13.6$\pm$0.3	 &   5.2    & 115.2    &  11.8 & 3   \\	 
 22879.9    & C$_{8}$H($\Pi_{3/2}$)	&  $J$=39/2--37/2, $F$=19--18, $l=e$    &	         &	    &	       &      &     \\ 
 22880.1    & C$_{8}$H($\Pi_{3/2}$)	&  $J$=39/2--37/2, $F$=20--19, $l=f$    &	         &	    &	       &      &     \\ 
 22880.1    & C$_{8}$H($\Pi_{3/2}$)	&  $J$=39/2--37/2, $F$=19--18, $l=f$    &	         &	    &	       &      &     \\	
 23123.0    & C$_{3}$S		        &  $J$=4--3			        & 13.9$\pm$0.1   &  10.5    & 237.1    &  11.8 &    \\ 
 23241.2    & HC$_{9}$N			&  $J$=40--39				& 15.9$\pm$0.1	 &   9.0    & 178.9    &  13.0 &    \\ 
 23340.1    & H$^{13}$CCCCCN		&  $J$=9--8				& 13.2$\pm$0.3	 &   7.7    & 136.5    &  13.0 & N   \\ 
 23565.2    & C$_{6}$H($^{2}\Pi_{3/2}$)	   &  $J$=17/2--15/2, $l=e$		&		 &  29.5    & 768.1    &  14.2 & 3, NS	  \\ 
 23567.2    & C$_{6}$H($^{2}\Pi_{3/2}$)	   &  $J$=17/2--15/2, $l=f$		&		 &          &          &      &     \\ 
 23600.2    & SiC$_{2}$		        &  $J_{K_{\rm a},K_{\rm c}}$=1$_{0,1}$--0$_{0,0}$& 13.9$\pm$0.0   &   55.9   & 1185.0   &  13.0  &     \\ 
 23687.9    & HC$_{7}$N		        &  $J$=21--20			        & 13.9$\pm$0.1   &  93.6    & 1922.2   &  13.0 &    \\ 
 23694.5    & NH$_{3}$			&  ($J$,$K$)=(1,1)			& 14.5$\pm$0.3	 &  19.8    & 474.4    &  13.0 &    \\ 
 23718.3    & HC$^{13}$CCCCN		&  $J$=9--8				&		 &  $<$20.7 & $<$693.5 &  13.0 & 4   \\ 
 23719.4    & C$_{6}$H($^{2}\Pi_{1/2}$)	   &  $J$=17/2--15/2, $l=f$		&		 &  $<$20.7 & $<$693.5 &  13.0 & N   \\
 23722.6    & NH$_{3}$			&  ($J$,$K$)=(2,2)			& 13.9$\pm$0.1	 &  17.6    & 345.9    &  13.0 &    \\
 23727.2    & HCCCC$^{13}$CN		&  $J$=9--8				& 15.1$\pm$0.5	 &  9.8	    & 138.6    &  13.0 &     \\
 23732.7    & U				&					&		 &  5.5	    & 135.1    &  11.8 &    \\
 23748.6    & C$_{6}$H($^{2}\Pi_{1/2}$)	   &  $J$=17/2--15/2, $l=e$		& 13.9$\pm$0.1	 &  16.7    & 277.3    &  13.0 & N   \\
 23822.3    & HC$_{9}$N			&  $J$=41--40				& 13.9$\pm$0.2	 &  10.6    & 217.2    &  13.0 &    \\ 
 23846.3    & U				&					&		 &  5.5	    & 118.1    &  11.8 &    \\
 23870.1    & NH$_{3}$			&  ($J$,$K$)=(3,3)			& 13.4$\pm$0.2	 &  16.2    & 338.8    &  11.8 & NS	\\
 23875.1    & MgNC			&  $N$=2--1, $J$=5/2--3/2		& 13.6$\pm$0.2	 &  10.2    & 171.0    &  11.8 & N   \\
 23939.0    & HCC$^{13}$CCCN		&  $J$=9--8				& 13.4$\pm$0.2	 &  10.7    & 145.9    &  13.0 &    \\ 
 23942.0    & HCCC$^{13}$CCN		&  $J$=9--8				& 13.4$\pm$0.3	 &  8.1	    & 175.1    &  13.0 &    \\ 
 23963.9    & HC$_{5}$N			&  $J$=9--8				& 13.8$\pm$0.0	 &  366.8   & 7494.0   &  13.0 &    \\ 
 24053.2    & C$_{8}$H($^{2}\Pi_{3/2}$)    &  $J$=41/2--39/2, $F$=21--20, $l=e$    & 16.6$\pm$0.5   &  6.1     & 209.1    &  11.8 &3,N  \\ 
 24053.2    & C$_{8}$H($^{2}\Pi_{3/2}$)    &  $J$=41/2--39/2, $F$=20--19, $l=e$    &		 &	    &	       &      &     \\ 
 24053.5    & C$_{8}$H($^{2}\Pi_{3/2}$)    &  $J$=41/2--39/2, $F$=21--20, $l=f$    &		 &	    &	       &      &     \\ 
 24053.5    & C$_{8}$H($^{2}\Pi_{3/2}$)    &  $J$=41/2--39/2, $F$=20--19, $l=f$    &		 &	    &	       &      &     \\ 
 24139.4    & NH$_{3}$	                &  ($J$,$K$)=(4,4)		        &		 & 6.8      & $<$125.0 &  11.8 &5, NS	 \\
 24403.3    & HC$_{9}$N			&  $J$=42--41				& 14.3$\pm$0.1	 & 13.5	    & 287.3    &  11.8 &    \\ 
 24540.2    & SiC$_{4}$			&  $J$=8--7				& 13.5$\pm$0.1	 & 14.2	    & 282.8    &  15.4 & N   \\ 
 24783.4    & C$_{6}$H$^{-}$		&  $J$=9--8				& 13.8$\pm$0.1	 & 12.3	    & 253.6    &  11.8 & N   \\ 
 24815.9    & HC$_{7}$N			&  $J$=22--21				& 14.0$\pm$0.0	 & 97.7	    & 1950.1   &  13.0 & NS	\\
 24862.7    & U				&					&		 & 5.1	    & 112.5    &  11.8 &    \\
 24901.4    & U				&					&		 & 7.9	    & 146.4    &  11.8 &    \\
 24984.3    & HC$_{9}$N			&  $J$=43--42				& 14.1$\pm$0.1	 & 13.5	    & 245.6    &  11.8 &    \\ 
 24991.3    & SiC$_{2}$		        &  $J_{K_{\rm a},K_{\rm c}}$=8$_{2,6}$--8$_{2,7}$ & 13.7$\pm$0.1  & 13.7	    & 268.5    &  11.8 &     \\
 25056.0    & NH$_{3}$	                &  ($J$,$K$)=(6,6)		        &	         & 5.5      & 95.2      & 11.8 & NS    \\
 25094.3    & U				&					&		 & 9.2	    & 158.5    &  13.0 &    \\
 25111.8    & U				&					&		 & 8.7	    & 191.5    &  13.0 &    \\
 25226.5    & C$_{8}$H($^{2}\Pi_{3/2}$)    &  $J$=43/2--41/2, $F$=22--21, $l=e$    & 16.1$\pm$0.3   & 10.6     & 249.2    &  13.0 & 3,N \\ 
 25226.5    & C$_{8}$H($^{2}\Pi_{3/2}$)    &  $J$=43/2--41/2, $F$=21--20, $l=e$    &	         &          &	       &      &     \\ 
 25226.8    & C$_{8}$H($^{2}\Pi_{3/2}$)    &  $J$=43/2--41/2, $F$=22--21, $l=f$    &	         &          &	       &      &     \\ 
 25226.8    & C$_{8}$H($^{2}\Pi_{3/2}$)    &  $J$=43/2--41/2, $F$=21--20, $l=f$    &	         &          &	       &      &     \\ 
 25565.3    & HC$_{9}$N		        &  $J$=44--43			        & 13.8$\pm$0.2	 & 13.0	    & 245.6    &  13.0 & N   \\ 
 25933.4    & H$^{13}$CCCCCN		&  $J$=10--9				& 14.5$\pm$0.2	 & 9.7	    & 189.4    &  15.4 & N   \\ 
 25943.9    & HC$_{7}$N			&  $J$=23--22				& 14.1$\pm$0.0	 & 90.8	    & 1998.5   &  15.4 &NS	\\
 25976.2    & U				&					&		 & 6.4	    & 162.2    &  17.7 &     \\
 25992.8    & U				&					&		 & 7.4	    & 246.4    &  17.7 &     \\
 26146.3    & HC$_{9}$N			&  $J$=45--44				& 13.5$\pm$0.2	 & 12.7	    & 216.1    &  17.7 & N   \\ 
 26254.8    & l-C$_{5}$H($^{2}\Pi_{1/2}$)  &  $J$=11/2--9/2, $F$= 6--5, $l=e$      & 13.1$\pm$0.3   & 23.0     & 377.4    &  27.2 & 3,N \\ 
 26255.2    & l-C$_{5}$H($^{2}\Pi_{1/2}$)  &  $J$=11/2--9/2, $F$= 5--4, $l=e$      &	         &	    &	       &      &     \\ 
 26258.7    & l-C$_{5}$H($^{2}\Pi_{1/2}$)  &  $J$=11/2--9/2, $F$= 6--5, $l=f$      & 15.1$\pm$0.4   & 24.2     & 477.6    &  27.2 & 3,N \\ 
 26259.1    & l-C$_{5}$H($^{2}\Pi_{1/2}$)  &  $J$=11/2--9/2, $F$= 5--4, $l=f$      &	         &          &	       &       &     \\	  
\hline
\end{longtable}      
\tablefoot{--(1) The blue-shifted component of the SiS $J$=1--0 line is stronger than the red-shifted one due to maser 
amplification \citep[18154.9~MHz;][]{1983ApJ...267..184H}. The peak intensity of its presumably also inverted red-shifted component 
is 198.5~mJy. (2) Blend with a U line at 18458.0~MHz. (3) The hyperfine structure is not resolved. (4) Blend with C$_{6}$H at 23719.4~MHz. 
(5) NH$_{3}$ (4,4) at 24139.4~MHz seems to be blended (see Fig.~\ref{Fig:zoomv1}). (N) Transitions that are detected for the first time outside the solar system are marked with ``N''. (NS) Transitions that are detected for the first time toward the source are marked with ``NS''. \tablefoottext{a}{The rms noise levels given here are for $\sim$28~\kms~wide channels.}}
                                                                                                                 
               }

\end{appendix}

\begin{appendix}\label{b}
\section{Zoom-in plots of observed spectra}

\begin{figure*}[!htbp]
\centering
\includegraphics[width = 0.8 \textwidth]{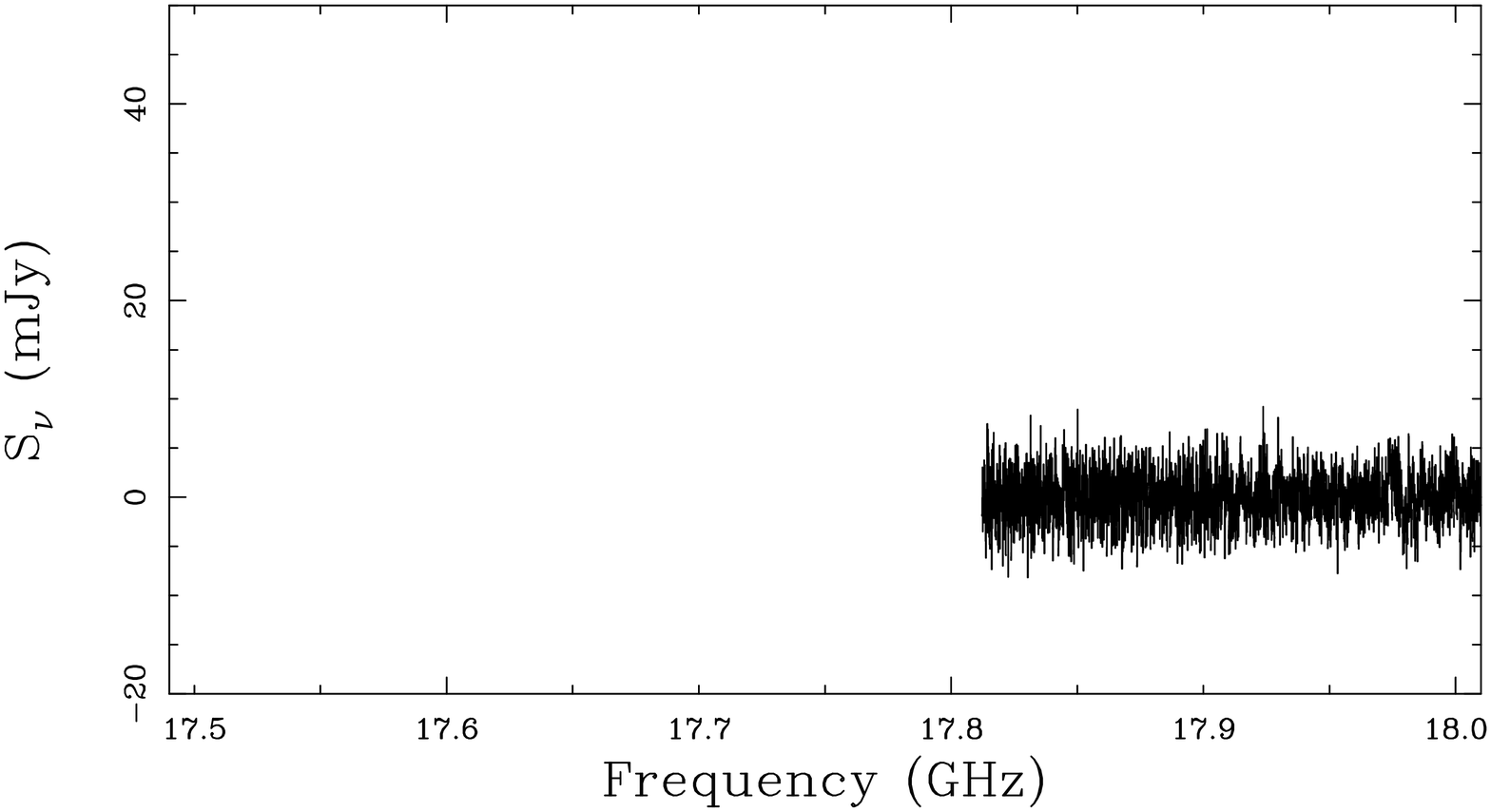}
\includegraphics[width = 0.8 \textwidth]{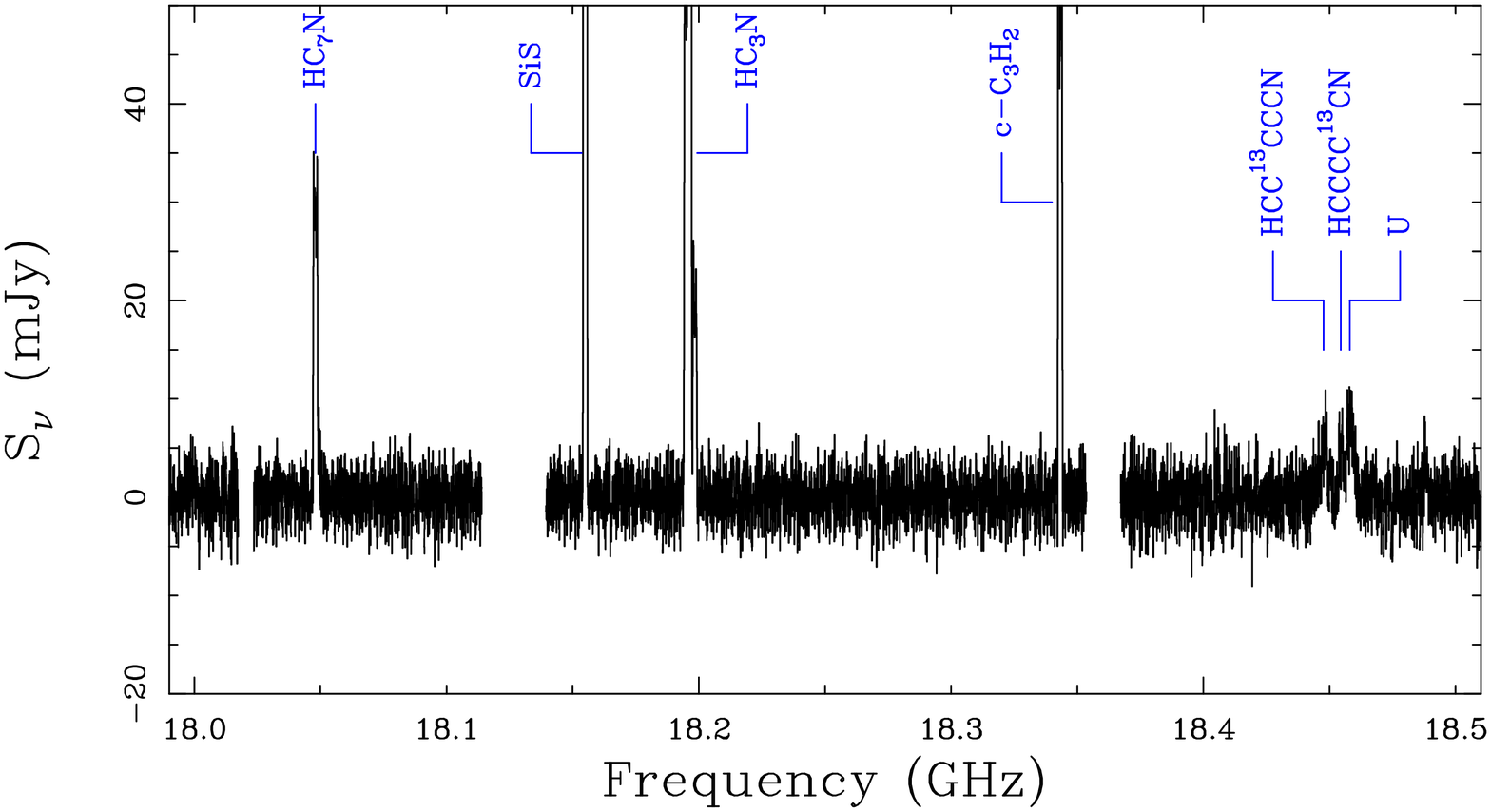}
\includegraphics[width = 0.8 \textwidth]{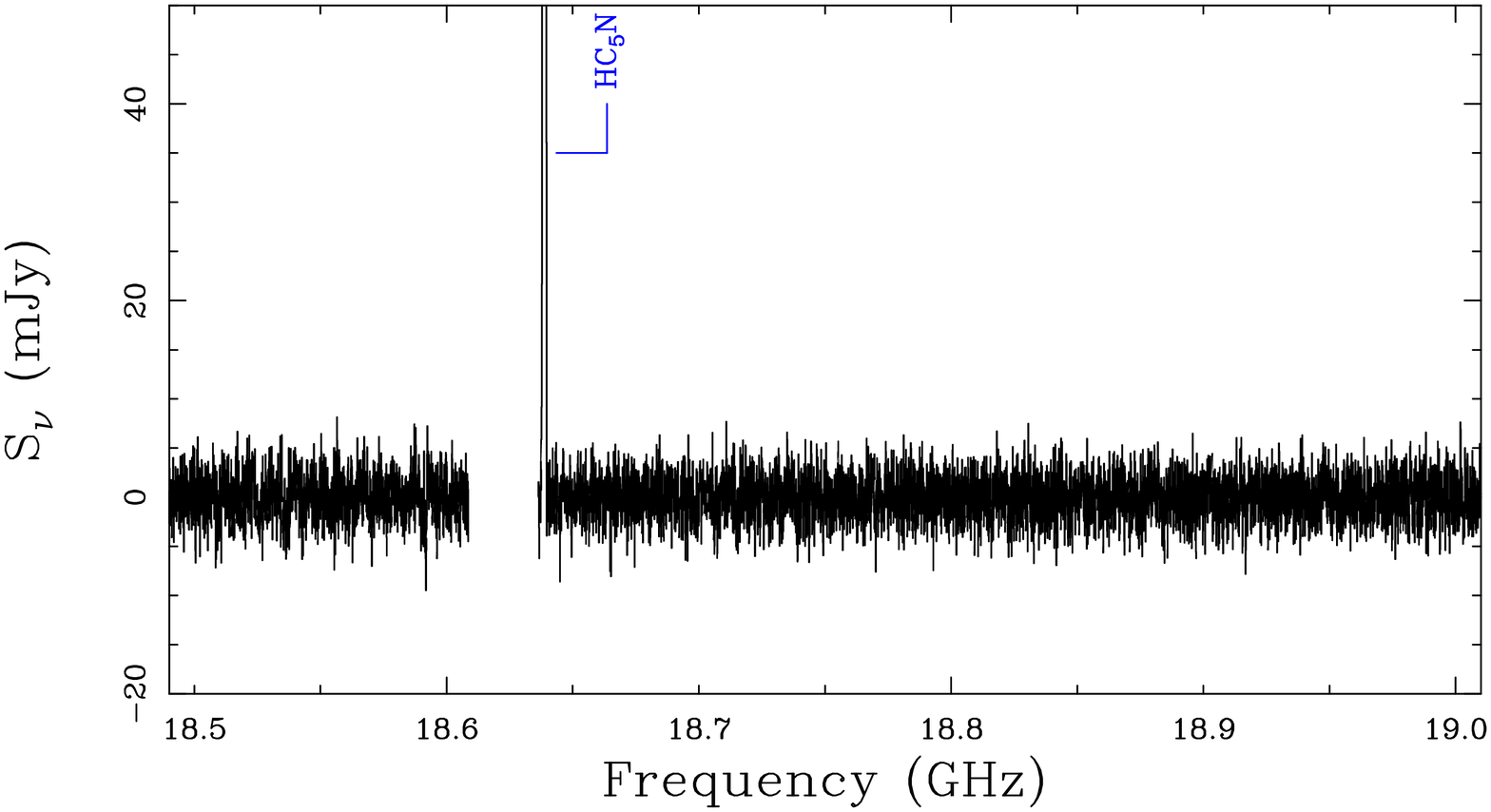}
\caption{{Observed spectrum of IRC $+$10216 from 17.8 to 26.3 GHz.} \label{Fig:irc}}
\end{figure*}

\begin{figure*}[!htbp]
\centering
\includegraphics[width = 0.8 \textwidth]{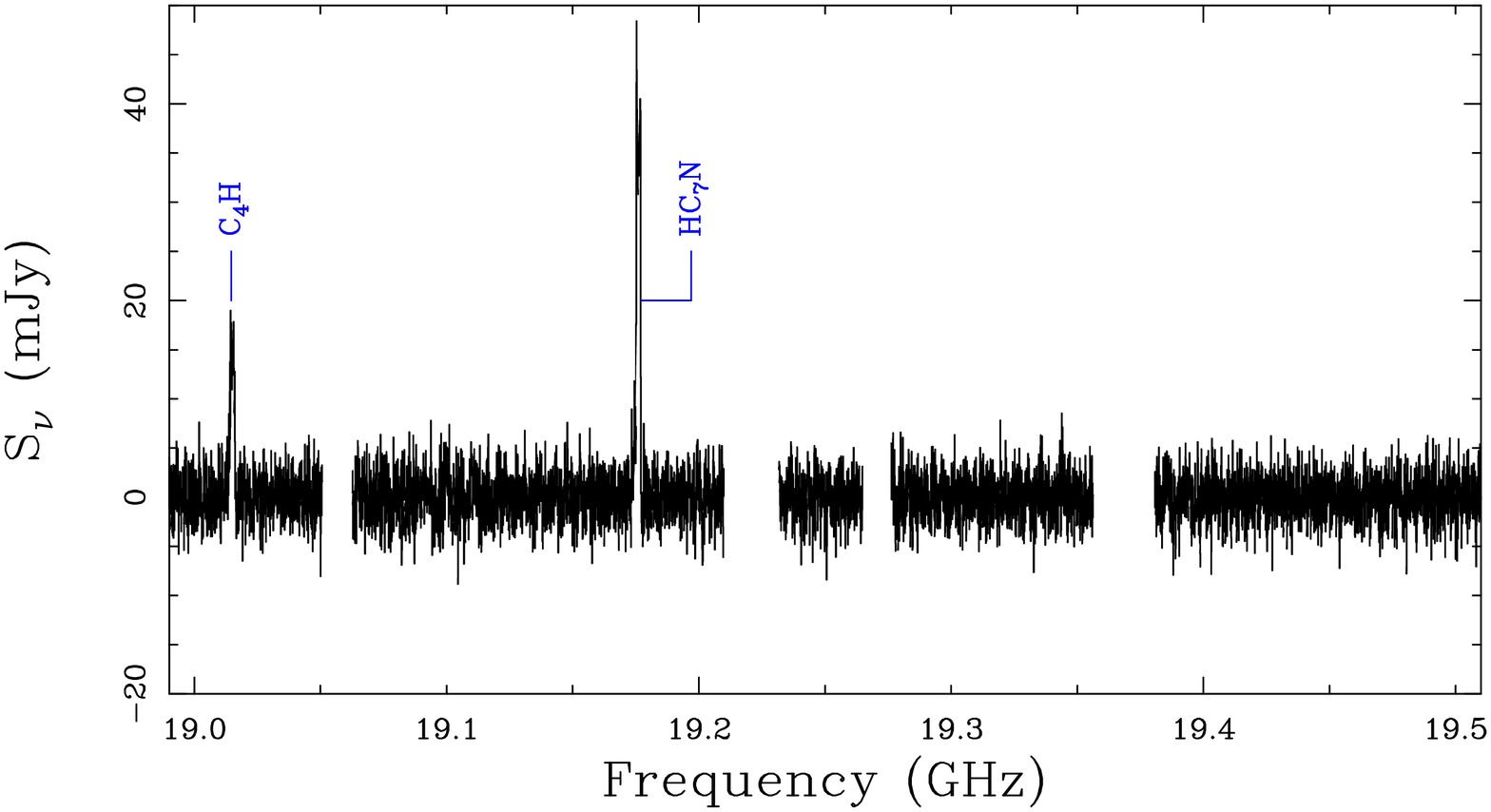}
\includegraphics[width = 0.8 \textwidth]{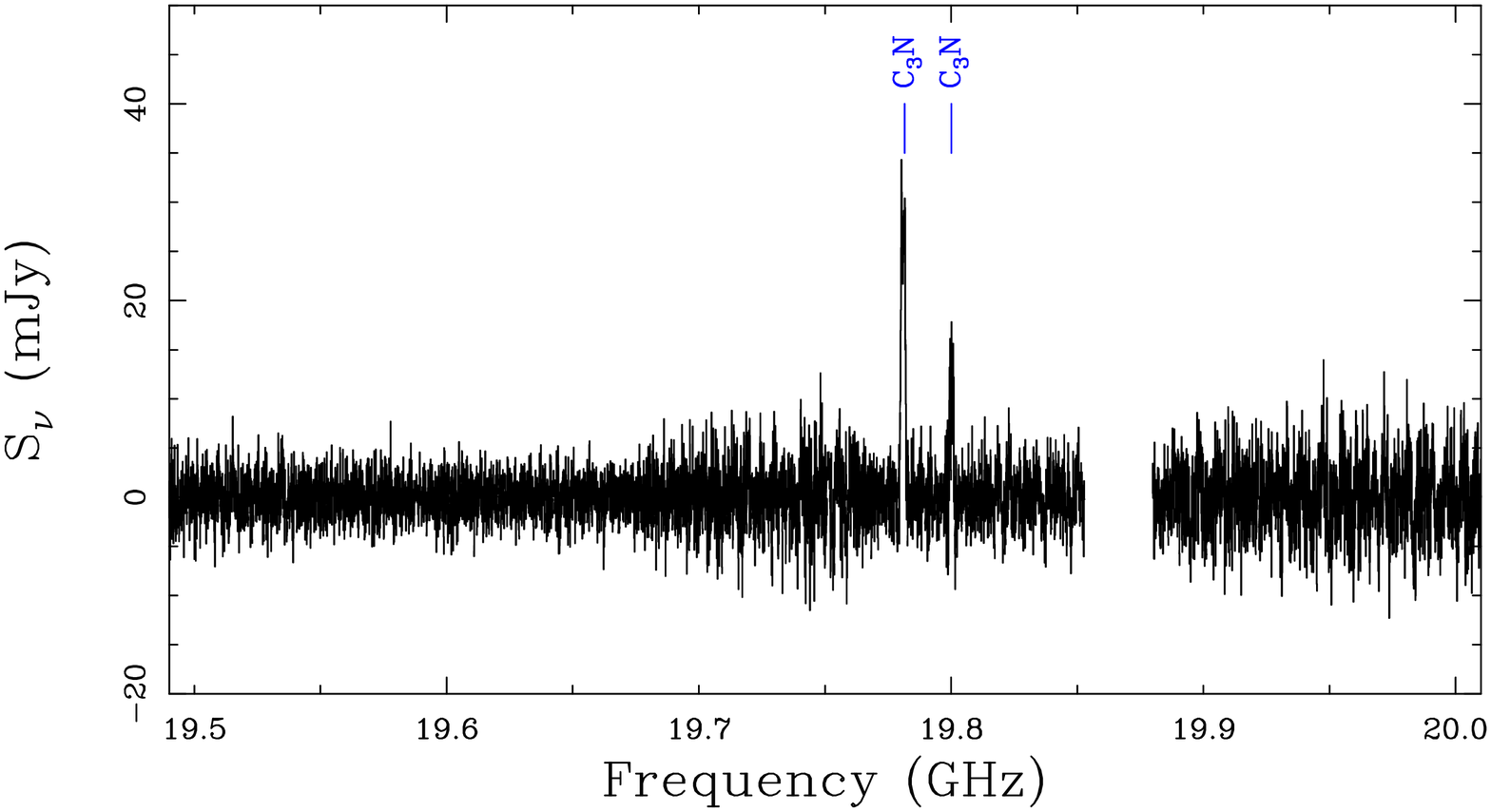}
\includegraphics[width = 0.8 \textwidth]{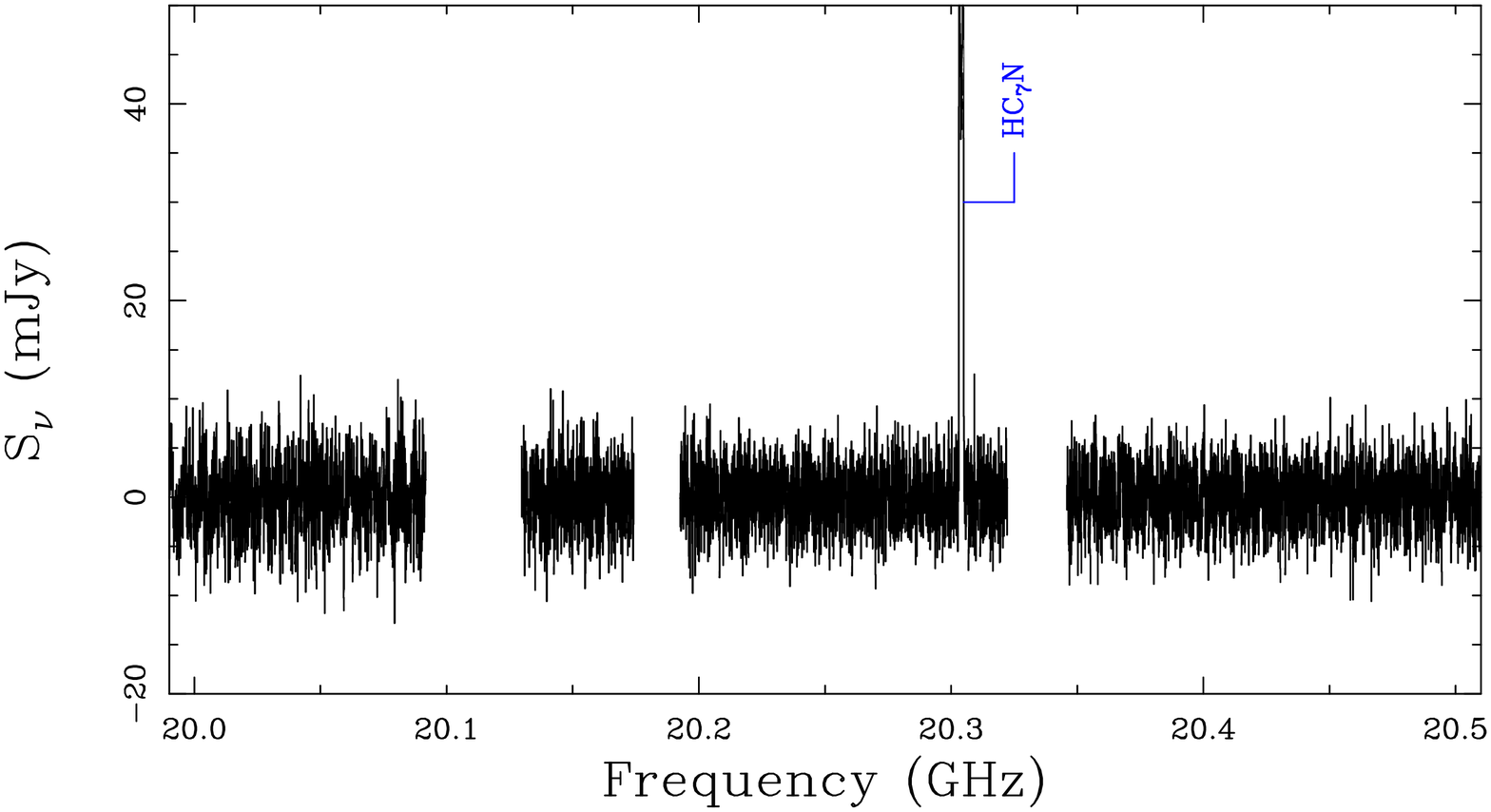}
\centerline{Fig. \ref{Fig:irc}. --- Continued.}
\end{figure*}

\begin{figure*}[!htbp]
\centering
\includegraphics[width = 0.8 \textwidth]{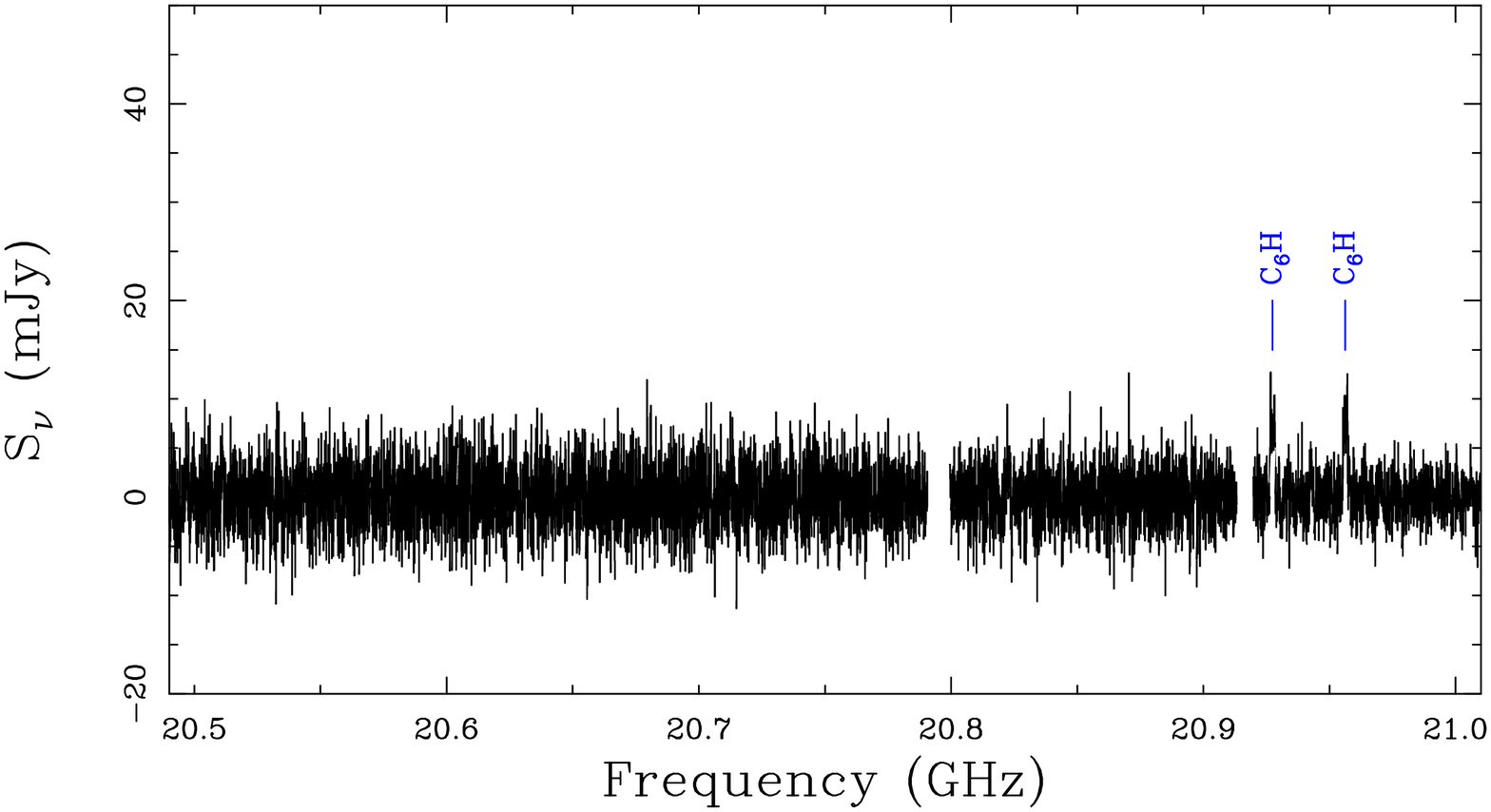}
\includegraphics[width = 0.8 \textwidth]{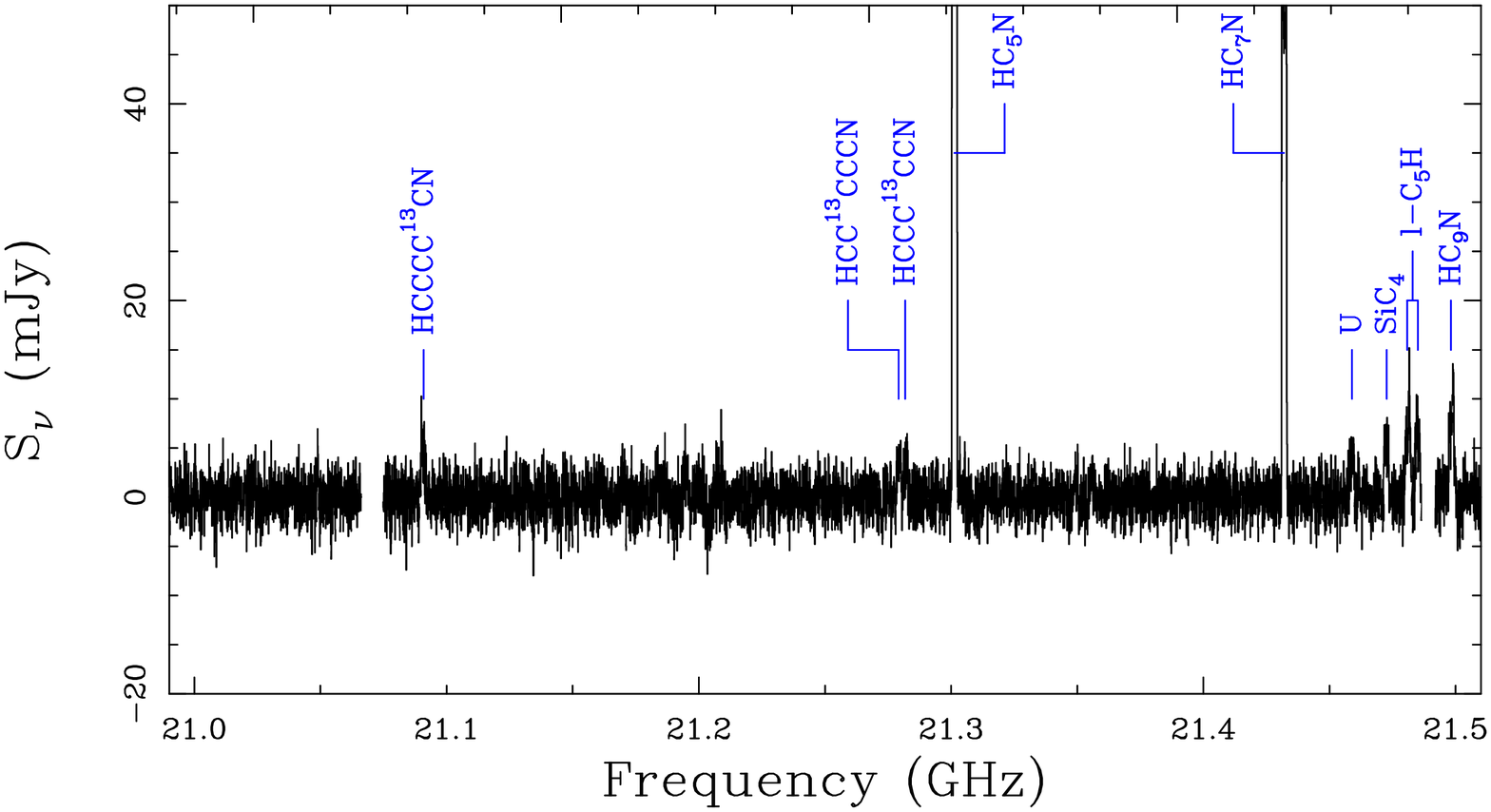}
\includegraphics[width = 0.8 \textwidth]{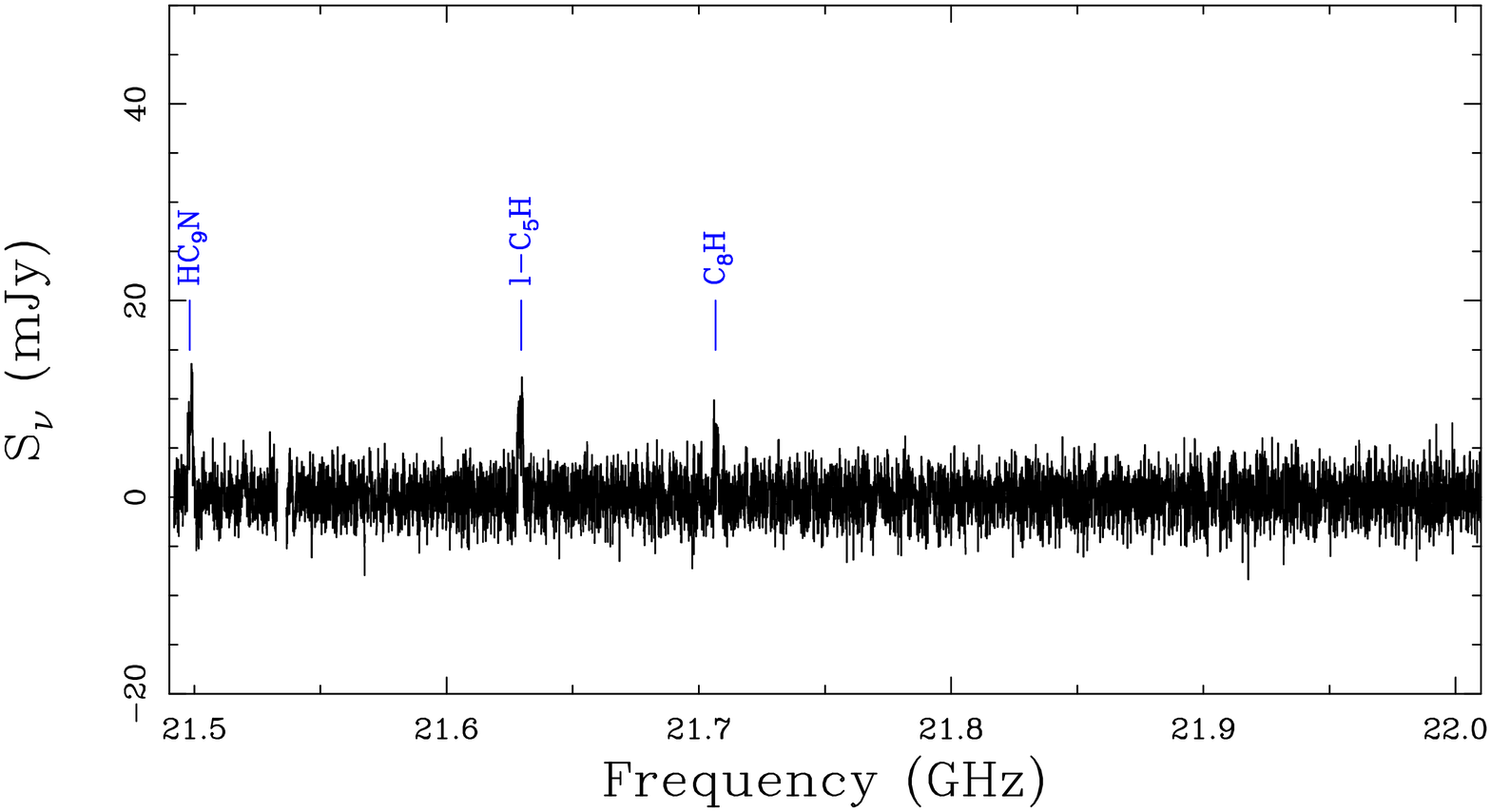}
\centerline{Fig. \ref{Fig:irc}. --- Continued.}
\end{figure*}

\begin{figure*}[!htbp]
\centering
\includegraphics[width = 0.8 \textwidth]{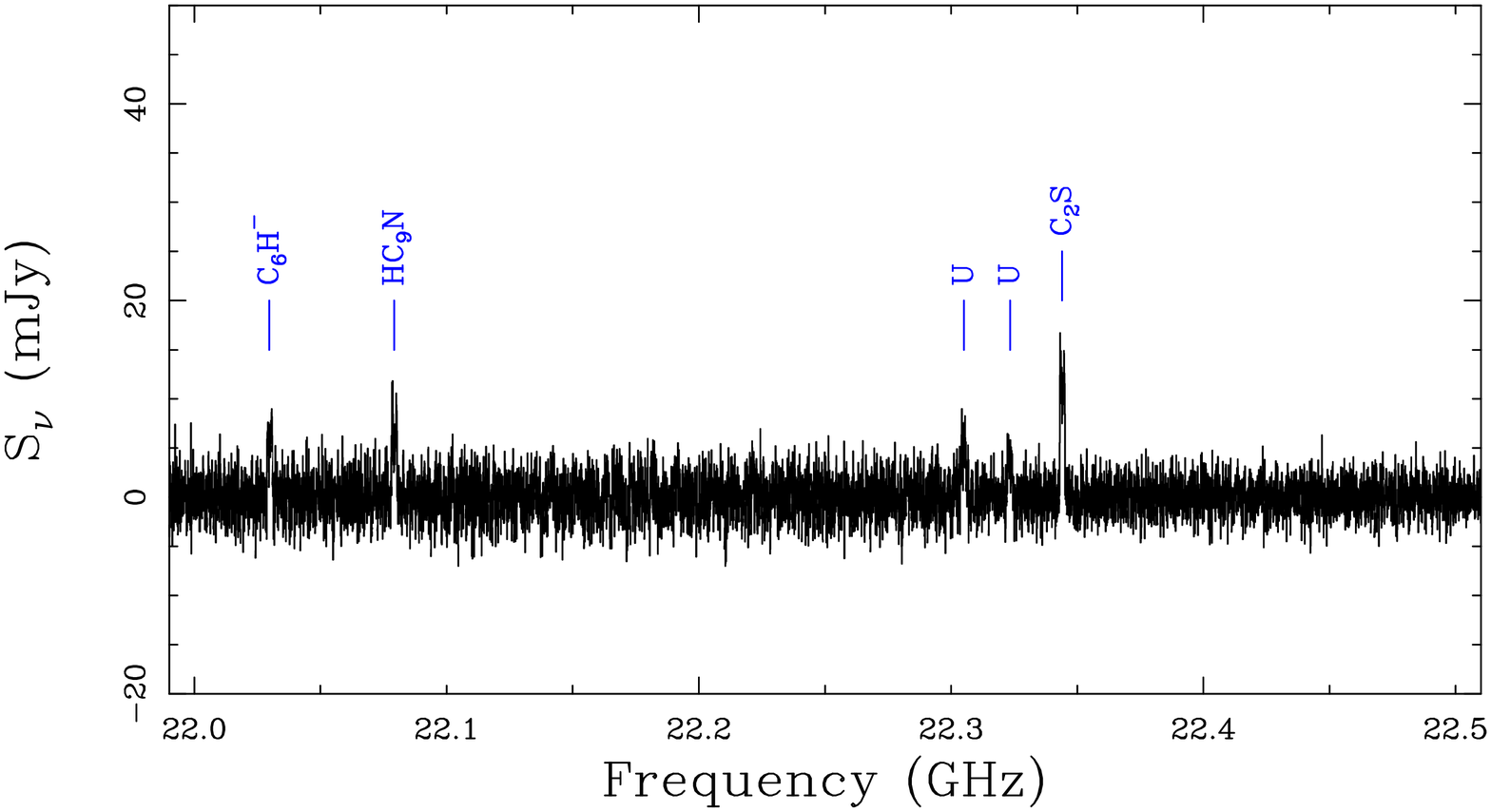}
\includegraphics[width = 0.8 \textwidth]{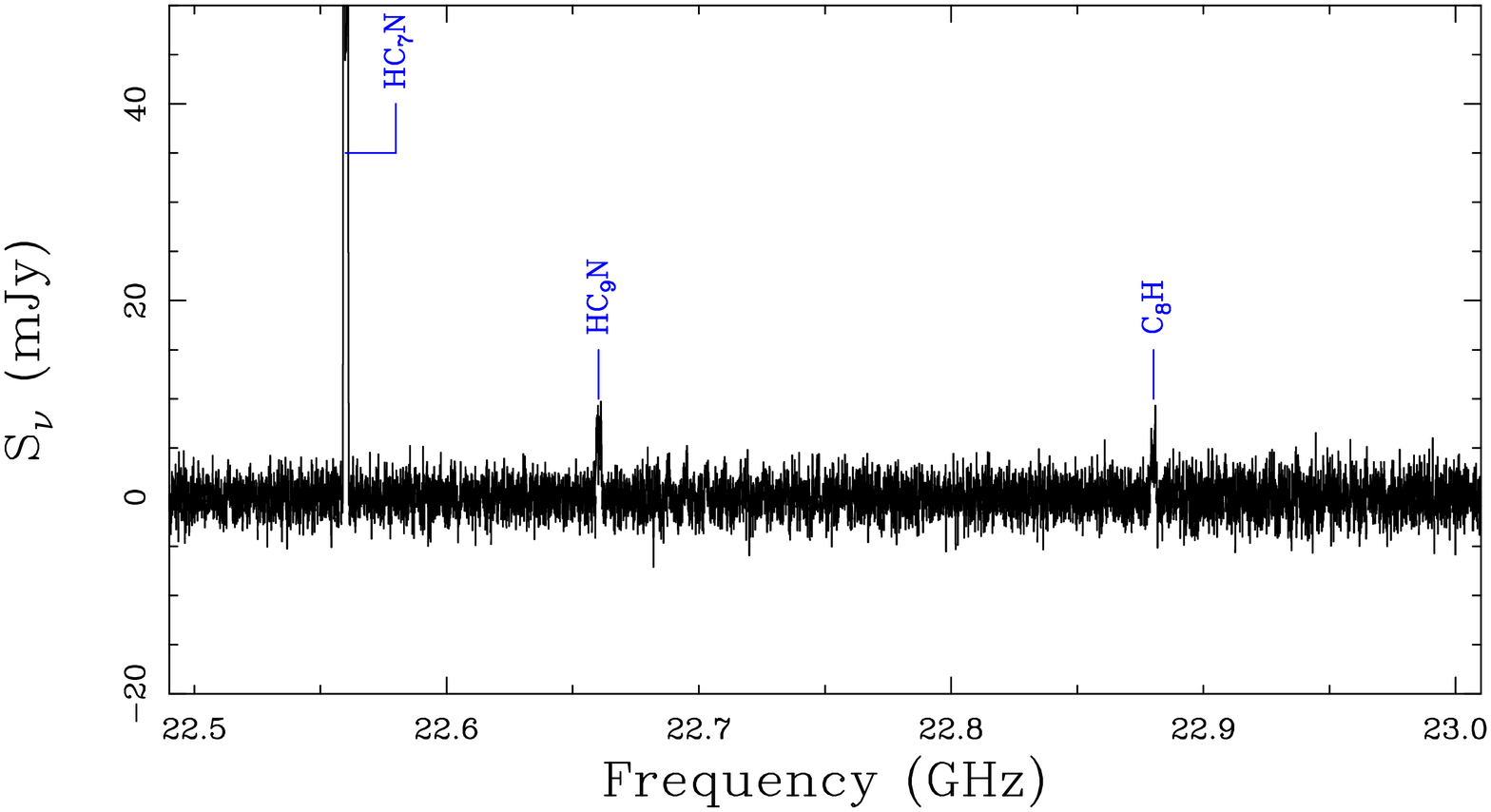}
\includegraphics[width = 0.8 \textwidth]{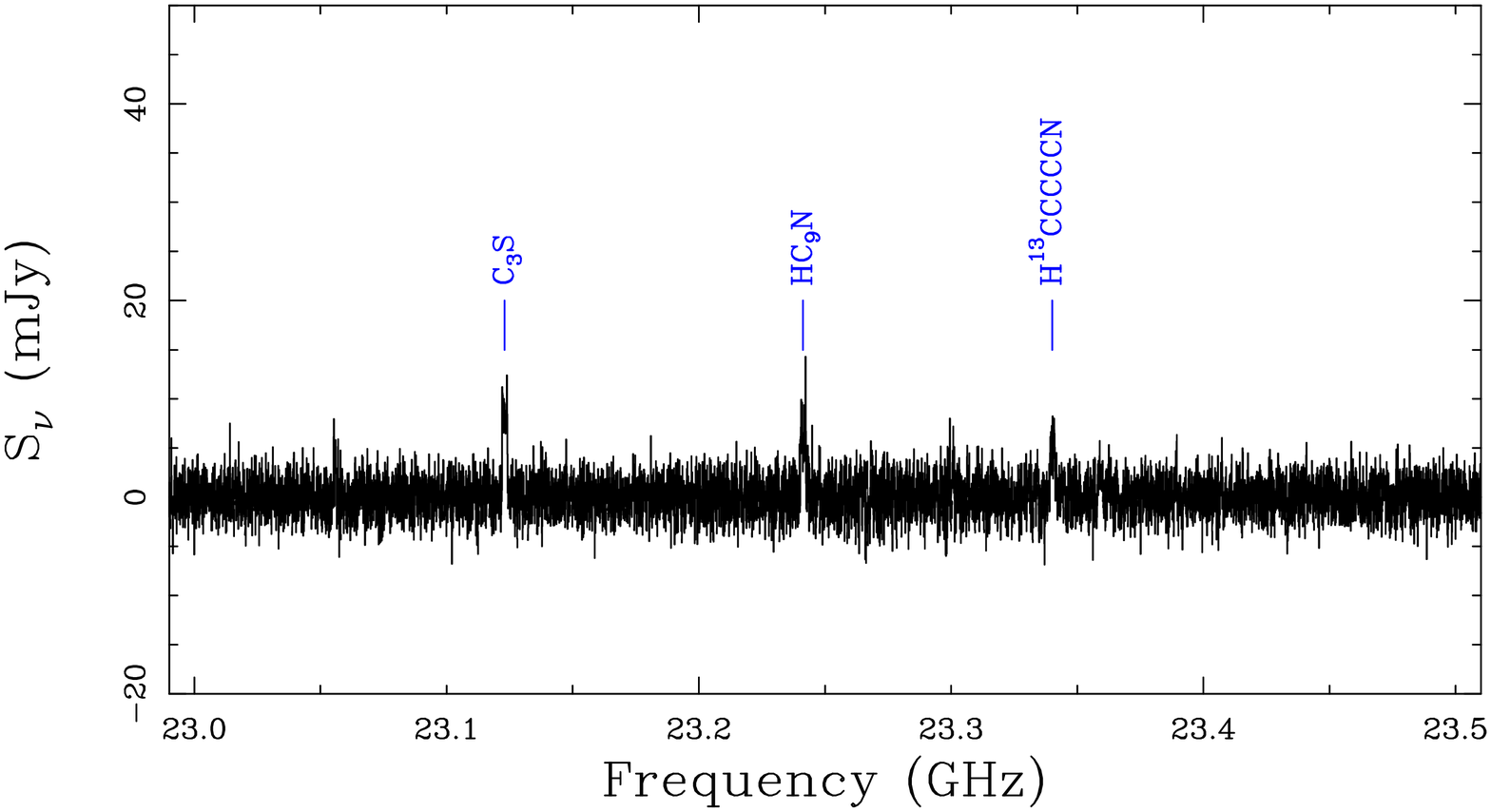}
\centerline{Fig. \ref{Fig:irc}. --- Continued.}
\end{figure*}

\begin{figure*}[!htbp]
\centering
\includegraphics[width = 0.8 \textwidth]{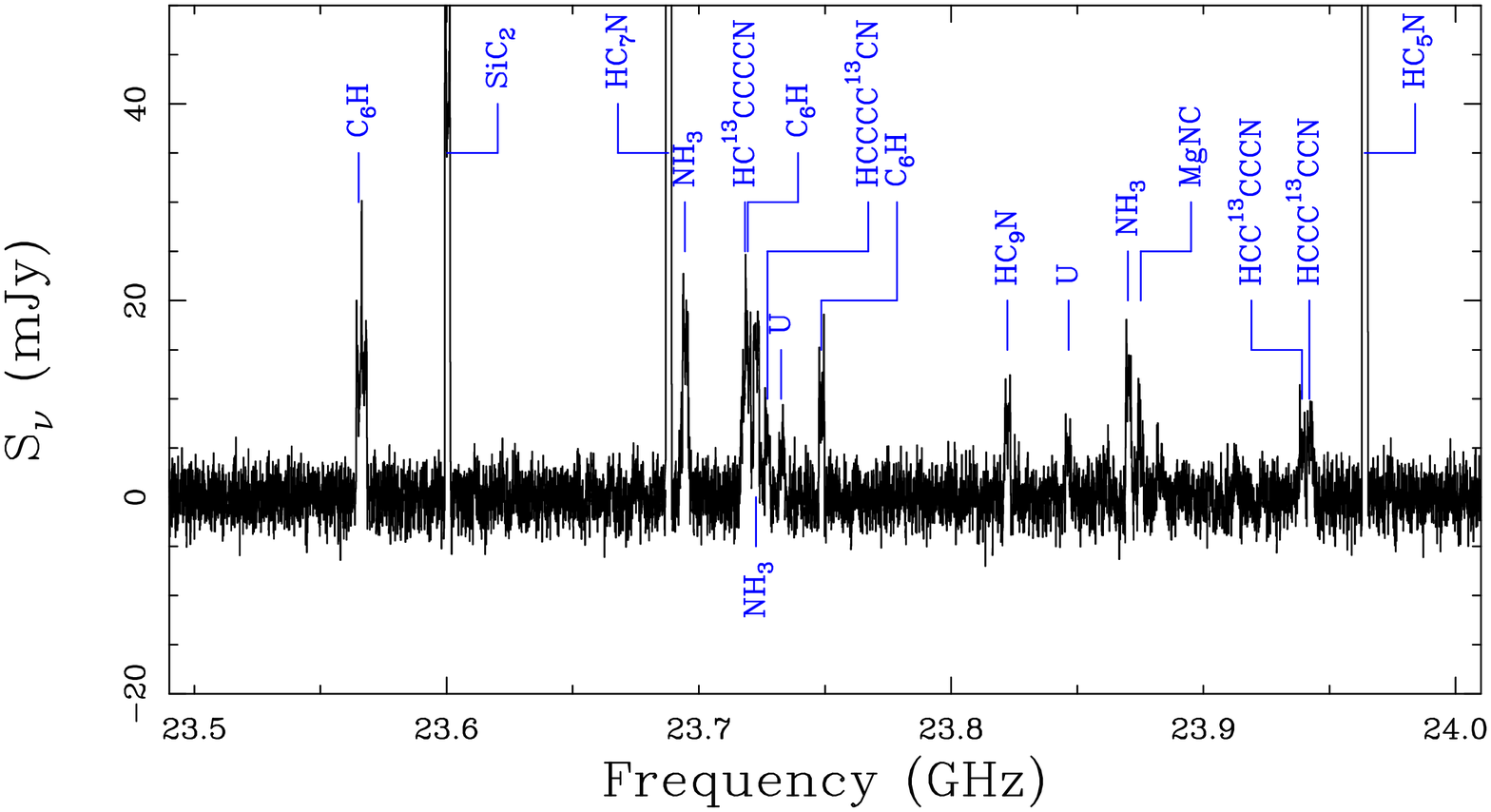}
\includegraphics[width = 0.8 \textwidth]{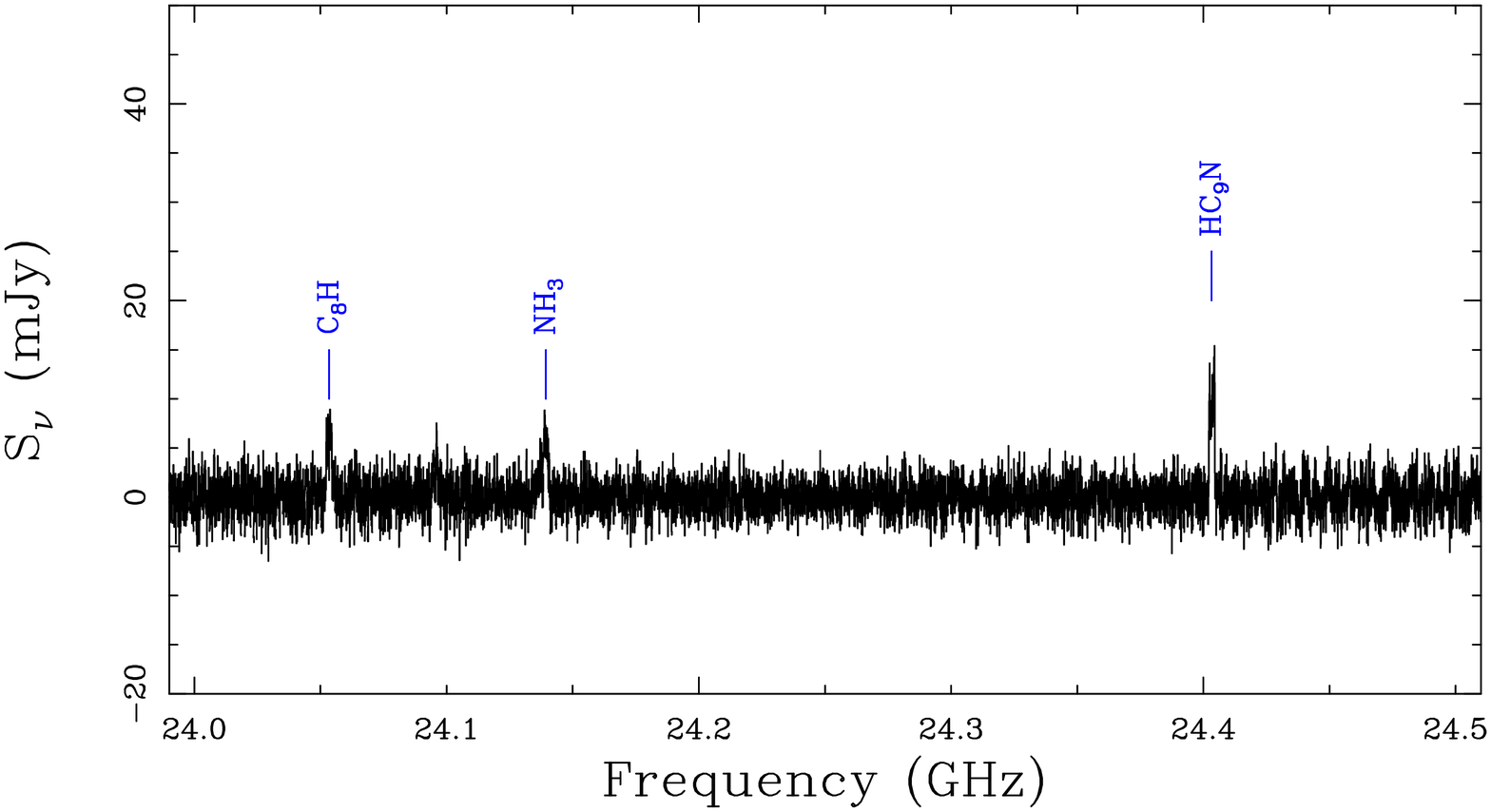}
\includegraphics[width = 0.8 \textwidth]{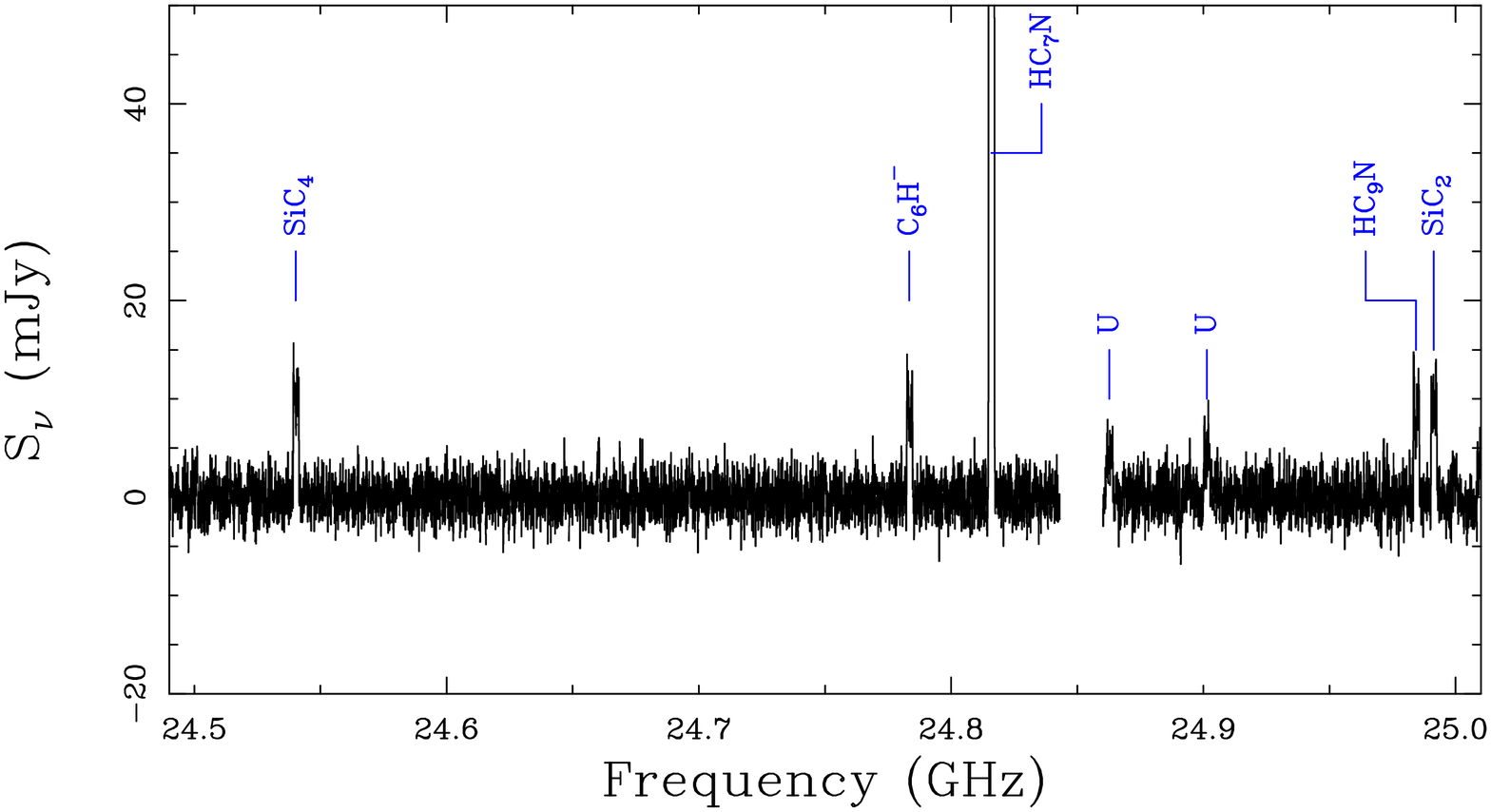}
\centerline{Fig. \ref{Fig:irc}. --- Continued.}
\end{figure*}

\begin{figure*}[!htbp]
\centering
\includegraphics[width = 0.8 \textwidth]{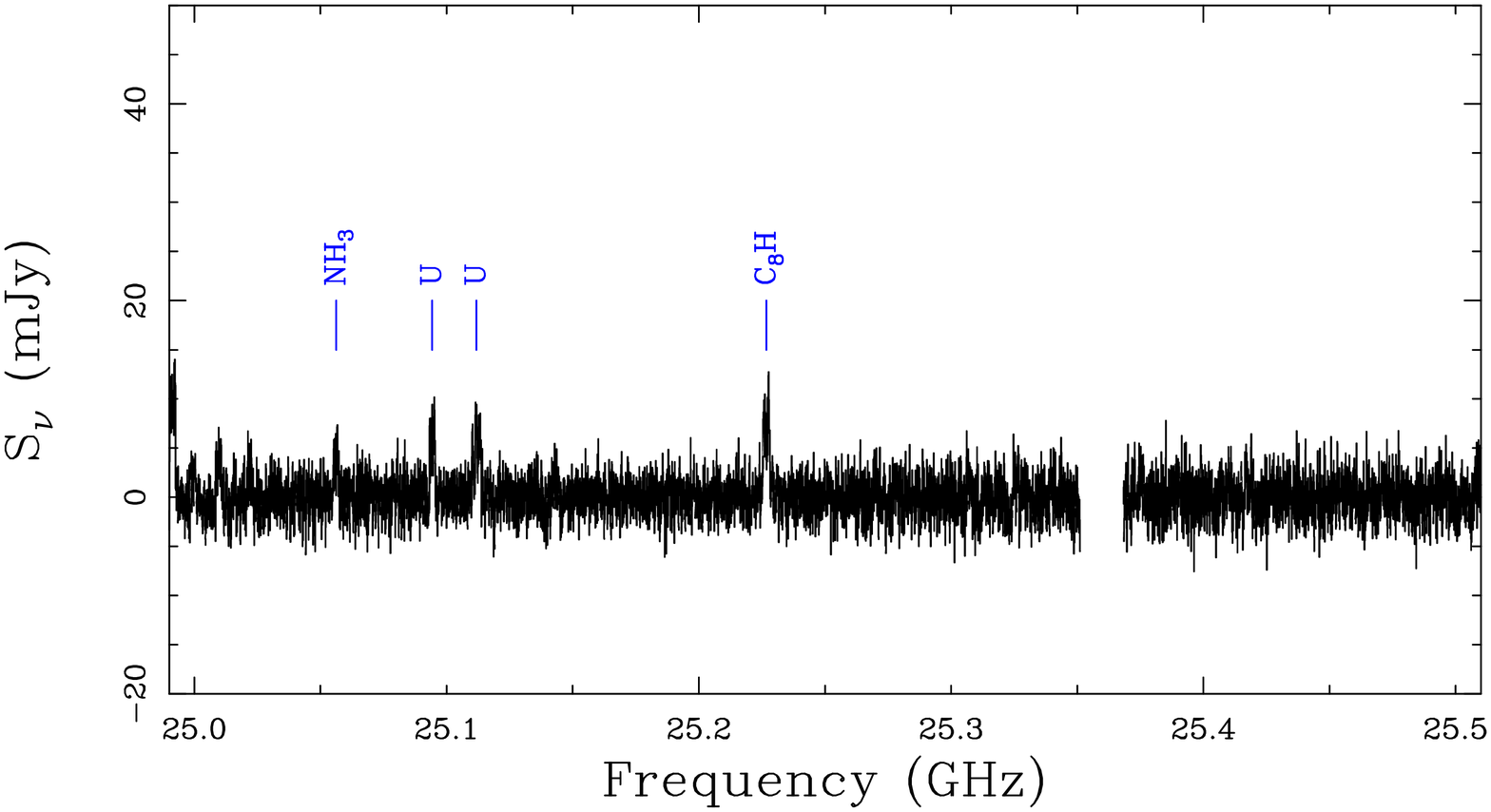}
\includegraphics[width = 0.8 \textwidth]{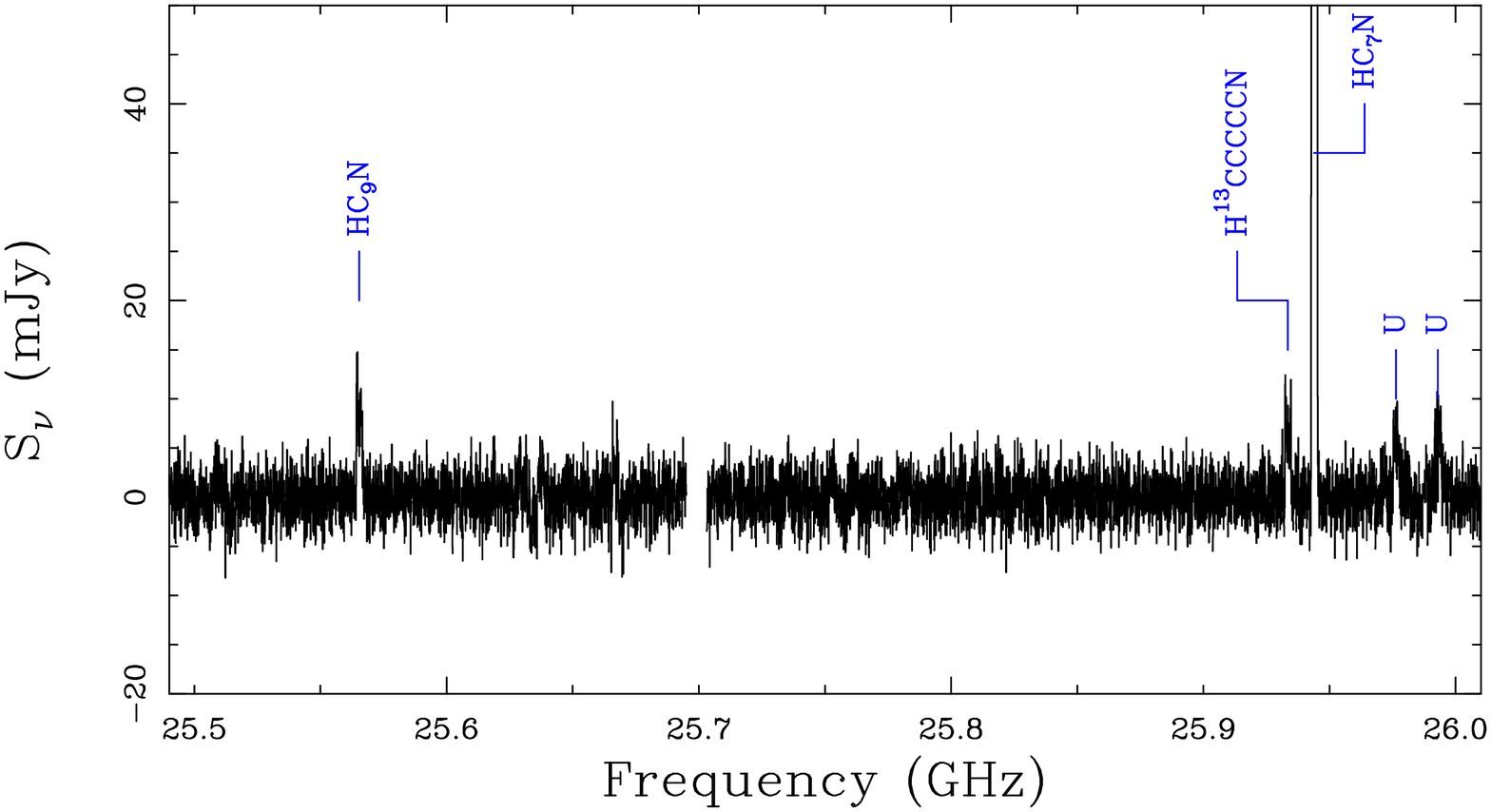}
\includegraphics[width = 0.8 \textwidth]{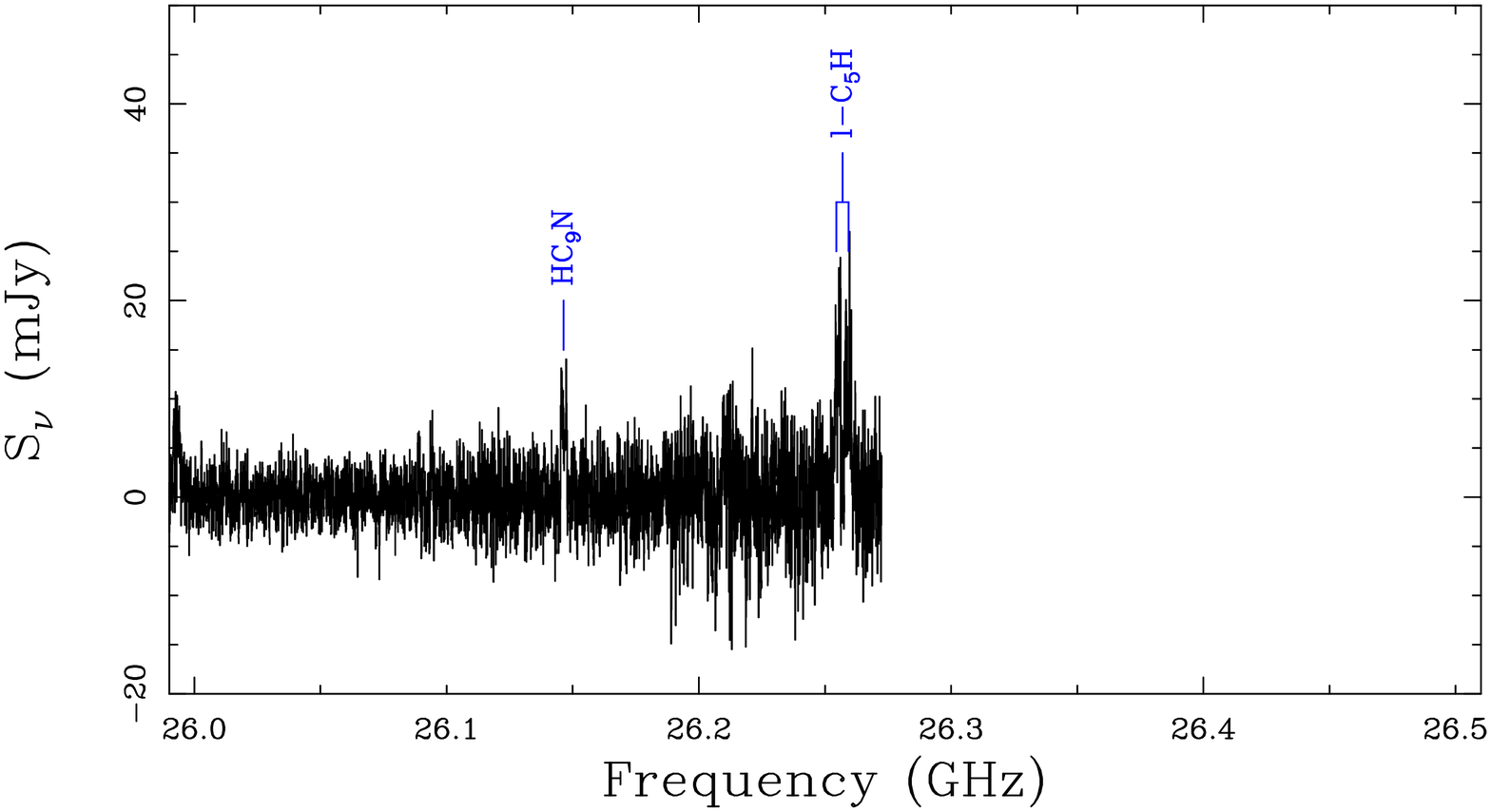}
\centerline{Fig. \ref{Fig:irc}. --- Continued.}
\end{figure*}

\begin{figure*}[!htbp]
\centering
\includegraphics[width = 0.9 \textwidth]{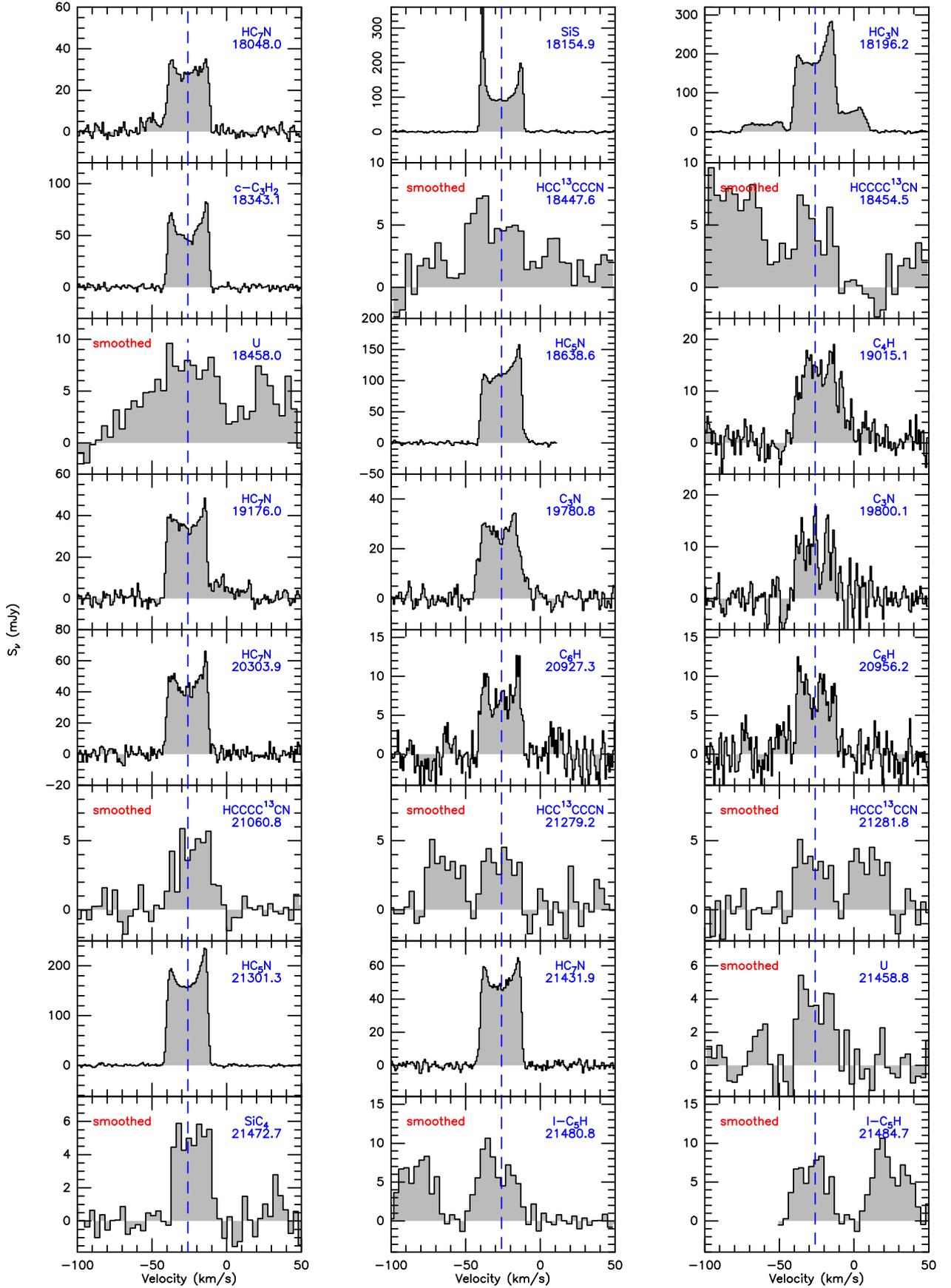}
\caption{{Zoom in all detected and tentative lines. The rest frequency in MHz and the assigned molecule of each line are shown in the upper right of each panel. Weak lines have been smoothed to have a channel width of $\sim$3.3~\kms, and they are marked with ``smoothed'' in the upper left of the corresponding panels. Otherwise, the channel width is $\sim$0.8~\kms (see Sect.~\ref{line}). The blue dashed line traces the systemic LSR velocity ($-$26.0~\kms) of IRC +10216. In the C$_{6}$H 23565.2~MHz panel, the other blue dashed line near $-$50~\kms\,refers to the C$_{6}$H's hfs component at 23567.2~MHz.}\label{Fig:zoomv1}}
\end{figure*}

\begin{figure*}[!htbp]
\centering
\includegraphics[width = 0.9 \textwidth]{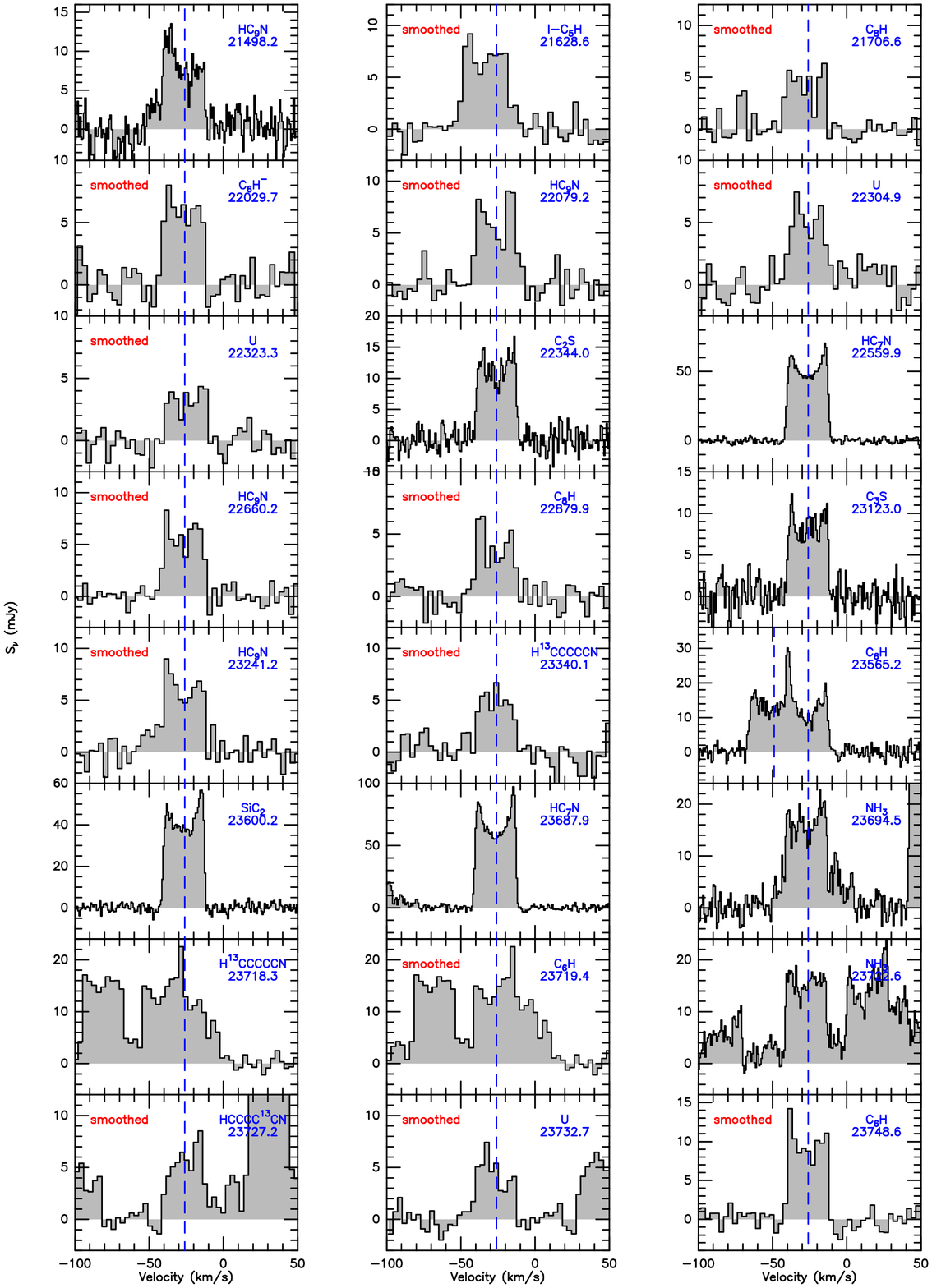}
\centerline{Fig. \ref{Fig:zoomv1}. --- Continued.}
\end{figure*}

\begin{figure*}[!htbp]
\centering
\includegraphics[width = 0.9 \textwidth]{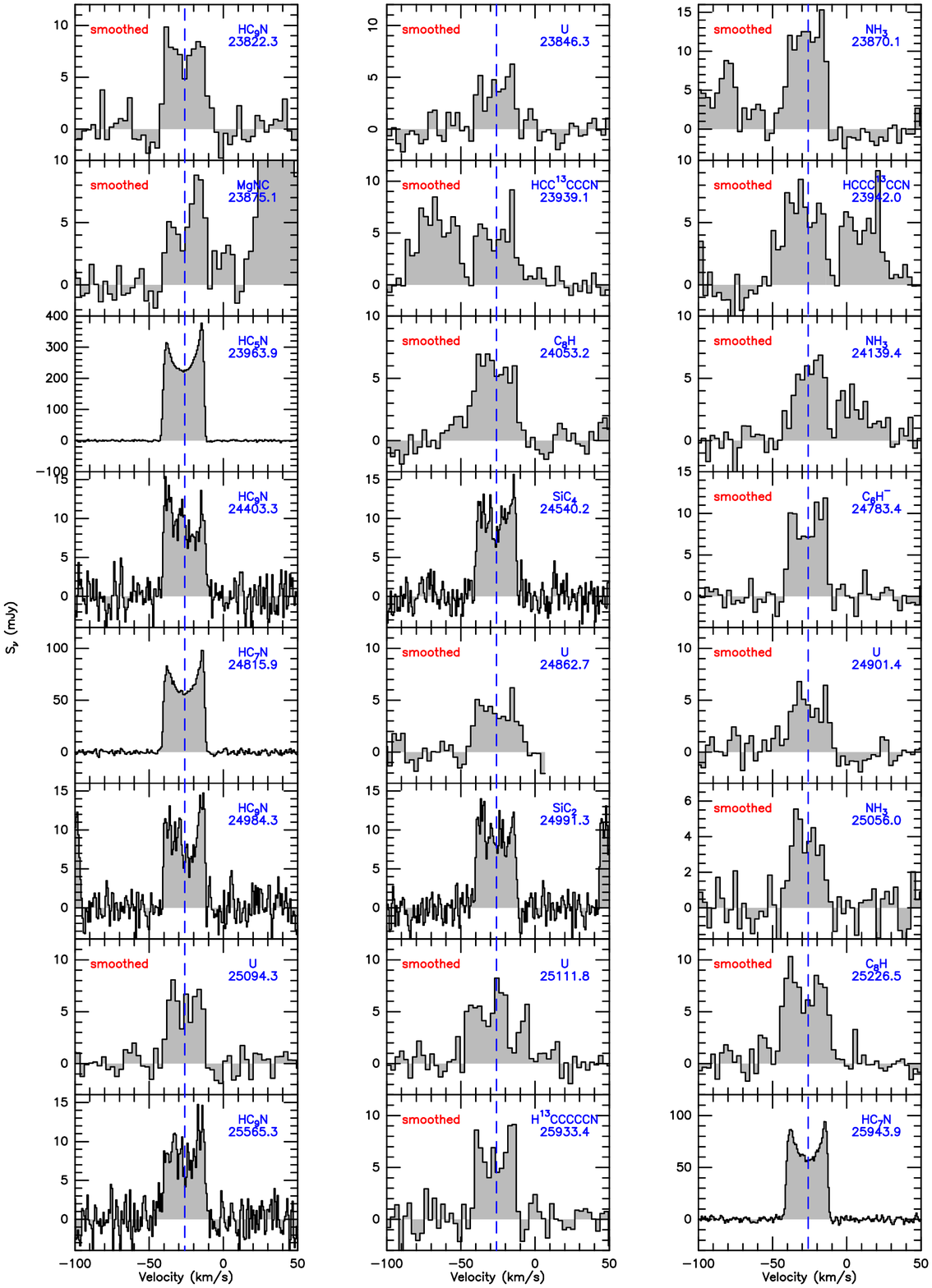}
\centerline{Fig. \ref{Fig:zoomv1}. --- Continued.}
\end{figure*}

\begin{figure*}[!htbp]
\centering
\includegraphics[width = 0.9 \textwidth]{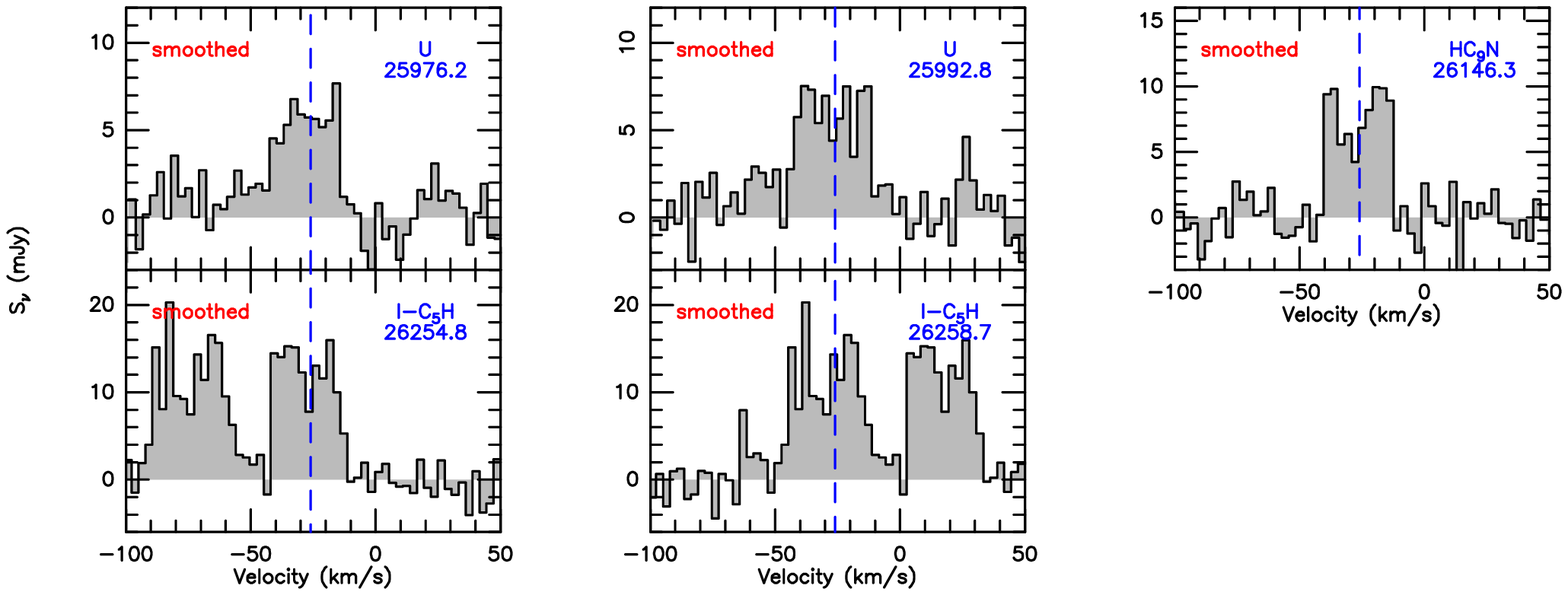}
\centerline{Fig. \ref{Fig:zoomv1}. --- Continued.}
\end{figure*}
\end{appendix}

\end{document}